\documentclass{aa}
\usepackage{natbib}
\bibpunct{(}{)}{;}{a}{}{,}
\usepackage{pifont}
\newcommand{\xmark}{\ding{55}}
\usepackage{orcidlink}
\usepackage{etoolbox}
\usepackage{graphicx}
\usepackage{txfonts}
\usepackage{hyperref}
\usepackage{caption}

\usepackage{subcaption}
\usepackage{lscape} 
\usepackage{placeins} 

\makeatletter
\renewcommand*\aa@pageof{, page \thepage{} of \pageref*{LastPage}}
\makeatother

\begin{document}

   \title{Revisiting the unification of tidal disruption events with polarimetry}

   \author{H.~C.~I.~Wichern \orcidlink{0009-0004-1442-619X}
   \inst{1}\fnmsep\thanks{hawic@space.dtu.dk} 
  \and
  G.~Leloudas \orcidlink{0000-0002-8597-0756}
  \inst{1}
    \and 
   M.~Pursiainen \orcidlink{0000-0003-4663-4300}
   \inst{2} 
    \and 
   A.~Cikota \orcidlink{0000-0001-7101-9831}
   \inst{3} 
    \and 
   G.~K.~Jaisawal \orcidlink{0000-0002-6789-2723}
   \inst{1} 
    \and 
   P.~Charalampopoulos \orcidlink{0000-0002-0326-6715}
   \inst{4} 
    \and 
   M.~Bulla \orcidlink{0000-0002-8255-5127}
   \inst{5,6,7} 
    \and 
   L.~Dai \orcidlink{0000-0002-9589-5235}
   \inst{8} 
     \and 
   J.~P.~Anderson \orcidlink{0000-0003-0227-3451}
   \inst{9} 
    \and 
   M.~Gromadzki \orcidlink{0000-0002-1650-1518}
   \inst{10} 
    \and 
   C.~P.~Guti\'errez \orcidlink{0000-0003-2375-2064}
   \inst{11,12} 
    \and 
   T.~E.~Müller-Bravo \orcidlink{0000-0003-3939-7167}
   \inst{13,14} 
    \and 
   M.~Nicholl \orcidlink{0000-0002-2555-3192}
   \inst{15} 
    }

   \institute{DTU Space, Technical University of Denmark, Elektrovej 327/328, DK-2800 Kongens Lyngby, Denmark
    \and
    Department of Physics, University of Warwick, Gibbet Hill Road, Coventry CV4 7AL, UK
    \and
    Gemini Observatory / NSF's NOIRLab, Casilla 603, La Serena, Chile
    \and
    Department of Physics and Astronomy, University of Turku, FI-20014 Turku, Finland
    \and
    Department of Physics and Earth Science, University of Ferrara, via Saragat 1, I-44122 Ferrara, Italy
    \and
    INFN, Sezione di Ferrara, via Saragat 1, I-44122 Ferrara, Italy 
    \and 
    INAF, Osservatorio Astronomico d’Abruzzo, via Mentore Maggini snc, 64100 Teramo, Italy
    \and 
    Department of Physics, University of Hong Kong, Pokfulam Road, Hong Kong
    \and 
    European Southern Observatory, Alonso de C\'ordova 3107, Casilla 19, Santiago, Chile
    \and
    Astronomical Observatory, University of Warsaw, Al. Ujazdowskie 4,00-478 Warszawa, Poland
    \and
    Institut d'Estudis Espacials de Catalunya (IEEC), Edifici RDIT, Campus UPC, 08860 Castelldefels (Barcelona), Spain
    \and
    Institute of Space Sciences (ICE, CSIC), Campus UAB, Carrer de Can Magrans, s/n, E-08193 Barcelona, Spain
    \and
    School of Physics, Trinity College Dublin, The University of Dublin, Dublin 2, Ireland
    \and
    Instituto de Ciencias Exactas y Naturales (ICEN), Universidad Arturo Prat, Chile
    \and
    Astrophysics Research Centre, School of Mathematics and Physics, Queens University Belfast, Belfast BT7 1NN, UK
    }

   \date{Received ... XX, YYYY; accepted ... XX, YYYY}
 
  \abstract
   {\textit{Aims}. Tidal disruptions of stars by supermassive black holes produce emission at different wavelengths, of which the optical emission is of ambiguous origin. A unification scenario of tidal disruption events (TDEs) has been proposed to explain the different classes of X-ray and optically selected events by introducing a dependence on the viewing angle and geometry. This work aims to test the unification scenario among optically bright tidal disruption events using polarimetry.\\
    \textit{Methods}. By studying the optical linear polarisation of nineteen tidal disruption events (of which nine are newly analysed in this work), we place constraints on their photosphere geometry, inclination, and the emission process responsible for the optical radiation. 
    We also investigate how these properties correlate with the relative X-ray brightness of the events, quantified by the $L_X / L_g$ ratio.\\
    \textit{Results}. We find that 14/16 non-relativistic events can be accommodated by the unification model. Continuum polarisation levels of non-relativistic TDEs lie most often in the range $P\sim 1-2$\% (13 events), and for all except one event, remain below 6\%. For those optical TDEs that have multiple epochs of polarimetry, the continuum polarisation levels decrease with time after peak light for 5/10 events, increase for 3/10 events, and stay approximately constant for 2/10 events. When observed after +70 days (7/16 events), they become consistent with zero polarisation within uncertainties (5/7 events). This implies the photosphere geometries of tidal disruption events are at least initially asymmetric and evolve rapidly which, if tracing the formation of the accretion disk, suggests efficient circularisation. 
    The polarisation signatures of emission lines of seven TDEs directly support a scenario in which optical light is reprocessed in an electron-scattering photosphere. TDEs are most often weak in X-rays when significantly polarised. 
    However, a subset of events deviates from the unification model to some extent, suggesting this model may not fully capture the diverse behaviour of TDEs.
    Multi-epoch polarimetry plays a key role in understanding the evolution and emission mechanisms of tidal disruption events.}

   \authorrunning{Wichern et al.}
   \titlerunning{Revisiting the unification of tidal disruption events with polarimetry}
   \maketitle

\section{Introduction}

During a tidal disruption event (TDE), a star in the vicinity of a supermassive black hole gets torn apart by the tidal forces that act on it \citep{1975Hills, 1988Rees}. Radiation is produced at several stages: from the early nozzle shock at the pericentre \citep{2022Bonnerot}, to self-intersection shocks as the stellar debris forms a stream and falls back onto the black hole \citep{2016Jiang}, and the formation of an accretion disk as the stream finally circularises \citep{2009Strubbe}. This radiation is emitted at X-ray and optical/ultraviolet (UV) wavelengths, and is visible from months to years after the event takes place. Strong outflows, winds, and relativistic jets may be launched in addition, capable of powering the radio emission (e.g, \citealt{2025Alexander} and references therein). While their X-ray emission is thought to be accretion-powered, the origin of the early optical/UV emission of TDEs has been a topic of debate for the past decade. In optically selected TDEs, the early optical/UV radiation appears to be emitted from more extended characteristic radii compared to the thermal soft X-rays, \citep{2021Gezari}, whose emission region is spatially consistent with the inner regions of an accretion disk \citep{2024Guolo}. At late times, also the optical/UV component can also be ascribed to a spreading and cooling disk (e.g., \citealt{2019VanVelzen, 2020Jonker, 2023Wen}). For the early component, several origins have been proposed, which can be broadly grouped in two main hypotheses. The first hypothesis is that the optical/UV emission is due to reprocessing of extreme UV (EUV)/X-rays in an envelope of optically thick ejecta surrounding a rapidly formed accretion disk (e.g., \citealt{1997Loeb, 2014Guillochon, 2014Coughlin, 2015Miller, 2016Roth, 2018Dai, 2022Thomsen}). The second hypothesis suggests that the optical/UV photons are emitted in self-intersection shocks as the debris stream circularises \citep{2015Piran, 2016Bonnerot, 2016Jiang, 2017Svirski, 2021Hayasaki}. Variations of these include an origin in collision-induced outflows \citep{2019LuBonnerot, 2023Charalampopoulos}. In this model, X-ray photons are produced by accreting matter falling from the self-intersection point towards the black hole, and subsequently reprocessed and re-emitted as optical/UV photons inside the collision-induced outflow (CIO) that is launched from the intersection point.
\\ 
\indent
In the first scenario, variations in X-ray brightness and the evolution thereof can be attributed to viewing angle effects and a time-dependent opacity of the envelope, whose density is highest near the plane of the accretion disk. X-ray bright TDEs are seen along the unobscured pole of the disk, while TDEs that are observed edge-on first show ultraviolet or optical light, and it is only later on that X-rays are fully revealed as the ejecta become transparent. Such a late-time X-ray (re)brightening has been observed in many optically selected TDEs (e.g., \citealt{2017Gezari, 2020Hinkle, 2020Jonker, 2020Kajava, 2020Shu, 2022Liu, 2024Guolo}). Besides these face-on and edge-on cases, in-between inclinations allow for intermediate variants. This unification model for TDEs was proposed by \citet{2018Dai}, and was later extended to include a time dependence by varying accretion rates \citep{2022Thomsen}.

In the second scenario, the late-time X-rays are caused by delayed accretion disk formation due to inefficient circularisation (e.g., \citealt{2015Shiokawa,2024SteinbergStone}). As the emission region of the UV/optical radiation in TDEs seems to coincide with the self-intersection radius (e.g., \citealt{2017Wevers}), it is plausible that this radiation is produced in self-intersection shocks - independently from the accretion-powered X-ray emission. 

Polarimetry can provide important clues to the origin of the optical emission and of the photosphere geometry of TDEs. Previously, the sample of TDEs that have polarimetric data and which have been published in the literature consisted of seven optical TDEs (OGLE16aaa; \citealt{2018Higgins}, AT~2018dyb; \citealt{2020Holoien, 2022Leloudas}, AT~2019azh; \citealt{2022Leloudas}, AT~2019dsg; \citealt{2020Lee, 2022Leloudas}, AT~2019qiz; \citealt{2022Patra}, AT~2020mot; \citealt{2023Liodakis}, AT~2023clx; \citealt{2024Charalampopoulos, 2025Uno, 2025Koljonen}) and three relativistic TDEs (Swift J164449.3+573451; \citealt{2012Wiersema}, Swift J2058+0516; \citealt{2020Wiersema}, AT~2022cmc; \citealt{2023Cikota}). \citet{2022Leloudas} identified the following similarities between three TDEs: (1) non-zero polarisation degrees at peak light imply aspherical photospheres at early times, (2) at later times the polarisation drops and the photospheres attain an axial symmetry, and (3) the polarisation of the spectral continuum and emission lines are fully consistent with an electron scattering origin. Specifically, the emission lines exhibit a  depolarised core, smoothly connecting to the continuum polarisation level along the wings, as was also observed in two other optical TDEs \citep{2022Patra,2025Uno}. Relatively high levels of polarisation ($\sim$7--8\%) were observed for two of the relativistic TDEs, which could be attributed to synchrotron radiation from forward shocks and the (poorly understood) effects of dust \citep{2012Wiersema,2020Wiersema}. The highest level of polarisation observed was $\sim$25\% \citep{2023Liodakis}; according to the authors, this ruled out a reprocessing origin, but could instead have originated from synchrotron emission due to stream-stream collision shocks. 

There exist more nuclear transients that have shown TDE-like properties. Examples are the extremely luminous ASASSN-15lh \citep{2016Dong,2016Leloudas}, which presented a double-peaked light curve and which showed a sudden increase in polarisation in-between the two peaks (from $\sim$0.5\% to $\sim$1.2\%; \citealt{2020Maund}), and the extreme coronal line emitter AT~2022fpx which showed variable polarisation at peak light \citep{2024Koljonen}. However, even if TDEs, these events are different than classical optical/UV TDEs \citep{2021VanVelzen} or the established relativistic TDEs, and we therefore do not consider them here.

The sample of just ten TDEs already alludes to a diversity in their observed properties. And very recently, new sample studies have been carried out, which favour \citep{2025Jordana-Mitjans} or tentatively disfavour \citep{2025Floris} a reprocessing scenario. To draw more compelling conclusions about the origin of polarisation in TDEs, and if and how it connects to the TDE unification model, a larger sample needs to be studied. 
\\ \\
In this work, we build on an existing sample of TDEs with optical linear polarimetry data by including nine more events. When interpreting the data, we focus specifically on the unification scenario, for which most polarisation predictions exist. In Section \ref{sec:observations} we discuss the reduction of the broadband polarimetry and spectropolarimetry data, and we give an overview of the available multi-wavelength photometry. We then interpret these data in Section \ref{sec:results}, and we discuss them in the broader context of the sample of TDEs with polarimetric data in Sections \ref{sec:Discussion-Sample} and \ref{sec:Discussion-Indiv}. We provide our conclusions and a summary in Section \ref{sec:Conclusion}. 

\section{Observations, data reduction, and analysis}
\label{sec:observations}

We expand on previous samples by analysing data of nine TDEs. We complement this new dataset with data from the literature. For a subset of the literature data the analysis has been conducted by our team in a similar manner (including corrections for the interstellar polarisation (ISP) and host dilution; see below), while for a few TDEs we include the data as published. 

Table \ref{tab:obslog} shows the observation log of the polarimetry of TDEs newly analysed in this work. The phase (column 2) was calculated with respect to peak light, determined based on fits (either a polynomial fit or a Gaussian Rise, Exponential Decay model fit; \citealt{2021VanVelzen}) to the single- or multi-band optical light curves, depending on the sampling of the light curve in each band. Light curves were constructed using forced photometry \citep{2018Masci,2021Shingles} from the Asteroid Terrestrial-impact Last Alert System (ATLAS; \citealt{2018Tonry,2020Smith}) in the $c,o$ bands and the Zwicky Transient Facility (ZTF; \citealt{2019Bellm,2019Graham}) in the $g,r$ bands.

\begin{table*}
\centering
\caption{Observing log of polarimetry and of photometry and spectra used to perform the host corrections.}
    \begin{tabular}{lllll|ll} \hline
     Date & Phase & Telescope / & Grism / filter &  Exposure time & Light curve & Closest TDE spectra \\
      & (days) & instrument &  &  per HWP (s) & photometry & and/or host spectra: \\
      & & & & & & telescope / instrument \\ 
      & & & & & & and observation date \\ \hline \hline

     \textbf{AT~2018hyz} &  &  &  &  & &  \\
      58455 (2018-12-03) & +9 & NOT/ALFOSC & $V$ &  350  & ... & NTT/EFOSC2\\
      & & & & & & (2018-12-03)\\ \hline

     \textbf{AT~2019lwu} &  &  &  &  &  &   \\
     58747 (2019-09-21) & +51 & VLT/FORS2 & $B,V$ &  950, 400 & ZTF & ... \\ \hline

     \textbf{AT~2020zso} &  &  &  &  & &    \\
     59171 (2020-11-18) & --24 & VLT/FORS2 & $B,V,R$ & 150, 100, 100 & ZTF & NTT/EFOSC2\\
       & & & & & & (2020-11-16, 2024-06-06) \\
     59178  (2020-11-25) & --17 & VLT/FORS2 & $B,V,R$  & 300, 300, 180 & ZTF & NTT/EFOSC2\\
       & & & & & & (2020-11-22, 2024-06-06) \\
     59186 (2020-12-03) & --9 & VLT/FORS2 & $B,V,R$  & 200, 150, 150 & ZTF & NTT/EFOSC2 \\ 
       & & & & & & (2020-12-08, 2024-06-06) \\
     59194 (2020-12-11) & --1 & VLT/FORS2 & $B,V,R$  & 150, 100, 75 & ZTF & NTT/EFOSC2\\
       & & & & & & (2020-12-08, 2024-06-06) \\ \hline

     \textbf{AT~2021blz} &  &  &  &  & &  \\
     59252 (2021-02-07) & +5 & VLT/FORS2 & $B,V$ &  120, 80 & ATLAS & NTT/EFOSC2 \\
       & & & & & & (2021-02-04, \\
       & & & & & & 2023-12-19, 2023-12-20) \\
     59257 (2021-02-12) & +10 & VLT/FORS2 & 300V &  2$\times$1000 & ATLAS & NTT/EFOSC2 \\
       & & & & & & (2023-12-19, 2023-12-20) \\
     59263 (2021-02-18) & +16 & VLT/FORS2 & $B,V,R$ & 220, 80, 120 & ATLAS & NTT/EFOSC2 \\
       & & & & & & (2021-02-15,  \\
       & & & & & & 2023-12-19, 2023-12-20) \\
     59279 (2021-03-06) & +32 & VLT/FORS2 & 300V & 2$\times$1000 & ATLAS & NTT/EFOSC2 \\ 
       & & & & & & (2023-12-19, 2023-12-20)\\ \hline

     \textbf{AT~2022bdw} &  &  &  &  &  &  \\
     59640  (2022-03-02) & +2 & VLT/FORS2 & 300V &  3$\times$780  & ATLAS, ZTF & NTT/EFOSC2 (2024-03-13) \\
     59905 (2022-11-22) & +267 & VLT/FORS2 & $B,V,I$  & 160, 110, 110 & ... & ... \\ \hline
     
     \textbf{AT~2022dsb} &  &  &  &  &  &  \\
     59645 (2022-03-07) & +2 & VLT/FORS2 & 300V & 3$\times$780 & ATLAS, ZTF & NTT/EFOSC2 (2024-03-12) \\
     59666 (2022-03-28) & +23 & VLT/FORS2 & 300V & 3$\times$780 & ATLAS, ZTF & NTT/EFOSC2 (2024-03-12) \\
     59758 (2022-06-28) & +115 & VLT/FORS2 & $B,V,R,I$ & 150 100, 100, 100 & ... & ... \\ \hline
     
     \textbf{AT~2022exr} &  &  &  &  & &  \\
     59844 (2022-09-22) & +169 & VLT/FORS2 & $V$ & 400 & ZTF & ... \\ \hline

     \textbf{AT~2022hvp} &  &  &  &  &  &  \\
     59712 (2022-05-13) & +16 & NOT/ALFOSC & $V$ & 400 & ATLAS, ZTF & ... \\ \hline

     \textbf{AT~2023mhs} & &  &  &  & &  \\
     60139 (2023-07-14) & +4 & NOT/ALFOSC & $V,R$ & 200 & ATLAS, ZTF & Keck/LRIS (2023-07-18)\\
      & &  & &  &  & APO/SDSS (2007-05-12) \\
     60144 (2023-07-19) & +9 & VLT/FORS2 & 300V & 2$\times$660 & ATLAS, ZTF & APO/SDSS (2007-05-12) \\ \hline

    \end{tabular}
\tablefoot{Columns: (1) Modified Julian Date (Universal Time) of observations, (2) phase with respect to observed maximum light, (3) telescope and instrument used for polarimetry observations, (4) grism or filter used, (5) exposure time per HWP angle (corresponding to the respective filter bands in column (4) in case of broadband polarimetry), (6) photometry used to scale the flux spectrum and host spectra to perform the host correction, and (7) telescope and instrument used to obtain the transient and host spectra, along with the observation date. For the spectropolarimetric observations, only the host spectrum is given. The referenced spectra of AT~2018hyz were already scaled and host-subtracted by \citet{2020Short,2020Gomez}. In the case of AT~2022exr, the phase corresponds to the first peak in its light curve.}
\label{tab:obslog}
\end{table*}

Polarimetry of seven TDEs was obtained with the 8.2-m Very Large Telescope Focal Reducer and Low Dispersion Spectrograph (VLT/FORS2; \citealt{1998Appenzeller}); four of these have both broadband polarimetry and spectropolarimetry data, the others have broadband polarimetry data only. In addition, single epochs of broadband polarimetry with the Alhambra Faint Object Spectrograph and Camera mounted on the 2.56-m Nordic Optical Telescope (NOT/ALFOSC) were obtained for three TDEs (58-005; PI: Leloudas, 65-030; PI: Charalampopoulos, 67-009; PI: Pursiainen).

\subsection{Data reduction -- broadband polarimetry}
\label{sec:red_impol}
VLT/FORS2 and NOT/ALFOSC broadband polarimetry data were reduced using custom pipelines written in Python (see \citealt{2023Pursiainen} for details of a similar reduction routine). The pipeline utilises the \textsc{photutils} \citep{2024Bradley} package to perform aperture photometry in the ordinary and extraordinary beams. We varied the aperture radii to be 1.0, 1.5, 2.0, 2.5, 3 times the average full width at half-maximum (FWHM) of the identified field stars in each image (e.g., \citealt{2015Leloudas, 2017Leloudas}). Generally, we find that varying the aperture size causes small differences in the normalised Stokes $q,u$ parameters, however, their respective values are still consistent within the error margins. The signal-to-noise ratio (S/N) tends to be highest for aperture sizes in the range 1.5--2.5$\times$FWHM, and we thus adopt an aperture size of 2$\times$FWHM as a default. The background is removed using the \texttt{Background2D} class in \textsc{photutils}.

From the extracted fluxes of the ordinary and extraordinary beams in all images corresponding to the four HWP angles $\theta_{\text{HWP}} = 0.0^{\circ}, 22.5^{\circ}, 45.0^{\circ}$ and $67.5^{\circ}$, the normalised Stokes parameters $q,u$ and, subsequently, the polarisation $P$ and polarisation angle $\theta$ and their uncertainties are calculated following the equations given by \cite{2006PatatRomaniello}. For all measurements of the Stokes $q,u$ parameters, we correct for the chromatism of the half-wave plate (see also the FORS2 manual\footnote{\url{https://www.eso.org/sci/facilities/paranal/instruments/fors/doc.html}}) and we apply an instrumental correction following \cite{2020GonzalezGaitan}. Finally, we correct for the polarisation bias using the generalised modified asymptotic estimator proposed by \citet[][; see their equation 37]{2014Plaszczynski}.

\subsection{Data reduction -- spectropolarimetry}
The spectropolarimetry data were reduced using a custom Python pipeline. The pipeline performs basic reductions such as bias correction, background removal (using a low-order polynomial fit to regions adjacent to the extraction region), and cosmic ray removal (using the \textsc{Astro-SCRAPPY} package; \citealt{2018McCully}). Flat fields are omitted, as the effects of pixel-to-pixel variations are resolved for the most part due to the redundancy of multiple HWP angles (e.g., \citealt{2006PatatRomaniello, 2020GonzalezGaitan}). The pipeline carries out wavelength calibration using arc lamp spectra obtained during the same night, and achieves a typical calibration accuracy of $\lesssim$0.3\,Å (for the 300V grism used here). The wavelength solution is obtained by combining the solutions for the ordinary and extra-ordinary beam within one image and interpolating them to a common wavelength grid. The spectra from the ordinary and extraordinary beam are typically extracted within a region of 2$\times$FWHM from the centre of each beam (which changes along the dispersion axis). Extraction of these spectra is done through optimal extraction (e.g., \citealt{1986Horne}) within the same aperture region for both beams. The extracted spectra are subsequently binned by 25\,Å to increase the S/N per bin, and a wavelet transform is applied to minimise noise (see Appendix \ref{sec:AppendixB}). Such wavelet transforms have been successfully applied to spectra of supernovae in the past \citep{2010Wagers, 2019Cikota}. Our pipeline has been tested on both polarised and unpolarised standard stars \citep{2016Cikota} and previous datasets of TDEs \citep{2022Leloudas}, yielding good agreement with results obtained via e.g., standard \texttt{IRAF} \citep{1986Tody} procedures. 
\\ \\ 
The TDEs in our sample have multiple cycles over $\theta_{\text{HWP}}$ in a single night; to enhance the signal-to-noise ratio, we stack these cycles together by calculating the variance-weighted average of the extracted spectra from all cycles for a given angle $\theta_{\text{HWP}}$ (0$^{\circ}$, 22.5$^{\circ}$, 45$^{\circ}$, or 67.5$^{\circ}$). The normalised $q,u$ spectra are then calculated following the equations by \citet{2006PatatRomaniello}, and the chromatic correction and polarisation bias correction are applied as described in Sect. \ref{sec:red_impol}\footnote{For spectropolarimetry data, we calculate the polarisation bias using the correlation $\rho$ between the uncertainties on $q$ and $u$ \citep{2014Plaszczynski}. We note that $\sigma_{q}$ and $\sigma_{u}$ were found to be strongly correlated in most cases.}.

\subsection{Correction for interstellar polarisation}
We corrected our polarisation measurements for the Galactic (Milky Way) ISP due to dust grains along the line of sight in the direction of the TDE. In doing so, we assume the stars in the (projected) vicinity of the target used to estimate the ISP are intrinsically unpolarised. We verified that the sources identified by the \textsc{DAOFIND} algorithm \citep{1987Stetson} implemented in \textsc{photutils} were in fact stars by cross-matching them with the \textit{Gaia} Data Release 2 \citep{GaiaDR2, Gaia} catalogue. We calculated a weighted average of the polarisation values of all field stars with sufficient signal-to-noise ratio (S/N$\geq200$), yielding an estimate of $q_{\text{ISP}}$ and $u_{\text{ISP}}$ in each filter. In the case of broadband polarimetry, we then correct the polarisation of the TDE by subtracting $q-q_{\text{ISP}}$ and $u-u_{\text{ISP}}$. For the TDEs that have both broadband polarimetry and spectropolarimetry data, we correct the $q(\lambda),u(\lambda)$ spectra by fitting the broadband ISP measurements with a Serkowski law $P(\lambda) = P_{\text{max}}\cdot\text{exp}\left({-K\text{ln}^2(\lambda_{\text{max}}/\lambda)}\right)$, where $P_{\text{max}}$ is the maximum polarisation at wavelength $\lambda_{\text{max}}$ and $K$ is an empirical constant \citep{1975Serkowski}. Our ISP estimates, more details on the Serkowski law fits, and additional verifications are provided in Appendix \ref{sec:AppendixA}. The ISP spectrum $P_{\text{ISP}}(\lambda)$ is then converted to $q_{\text{ISP}}(\lambda)$ and $u_{\text{ISP}}(\lambda)$ as
\begin{equation}
\begin{split}
    q_{\text{ISP}}(\lambda) = \text{cos}(2\bar{\theta})\cdot P_{\text{ISP}}(\lambda) \\
    u_{\text{ISP}}(\lambda) = \text{sin}(2\bar{\theta})\cdot P_{\text{ISP}}(\lambda)
\end{split}
\end{equation}
where $\bar{\theta}$ is the weighted average ISP position angle in radians. 

\subsection{Correction for host galaxy contamination}
Another correction required is that of the underlying host galaxy. We define the host contribution to the flux as the flux ratio of the host light and the total light of both the TDE and its host (e.g., \citealt{2022Leloudas}):
\begin{equation}
\alpha(\lambda) = I_{\text{host}}(\lambda)/I_{\text{tot}}(\lambda) = I_{\text{host}}(\lambda)/(I_{\text{TDE}}(\lambda) + I_{\text{host}}(\lambda))
\end{equation}
The ISP-corrected polarisation spectra $q(\lambda),u(\lambda)$ can be corrected for the host contribution $\alpha(\lambda)$ by dividing them by a factor $(1-\alpha(\lambda))$; $q(\lambda)/(1-\alpha(\lambda))$, $u(\lambda)/(1-\alpha(\lambda))$. Equivalent expressions can be used to obtain the broadband host contribution $\alpha(B,V,R,I)$, where $B,V,R,I$ are the broadband filters, to correct the broadband polarimetry data. When using these formulas, it is assumed that the host is intrinsically unpolarised (e.g., \citealt{2008Andruchow}); for AT~2022bdw and AT~2022dsb, which have late-time broadband polarimetry that likely consists of almost purely the ISP and host light\footnote{We note that for this reason, we do not attempt to correct these epochs of AT~2022bdw and AT~2022dsb for any host contamination.}, we could verify that the ISP-corrected polarisation degrees are indeed consistent with zero.
\\ \\
When estimating the host contribution $\alpha$, we distinguish three cases:
\\
Case I: a TDE has spectropolarimetry data: to obtain $\alpha(\lambda)$, we divide a spectrum of the host galaxy by the (host-contaminated) TDE flux spectrum from the spectropolarimetry data itself. This is the case for four TDEs.
\\
Case II: a TDE has broadband polarimetry data, and there are spectra of the host and of the TDE (close in time to the epoch of polarimetry) available: we obtain $\alpha(\lambda)$ similar to Case I, and then convert this into $\alpha(B,V,R,I)$ by convolving with the respective filter response function. This is the case for two TDEs.
\\
Case III: a TDE has only broadband polarimetry, and no suitable (host) spectra to estimate $\alpha(\lambda)$: we obtain $\alpha$ in other broadband filters (such as $g$, $r$, $c$, and $o$) using photometry of the TDE and host, and convert this into $\alpha(B,V,R,I)$. This is the case for three TDEs (AT~2019lwu, AT~2022exr, and AT~2022hvp).
\\ \\
In all three cases, we require reliable estimates of the host magnitude, and of the magnitude of the pure TDE light at the time of polarimetry. Spectra in Cases I and II need to be scaled with photometry to account for (differential) slit losses. For the pure-TDE photometry, multi-band differential Point Spread Function (PSF) magnitudes are available from the ATLAS \citep{2021Shingles} and ZTF \citep{2018Masci} forced photometry servers, and we use these since most TDE light curves are best sampled by observations from these two surveys. In some cases, we interpolate to the epochs of polarimetry by fitting the light curves. Obtaining reliable photometry of the host galaxies is more complicated, however, due to their extended nature. More details are provided in Appendix \ref{sec:AppendixA}. 

\subsection{Analysis in the q,u plane}
\label{sec:intermezzo}

One way to study the geometry and (spherical or axial) symmetry of the TDE photosphere, is to study the structure in the Stokes $q,u$ plane. As discussed in the review by \citet{2008WangWheeler} (see also \citealt{2001Leonard}), one can decompose the $q,u$ parameters into two components: one component that lies along the so-called dominant axis, and one component perpendicular to the dominant axis (referred to as the orthogonal axis). The polarisation projected onto the dominant axis carries information on the global deviations from spherical symmetry, while the orthogonal axis encodes the deviations from axial symmetry. In the $q,u$ plane, a global axial symmetry presents itself as a line fit to the dominant axis passing through the origin. Conversely, if the dominant axis does not pass through the origin, a global deviation from axial symmetry is implied. 
\\
\indent
Following \citet{2010Maund}, we carry out a principal component analysis in addition to a linear least-squares fit to the dominant axis, which are two complementary techniques. From the principal component analysis we infer the rotation angle between the dominant axis and the $q$-axis $\theta_{\text{rot}}$, the weighted centroid values $\bar{Q},\bar{U}$, and the axial ratio $b/a$ of the ellipse that captures difference in spread along the dominant axis and spread along the orthogonal axis, where null polarisation (consistent with a perfectly spherical geometry) corresponds to a ratio $b/a$~=~0. We note that the data do not always allow for a good linear fit. These cases are shown without a dominant axis fit in Appendix \ref{sec:AppendixC}.

\subsection{X-ray observations}
To aid the interpretation of our results in a broader context, we have collected the available X-ray data of the TDEs in our sample. These include data from literature (referenced in Sect. \ref{sec:Discussion-multi-wavelength}), but also observations that have not yet been published.
The latter include observations from the Neutron Star Interior Composition Explorer (NICER; \citealt{2012Gendreau,2014Arzoumanian}), the Chandra X-ray Observatory (Chandra; \citealt{2002Weisskopf}), the Neil Gehrels Swift Observatory (Swift; \citealt{2004Gehrels}), and XMM-Newton \citep{2001Jansen}, whose data reductions are described in Appendix \ref{sec:AppendixD}. The data are presented in Table \ref{tab:X-rays}.

\section{Results}
\label{sec:results}

In this section, we present our results for new TDEs studied in this work. We first focus on TDEs for which spectropolarimetry is available in Section \ref{sec:specpol}, as these data allow us to extract more detailed information on the geometry and emission properties of the TDEs. Events with broadband polarimetry only are presented in Section \ref{sec:impol}. We save the interpretation of these results for the discussion section, in the context of the entire TDE population.

\subsection{TDEs with Spectropolarimetry }
\label{sec:specpol}

\subsubsection{AT~2021blz}

\begin{figure*}
\centering
	\includegraphics[width=0.9\columnwidth]{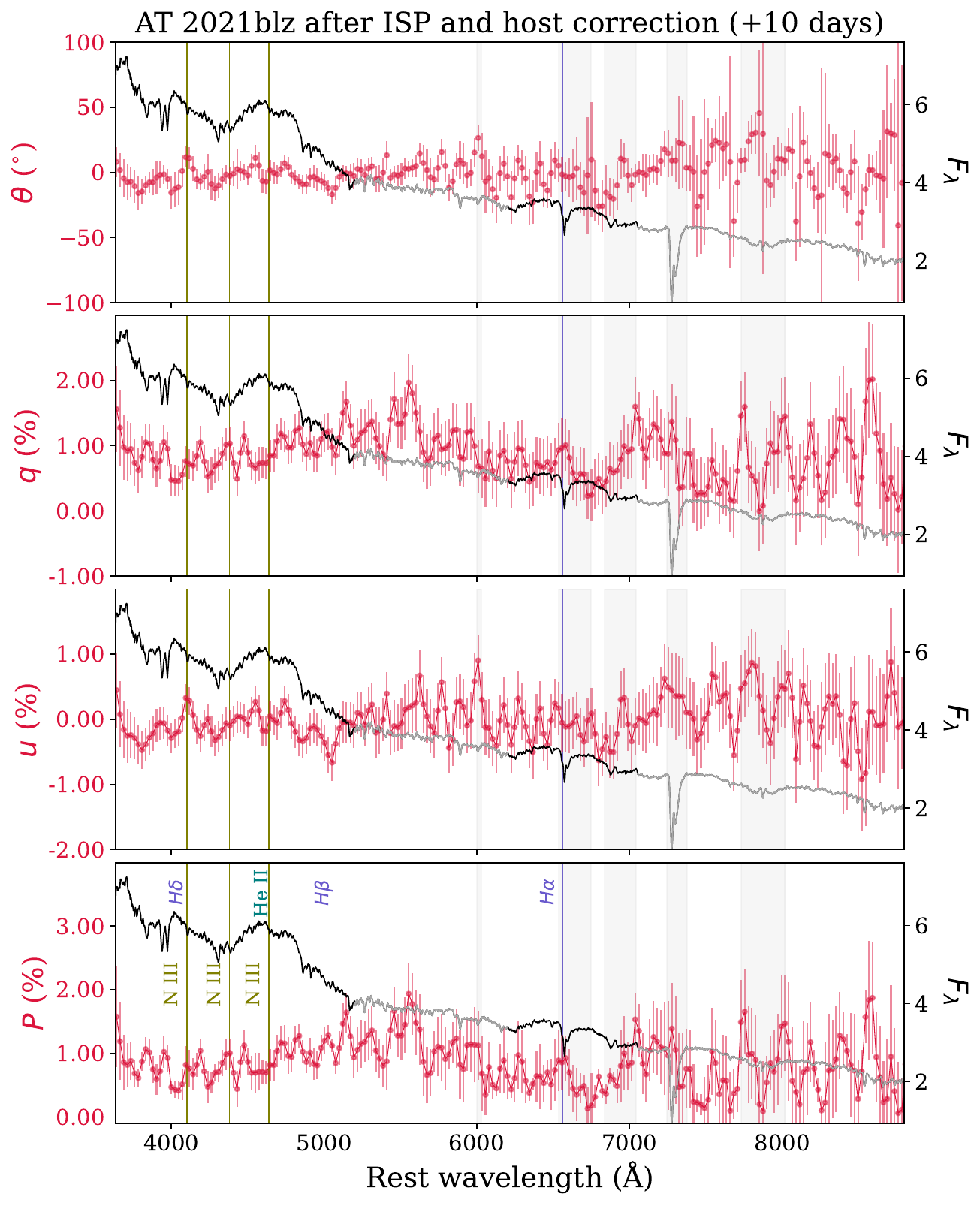}
	\includegraphics[width=0.9\columnwidth]{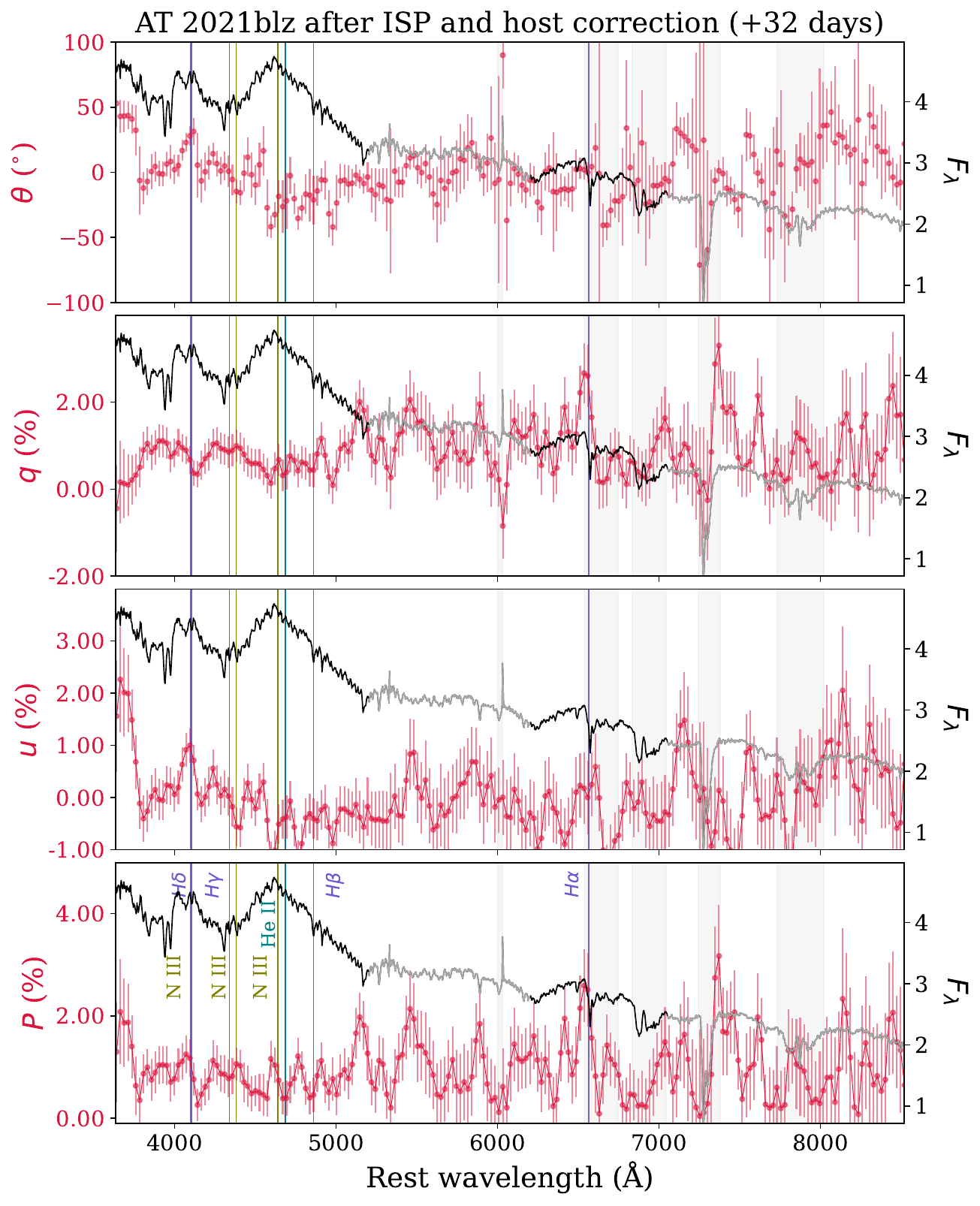}
\caption{Polarisation spectra of two epochs of AT~2021blz after correcting for the ISP and the host galaxy dilution. In each panel, left-hand axes refer to the polarisation angle $\theta$, the Stokes parameters $q$ and $u$, and the polarisation degree $P$ (corrected for the polarisation bias), all shown in red. The right-hand axes refer to the flux spectra at the time of polarimetry (in units of $10^{-16}$ erg cm$^{-2}$ s$^{-1}$ Å$^{-1}$), which are shown in black across emission lines and in grey across the assumed continuum. Regions of strong telluric absorption are marked as grey shaded regions. The identified line species are shown as vertical coloured lines, e.g., the Balmer lines (dark blue), N~III emission lines (olive), and the He~II emission line (green).}
\label{fig:specpol_AT2021blz}
\end{figure*}

We have analysed two epochs of spectropolarimetry at +10 and +32 days after the TDE peak, supplemented by two epochs of broadband polarimetry in the $B,V (R)$ bands at +5 and +16 days after peak. During the first epoch of spectropolarimetry, the flux spectrum of AT~2021blz contains broad H$\alpha$ and H$\beta$ emission lines, although the latter is strongly blended with the He~II~4686\,Å complex that also contains N~III lines, which appear blue-shifted (see also \citet{2021TNSTerwel}). In the polarisation spectrum, we see a broad but shallow depolarisation feature at the position of the H$\alpha$ line compared to the continuum polarisation (Fig.\ref{fig:specpol_AT2021blz}). At the position of the He~II 4686\,Å complex, we see a similar broad depolarised region, but polarisation features belonging to individual emission lines in this complex are not readily distinguishable. We measure a $V$-band polarisation (obtained by convolving the polarisation spectrum with the filter transmission) of $P_{V}$ = 1.50$\pm$0.09\%.

During the second epoch of spectropolarimetry at +32 days, the He~II 4686\,Å line has blended even more with the H$\beta$ line, which now appears as an extended red wing. Over the course of 22 days, the continuum polarisation has not decreased significantly, with the $V$-band polarisation now being 1.34$\pm$0.14\%, and there is no change in the polarisation angle. The apparent depolarised regions around the He~II complex and H$\alpha$ line persist, although a lower S/N and possible effects of telluric absorption affect the polarisation spectra during this epoch, complicating the identification of polarisation features.

Due to strong scatter in $q$ and $u$, especially during the second epoch, we do not find a good dominant axis fit during either epoch (Fig. \ref{fig:specpolpca_appendix}). A clear offset from the origin and values of $b/a>$~0 do present an absence of spherical symmetry. We find that the rotated Stokes parameter $u_{\text{rot}}$ is overall negative during both epochs, rather than evenly distributed around the axis $u_{\text{rot}}$ = 0\%, indicating a departure from axial symmetry.

\subsubsection{AT~2022dsb}
\begin{figure*}
\centering
	\includegraphics[width=0.9\columnwidth]{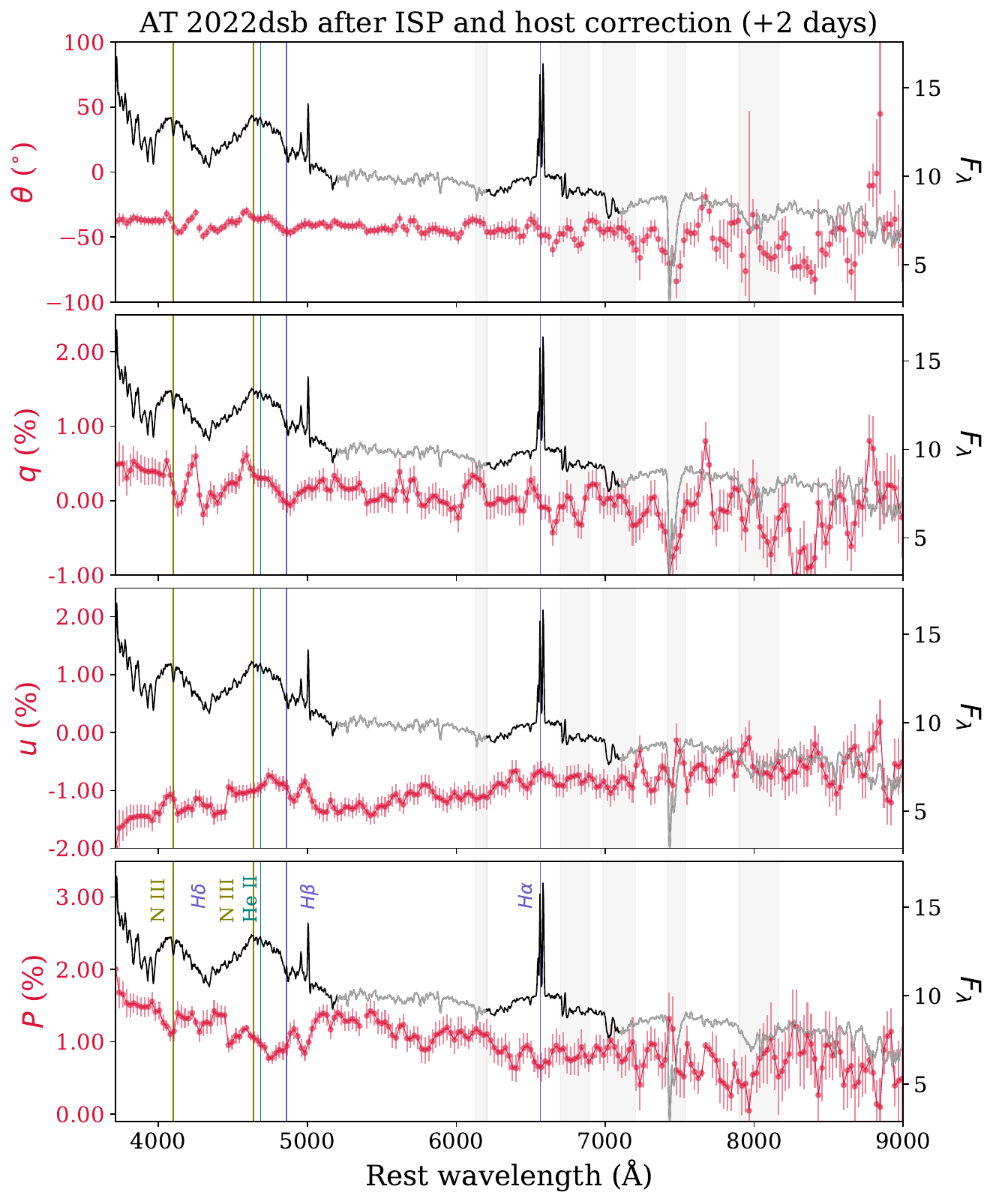}
	\includegraphics[width=0.9\columnwidth]{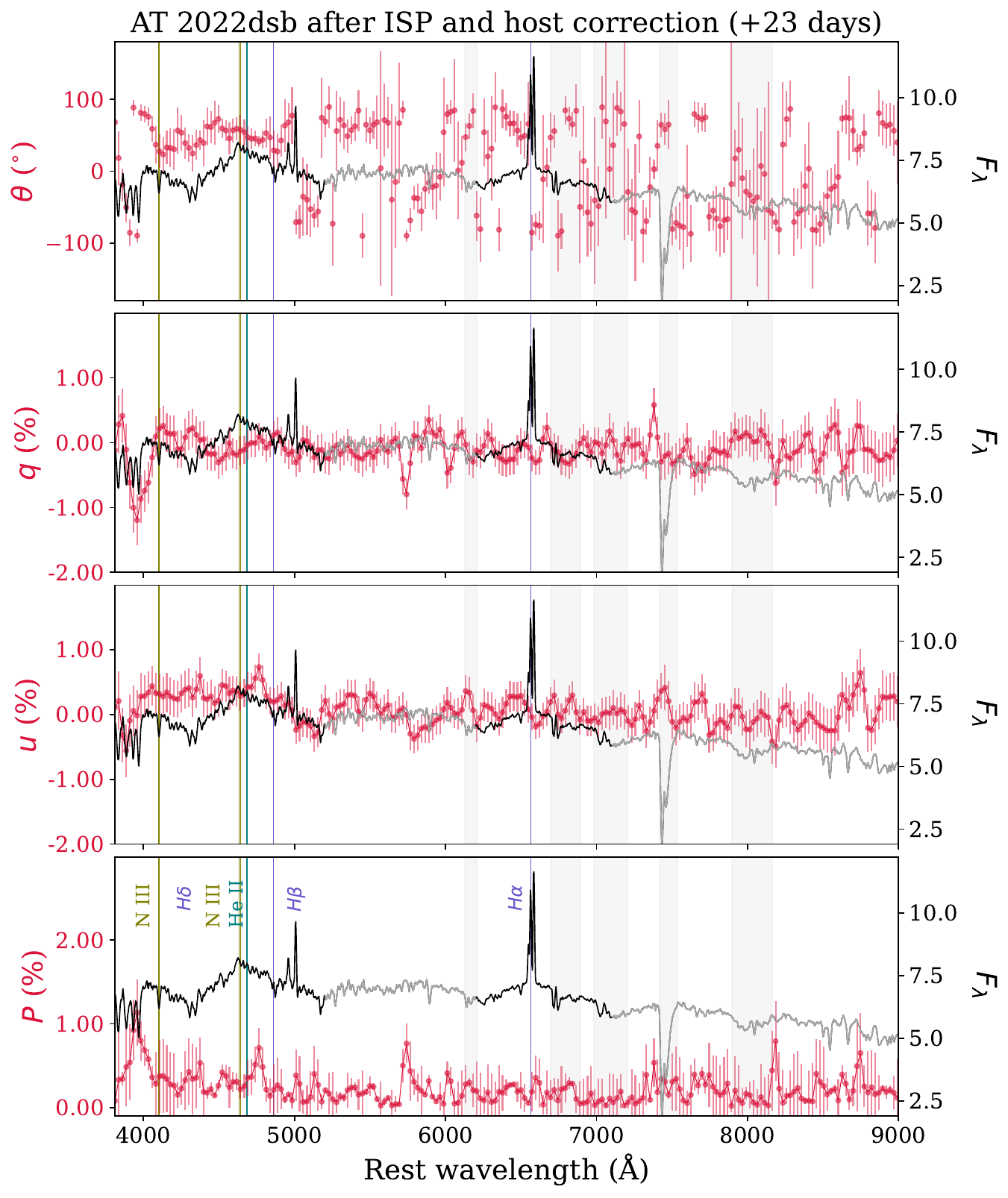}
\caption{Same as Figure \ref{fig:specpol_AT2021blz} but for AT~2022dsb.}
\label{fig:specpol_AT2022dsb}
\end{figure*}

AT~2022dsb is an H+He TDE characterised by several outflow signatures in it its early spectra, including a blue-shifted H$\alpha$ line --5 days before maximum light \citep{2024Malyali, 2022TNSFulton}.
We analyse two epochs of spectropolarimetry at +2 and +23 days, supplemented by one late-time epoch of broadband polarimetry ($B,V,R,I$ bands) +115 days post-peak. We assume the late-time broadband polarimetry does not present intrinsic polarisation of the TDE, and we note that it is consistent with null polarisation. The two epochs of spectropolarimetry are plotted in Figure \ref{fig:specpol_AT2022dsb}. During the first epoch, clear depolarisation is visible at the position of the N~III (or possibly H$\delta$), He~II, and H$\alpha$ emission lines in the flux spectrum. 

\begin{figure}[!hbt]
	\includegraphics[width=0.432\columnwidth]{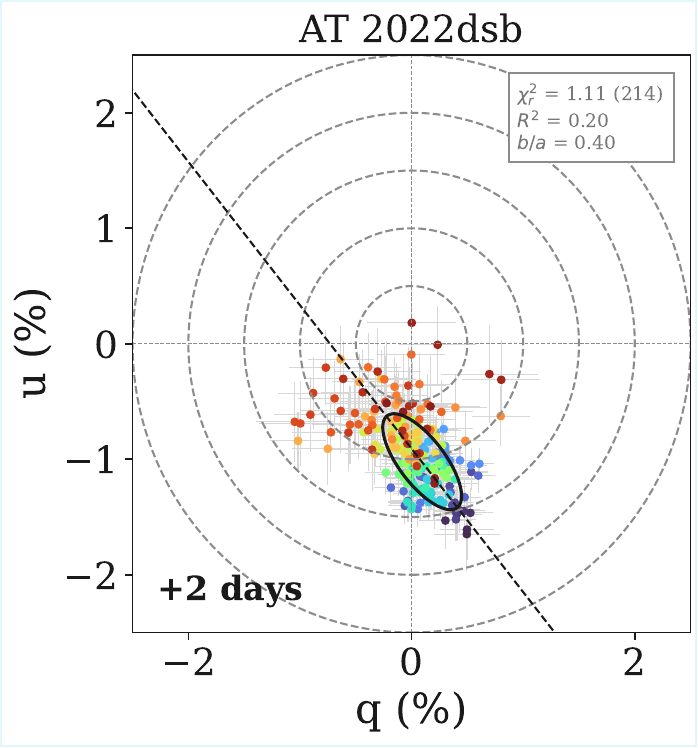}
    \includegraphics[width=0.568\columnwidth]{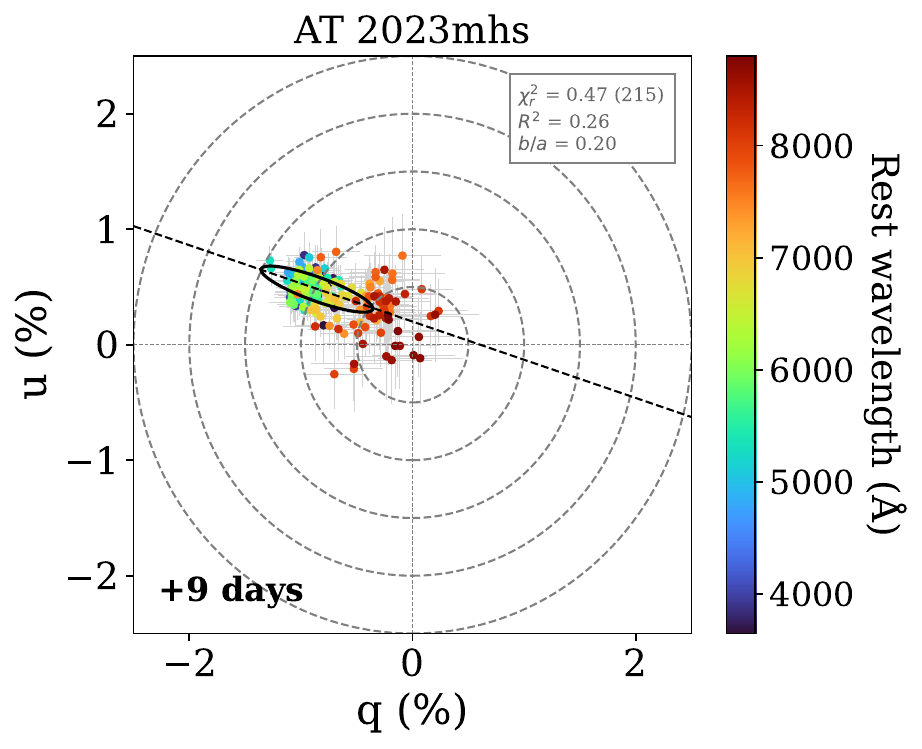}
\caption{Dominant axis fits to the data of AT~2022dsb at +2 days and AT~2023mhs at +9 days in the Stokes $q,u$ plane. For each fit, we include the reduced $\chi^2$ statistic followed by the number of degrees of freedom in parentheses, the coefficient of determination $R^2$, and the ellipse axial ratio $b/a$.}
\label{fig:specpolpca}
\end{figure}

The second epoch of spectropolarimetry, obtained three weeks later, no longer shows any clear line features, and the overall polarisation degree has decreased significantly (from $P_V = 1.24\pm$0.03\% to $P_V = 0.07\pm$0.04\%). Considering the $q,u$ plane, Figure \ref{fig:specpolpca} shows that the data of the first epoch can be fit quite well by a dominant axis, which steers clear from the origin, indicating a departure from axial symmetry. During the second epoch, the near-zero degree of polarisation does not allow for a good fit to the $q,u$ data points, which are clustered around the origin (Fig. \ref{fig:specpolpca_appendix}). 

\subsubsection{AT~2022bdw}
\begin{figure}
	\includegraphics[width=0.9\columnwidth]{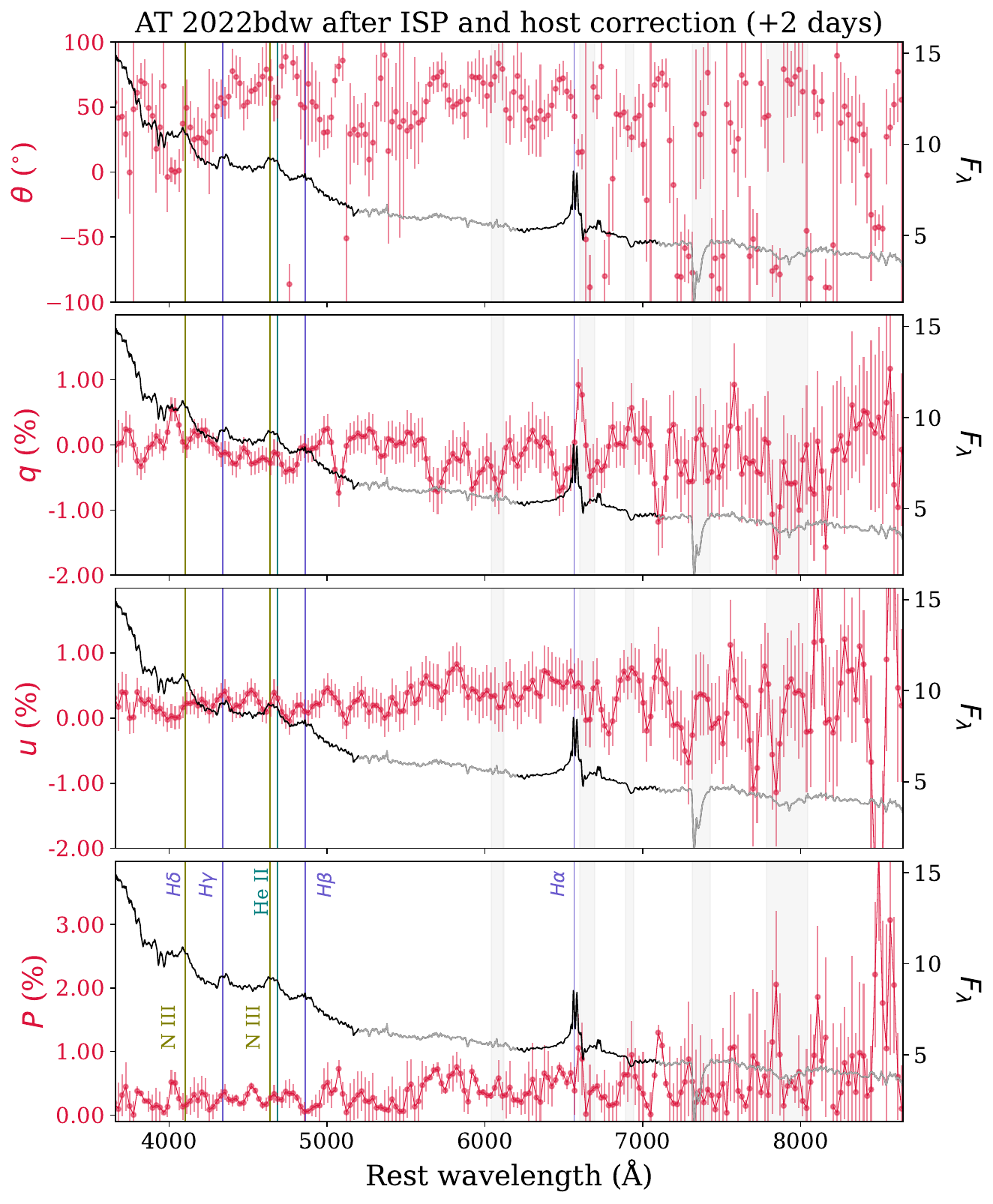}
\caption{Same as Figure \ref{fig:specpol_AT2021blz} but for AT~2022bdw.}
\label{fig:specpol_AT2022bdw}
\end{figure}

AT~2022bdw was classified as an H+He TDE due to the presence of Balmer lines and a He~II 4686\,Å complex \citep{2022TNSArcavi}, and we further identify N~III lines. We have analysed one epoch of spectropolarimetry and one epoch of broadband polarimetry in the FORS2 $B,V,I$ bands. The broadband polarimetry of AT~2022bdw was taken +267 days post-peak, and we assume it probes the contribution from the host galaxy and the ISP only. In all three bands, both the ISP and uncorrected transient polarisation are well below $\sim$~0.4\%. After correcting for the ISP the polarisation ranges from 0.05$\pm$0.10\% ($I$-band) to 0.16$\pm$0.17\% ($B$-band), consistent with zero polarisation. The polarisation spectrum of this TDE (Fig. \ref{fig:specpol_AT2022bdw}) obtained at +2 days post-peak shows an overall low polarisation (continuum level of 0.49$\pm$0.08\%) with few polarisation features. 

\begin{figure}
	\includegraphics[width=0.9\columnwidth]{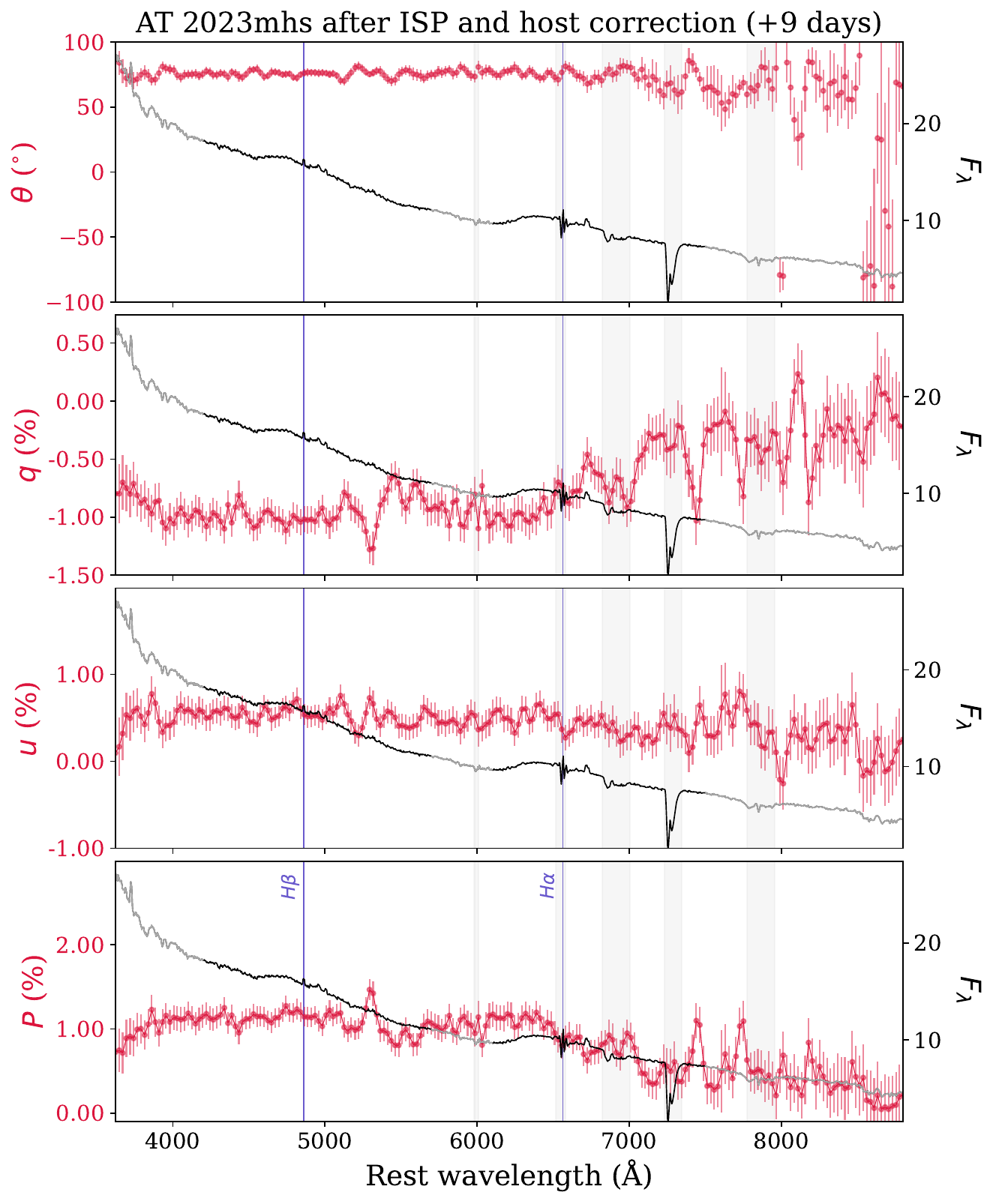}
\caption{Same as Figure \ref{fig:specpol_AT2021blz} but for AT~2023mhs.}
\label{fig:specpol_AT2023mhs}
\end{figure}

\subsubsection{AT~2023mhs}
\begin{figure}[!ht]
\centering
	\includegraphics[width=0.99\columnwidth]{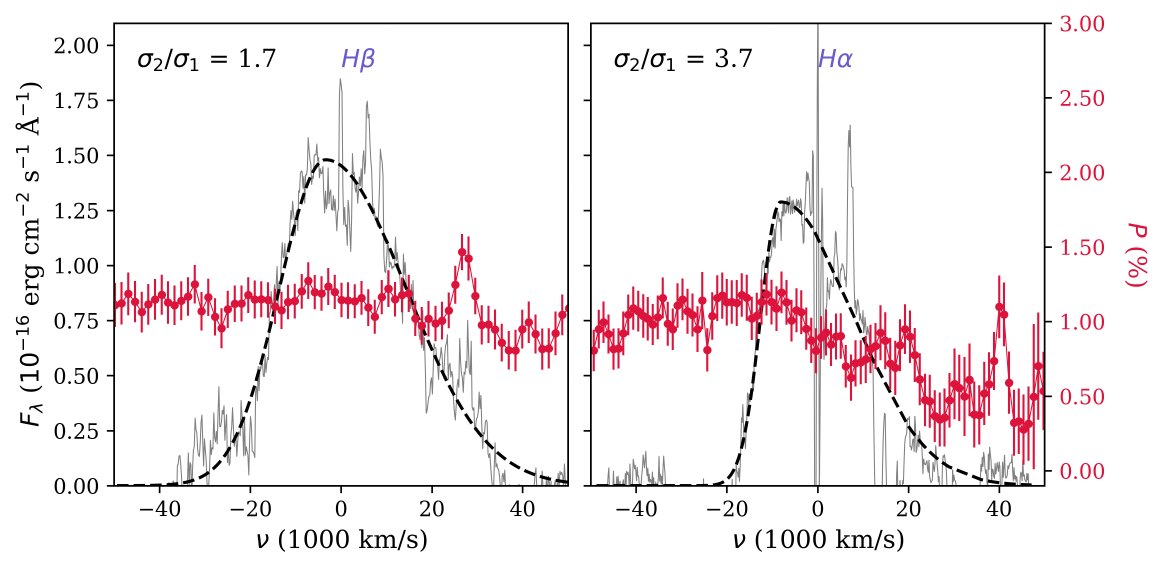}
\caption{Continuum-subtracted flux spectrum (grey) of AT~2023mhs, focussing on the H$\alpha$ and H$\beta$, which were fitted with an asymmetric Gaussian function (black dashed lines). The polarisation degree is shown in red. In the upper left corners, the ratio of the widths $\sigma_1$ (blue wing) and $\sigma_2$ (red wing) of the asymmetric Gaussian fit to each line are shown. A ratio $\sigma_2/\sigma_1 > 1$ for both lines indicates they are asymmetric with more broadened red wings.}
\label{fig:mhs_lines}
\end{figure}
The spectrum of AT~2023mhs is characterised by a blue continuum with broad Balmer lines. We present one epoch of NOT/ALFOSC $V,R$-band broadband polarimetry +4 days after maximum light, and one epoch of VLT/FORS2 spectropolarimetry at +9 days. These two measurements are consistent with each other within their uncertainties ($P_V$~=~0.84$\pm$0.19\% vs. $P_V$~=1.11$\pm$0.02\%). From Figure \ref{fig:specpolpca}, we infer that the geometry of AT~2023mhs at the epoch of spectropolarimetry was aspherical, but possibly axi-symmetric, as the best-fit dominant axis nearly crosses the origin.
\\
\indent
The polarisation spectrum of AT~2023mhs (Fig. \ref{fig:specpol_AT2023mhs}) shows a decrease with wavelength (the possible origin of which we discuss in Sect.\ref{sec:Discussion-Sample}). The polarisation maintains a roughly constant level up to the H$\alpha$ emission line, and then drops below 0.5\% in the reddest end of the spectrum. Although the flux spectrum is featureless aside from the broad H$\alpha$ and H$\beta$ lines, there is a dip in the $q$ spectrum on the right shoulder of the H$\beta$ line, producing a polarisation peak at the same location in the $P$ spectrum. We note that this feature is clearly present before applying any corrections (see Fig. \ref{fig:specpol_before} in Appendix \ref{sec:AppendixC}). Although similar features appear in the polarisation spectra of the other TDEs, they are mostly found in the red end where the signal is low, and are likely due to noise. In the case of AT~2023mhs, this peak stands out due to the high S/N. Polarised absorption lines, as opposed to depolarised emission lines, may arise from incomplete cancellation of the global continuum polarisation (especially for spherical geometry) by a partial obscuration at that wavelength (e.g., \citealt{2003Kasen}). However, at this wavelength we see no absorption feature in the flux spectrum. Instead, this feature might be a polarisation peak as a result of line broadening of the asymmetric H$\beta$ emission line. In TDEs, broadening of the wings of emission lines is due to the fact that the wings undergo more scatterings off electrons than the core \citep{2018Roth}. While the core of the line depolarises the continuum polarisation, the wings add to it, resulting in polarised peaks across the wings as also observed in e.g., AT~2018dyb \citep{2022Leloudas}. For this TDE, peaks were found on both sides of the symmetric broad H$\alpha$. However, in the case of AT~2023mhs, we find that both the H$\beta$ and H$\alpha$ line are asymmetric due to a wide red wing in the flux spectrum, which has been observed in TDEs before and is attributed to outflows \citep{2018Roth,2022Parkinson}. We highlight this by fitting an asymmetric Gaussian function
\begin{equation}
    f(x) = 
    \begin{cases}
    A\text{ exp}\left(-\frac{(x-\mu)^{2}}{2\sigma_{1}^{2}}\right) & x \leq \mu \\
    A\text{ exp}\left(-\frac{(x-\mu)^{2}}{2\sigma_{2}^{2}}\right) & x > \mu.
    \end{cases}
\end{equation}
to these emission lines through least-squares fitting, as shown in Figure \ref{fig:mhs_lines}. In velocity space, there is a polarisation peak at $\nu\sim$~30,000 km/s across the red wing of the H$\beta$ line. We observe that the short peak at $\nu\sim$~20,000 km/s or the peak at $\nu\sim$40,000 km/s might constitute a similar feature for the H$\alpha$ line, although this region in the polarisation spectrum has a more complex structure than the H$\beta$ line. For neither line does the blue wing show any of such peaks.

\subsection{TDEs with broadband polarimetry }
\label{sec:impol}

The remaining TDEs have broadband polarimetry only. 
The best observed event is AT~2020zso for which four epochs are available, and indeed before maximum light, while the remaining TDEs have a single epoch of observations. 
All broadband measurements are listed in 
Table \ref{tab:ImPol}, where we also provide synthetic measurements from spectropolarimetry by convolving with the filter response functions.

\begin{table*}
\centering
\caption{Broadband polarimetry of new TDEs presented in this work.}
    \begin{tabular}{cccccccc} \hline \hline
     TDE, & Phase & $q\pm\sigma_q$ & $u\pm\sigma_u$ & ($q\pm\sigma_q$)$_{\text{corr}}$ & ($u\pm\sigma_u$)$_{\text{corr}}$ & ($P\pm\sigma_P$)$_{\text{corr}}$ &  ($\theta\pm\sigma_{\theta}$)$_{\text{corr}}$ \\ 
      band & (days) & (\%) & (\%) & (\%) & (\%) & (\%) &  ($^{\circ}$) \\ \hline \hline
     AT~2018hyz & & & & & & &  \\
     $V$& +9 & --1.03$\pm$0.13 & --1.53$\pm$0.13 & --1.31$\pm$0.66 & --1.93$\pm$0.95 & 2.21$\pm$0.87 & --62.0$\pm$10.7 \\ \hline
     AT~2019lwu & &  &  &  &  &  &  \\
     $B$ & +51 & 0.02$\pm$0.30 & --0.37$\pm$0.15 & 0.14$\pm$0.84 & --0.77$\pm$0.43 & 0.31$\pm$0.27 & --40.0$\pm$16.4 \\
     $V$ & &  0.05$\pm$0.18 & --0.43$\pm$0.21 & --0.37$\pm$0.74 & --2.70$\pm$0.93 & 1.45$\pm$0.48 & --48.9$\pm$9.2 \\ \hline 
     AT~2020zso &  &  &  &  &  &  &  \\
     $B$& --24 & 1.69$\pm$0.10 & --0.07$\pm$0.10 & 2.03$\pm$0.20 & 0.88$\pm$0.16 & 2.21$\pm$0.20 & 11.7$\pm$2.5 \\
     * & --17 & 2.17$\pm$0.45 & 1.40$\pm$0.50 & 2.30$\pm$0.55 & 2.59$\pm$0.62 & 3.41$\pm$0.59 & 24.2$\pm$4.9 \\
     * & --9 & 1.18$\pm$0.10 & --0.83$\pm$0.13 & 0.92$\pm$0.12 & --0.23$\pm$0.15 & 0.94$\pm$0.12 & --7.1$\pm$3.7 \\
      & --1 & 1.45$\pm$0.09 & --0.62$\pm$0.09 & 1.27$\pm$0.11 & 0.08$\pm$0.11 & 1.27$\pm$0.11 & 1.8$\pm$2.5 \\
     $V$ & --24 & 1.29$\pm$0.10 & --0.42$\pm$0.10 & 2.53$\pm$0.45 & 0.52$\pm$0.25 & 2.57$\pm$0.45 & 5.8$\pm$4.9 \\
     * & --17 & 1.55$\pm$0.22 & --0.27$\pm$0.22 & 4.57$\pm$0.51 & --1.07$\pm$0.36 & 4.68$\pm$0.50 & --6.6$\pm$3.1 \\
      & --9 & 1.19$\pm$0.07 & --0.51$\pm$0.08 & 1.27$\pm$0.11 & --0.10$\pm$0.10 & 1.27$\pm$0.11 & --2.3$\pm$2.4 \\
      & --1 & 1.33$\pm$0.08 & --0.65$\pm$0.09 & 1.51$\pm$0.12 & --0.16$\pm$0.12 & 1.51$\pm$0.12 & --3.0$\pm$2.3 \\
     $R$ & --24 & 0.93$\pm$0.09 & --0.42$\pm$0.09 & 2.19$\pm$0.64 & 0.70$\pm$0.36 & 2.26$\pm$0.62 & 8.8$\pm$7.8 \\
     *  & --17 & 1.16$\pm$0.23 & --0.40$\pm$0.24 & 3.34$\pm$0.56 & --0.84$\pm$0.43 & 3.41$\pm$0.56 & --7.1$\pm$4.6 \\
      & --9 & 0.97$\pm$0.09 & --0.68$\pm$0.09 & 1.81$\pm$0.16 & --0.98$\pm$0.15 & 2.05$\pm$0.16 & --14.3$\pm$2.2 \\
      & --1 & 1.04$\pm$0.08 & --0.73$\pm$0.08 & 1.24$\pm$0.14 & --0.20$\pm$0.13 & 1.25$\pm$0.14 & --4.5$\pm$3.2 \\ \hline
     AT~2021blz & & & & & & & \\
     $B$ & +5 & 0.61$\pm$0.08 & 0.27$\pm$0.08 & 0.80$\pm$0.13 & 0.19$\pm$0.12 & 0.81$\pm$0.13 & 6.9$\pm$4.4 \\
      & +16 & 0.47$\pm$0.07 & 0.07$\pm$0.08 & 0.66$\pm$0.11 & --0.07$\pm$0.11 & 0.65$\pm$0.11 & --3.0$\pm$4.8 \\
     $V$ & +5 & 0.51$\pm$0.07 & 0.21$\pm$0.07 & 1.37$\pm$0.21 & 0.48$\pm$0.19 & 1.44$\pm$0.21 & 9.7$\pm$4.1 \\
      & +16 & 0.37$\pm$0.09 & 0.30$\pm$0.09 & 0.79$\pm$0.17 & 0.57$\pm$0.17 & 0.96$\pm$0.17 & 18.1$\pm$5.1 \\
     $R$ & +16 & 0.13$\pm$0.06 & 0.06$\pm$0.06 & 0.47$\pm$0.13 & 0.05$\pm$0.13 & 0.45$\pm$0.13 & 2.8$\pm$8.2 \\
     $V_s$ & +10 & 0.55$\pm$0.02 & --0.06$\pm$0.02 & 1.55$\pm$0.09 &  --0.17$\pm$0.09 &1.50$\pm$0.09 & --3.16$\pm$1.7 \\
      & +32 & 0.31$\pm$0.02 & --0.07$\pm$0.02 & 1.44$\pm$0.15 & --0.31$\pm$0.15 & 1.34$\pm$0.14 & --6.07$\pm$2.9 \\ \hline
     AT~2022bdw & & & & & & &  \\
     $B$ & +267 & --0.31$\pm$0.17 & 0.09$\pm$0.16 & --0.11$\pm$0.17 & 0.19$\pm$0.16 & 0.16$\pm$0.17 & 59.9$\pm$22.1 \\
     $V$ & +267 & --0.31$\pm$0.12 & --0.12$\pm$0.12 & --0.14$\pm$0.13 & --0.10$\pm$0.13 & 0.13$\pm$0.13 & --71.8$\pm$20.8 \\
     $I$  & +267 & --0.18$\pm$0.10 & 0.06$\pm$0.10 & --0.05$\pm$0.10 & 0.06$\pm$0.10 & 0.05$\pm$0.10 & 65.6$\pm$37.5 \\ 
     $V_s$ & +2 & 0.04$\pm$0.02 & --0.03$\pm$0.02 & 0.50$\pm$0.08 & --0.03$\pm$0.08 & 0.49$\pm$0.08 & --1.70$\pm$4.4 \\ \hline
     AT~2022dsb & & & & & & & \\
     $B$ & +115 & --0.07$\pm$0.14 & --0.21$\pm$0.14 & --0.12$\pm$0.14 & --0.08$\pm$0.14 & 0.10$\pm$0.14 & --73.9$\pm$28.5 \\
     $V$ & +115 & --0.06$\pm$0.10 & --0.20$\pm$0.11 & --0.08$\pm$0.10 & -0.05$\pm$0.11 & 0.06$\pm$0.10 & --75.2$\pm$1.2 \\
     $R$ & & 0.05$\pm$0.09 & --0.14$\pm$0.09 & 0.05$\pm$0.09 & 0.04$\pm$0.09 &  0.04$\pm$0.09 & 19.2$\pm$37.2 \\
     $I$ & +115 & --0.12$\pm$0.08 & --0.18$\pm$0.08 & --0.08$\pm$0.09 & --0.03$\pm$0.09 & 0.05$\pm$0.09 & --78.4$\pm$30.1 \\
     $V_s$ & +2 & 0.09$\pm$0.01 & --0.80$\pm$0.01 & 0.13$\pm$0.03 & --1.23$\pm$0.03 & 1.24$\pm$0.03 & --42.0$\pm$0.8 \\
     & +23 & --0.01$\pm$0.02 & --0.15$\pm$0.02 & --0.08$\pm$0.04 & 0.01$\pm$0.04 & 0.07$\pm$0.04 & 87.4$\pm$15.7 \\ \hline
     AT~2022exr & & & & & & &  \\ 
     $V$ & +169 & 0.04$\pm$0.08 & --0.06$\pm$0.08 & 0.13$\pm$0.34 & --0.01$\pm$0.33 & 0.07$\pm$0.34 & --2.5$\pm$73.7 \\ \hline
     AT~2022hvp & & & & & & &  \\
     $V$ & +16 & 0.19$\pm$32 & 1.13$\pm$0.85 & 0.24$\pm$0.41 & 1.45$\pm$1.09 & 1.40$\pm$1.08 & 40.2$\pm$21.0 \\ \hline
     AT~2023mhs & & & & & & & \\
     $V$ & +4 & --0.53$\pm$0.16 & 0.48$\pm$0.16 & --0.64$\pm$0.19 & 0.58$\pm$0.19 & 0.84$\pm$0.19 & 68.9$\pm$6.4 \\
     $R$ & +4 & --0.41$\pm$0.15 & 0.30$\pm$0.14 & --0.52$\pm$0.20 & 0.38$\pm$0.18 & 0.62$\pm$0.19 & 71.9$\pm$8.6\\
     $V_s$ & +9 & --0.81$\pm$0.01 & 0.44$\pm$0.02 & --0.98$\pm$0.02 & 0.54$\pm$0.02 & 1.11$\pm$0.02 & 75.6$\pm$0.5 \\ \hline
    \end{tabular}
\tablefoot{The first column gives the TDE name and bands (either VLT/FORS2 or NOT/ALFOSC) in which broadband polarimetry data were obtained, with multiple rows per band indicating multiple epochs of observations. The remaining columns contain the phase with respect to optical peak light (column 2), the values plus uncertainties of the normalised $q,u$ Stokes parameters (columns 3,4), the $q,u$ Stokes parameters corrected for the ISP and host galaxy light (columns 5,6), and the polarisation degree (column 7) and polarisation angle (column 8) after those same corrections. We also provide the synthetic $V$-band ($V_s$) polarisation degree derived from the polarisation spectrum for those TDEs that have spectropolarimetry.
\\
* An asterisk in the first column denotes datasets that were ignored in our analysis, as they were strongly affected by moonlight.}
\label{tab:ImPol}
\end{table*}
 
\textit{AT~2020zso}. The classification spectrum of AT~2020zso contained broad He~II 4686\,Å and Balmer lines \citep{2022Wevers}. Four epochs of broadband polarimetry in the $B,V,R$ bands were obtained prior to peak light (at times when double-peaked He~II and H$\alpha$ lines had already appeared in the spectrum, which are indicative of the presence of an accretion disk; \citealt{2022Wevers}). However, a bright Moon (80\% illumination fraction and angular distance of 30$^{\circ}$) during the second epoch might have affected the measured polarisation (see e.g., \citealt{2023Pursiainen}), especially in the $B,V$ bands. In addition, the polarisation of both unpolarised and polarised standard stars\footnote{\url{https://www.eso.org/sci/facilities/paranal/instruments/fors/inst/pola.html}} observed during the same night yielded inconsistent results. During the third epoch, the Moon was also bright, but farther away from the TDE. The polarisation degree of the standard star observed during this night was consistent with its tabulated value in the $V$-band, but not in the $B$-band. The more reliable values obtained during the first and final epoch are still relatively high, and show a drop in polarisation over a period of three weeks, up until roughly one day prior to maximum light (e.g., from $P_V$ = 2.57$\pm$0.45\% to $P_V$ = 1.51$\pm$0.12\%). These imply an aspherical photosphere.

\textit{AT~2019lwu and AT~2022hvp}. We encountered a similar issue for both AT~2019lwu (60\% lunar illumination and 90$^{\circ}$ from the Moon at time of observations) and AT~2022hvp (94\% lunar illumination and 75$^{\circ}$ from the Moon). We opted to use a circular annulus to subtract the local (elevated) background from each aperture flux, instead of the default 2D background map used for the other TDEs. In addition, for AT~2019lwu, we used a smaller source aperture of 1.5$\times$FWHM to let in as little background light as possible (although the result stays consistent within error bars when using our default aperture). This way, we obtain a $V$-band polarisation degree of 1.45$\pm$0.48\% for AT~2019lwu (Table \ref{tab:ImPol}). In the case of AT~2022hvp we find $P_{V}=1.40\pm$1.08\%; this value is corrected for the host contamination but not for the ISP, as no suitable field stars were present.

\textit{AT~2018hyz}. AT~2018hyz has a single epoch of NOT/ALFOSC $V$-band broadband polarimetry taken +9 days after its optical peak. Due to a lack of suitable field stars, we could not obtain an estimate of the ISP, but we place an upper limit of 0.39\% on the Galactic ISP based on the extinction in the direction of this transient (Table \ref{tab:ISP}). We infer a $V$-band polarisation degree of $P_V = 2.21\pm$0.87\%, which is just a 2.5$\sigma$ detection (Table \ref{tab:ImPol}). 

\textit{AT~2022exr}. Finally, for AT~2022exr, a single epoch of $V$-band VLT/FORS2 polarimetry was obtained. The light curve of this transient, which is well sampled by both ATLAS and ZTF data, shows two flares with very similar peak magnitudes separated by $\sim$115 days. The broadband polarimetry was obtained +169 days after the first peak in the optical light curve (at the moment also the second peak had already almost faded). Unsurprisingly, at such late times the polarisation has become consistent with zero ($P_V = 0.07\pm$0.34\%).

\section{Discussion: Sample properties}
\label{sec:Discussion-Sample}
In the previous section, we have presented and analysed the polarimetry data of nine new optical TDEs. We now combine these with previous results published in literature, and explore how this larger sample of now 19 TDEs fits within the unification scenario of tidal disruption events \citep{2018Dai}. In this section, we explore the properties of the sample, and we focus on individual events in Section \ref{sec:Discussion-Indiv}.

\subsection{Evolution of continuum polarisation of TDEs}

\begin{figure*}[!ht]
\centering
	\includegraphics[width=1.0\textwidth]{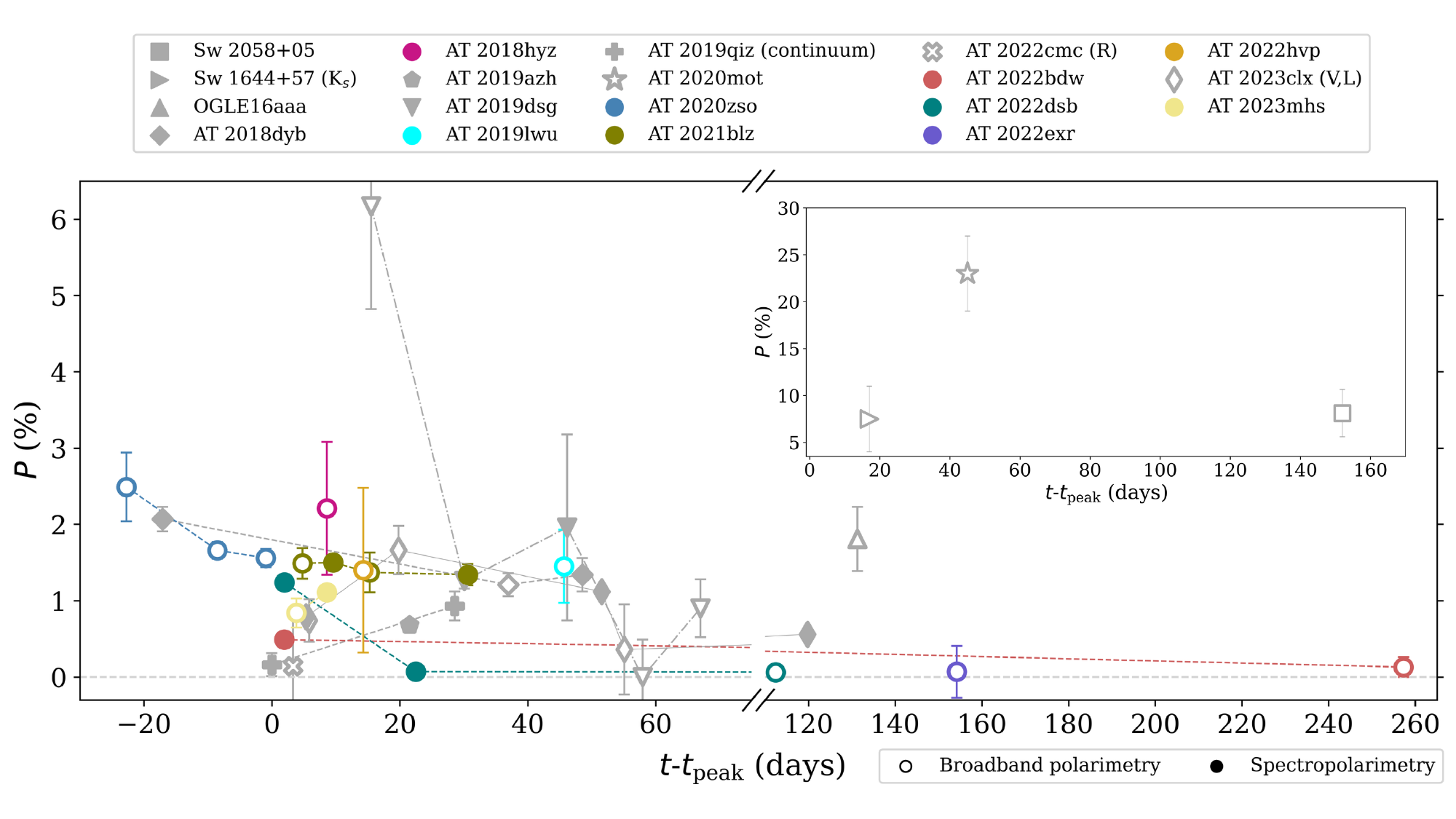}
\caption{Polarisation degree as a function of time since peak light (in the rest frame) of all TDEs with polarimetry data. Unless indicated otherwise in the legend, these are all $V$-band values (for AT~2023clx, broadband polarimetry was obtained in the LT/MOPTOP $L$-band; \citealt{2024Charalampopoulos}). Open markers correspond to broadband polarimetry measurements, and closed markers to spectropolarimetry measurements (converted to a broadband value). Data collected from the literature are shown in grey, while coloured symbols indicate data analysed in this study. The inset shows the TDEs AT~2020mot, Swift J164449.3+573451, and Swift J2058+0516 (abbreviated to Sw 1644+57 and Sw 2058+05) for clarity.}
\label{fig:mainplot}
\end{figure*}

In Figure \ref{fig:mainplot}, we plot the $V$-band polarisation degree $P_V$ of all 19 TDEs as a function of time since optical maximum light. The $V$-band polarisation degree is assumed to best probe the continuum polarisation, because (1) in the rest frame, the $V$-band typically covers the $\sim$~5000--6000\,Å region between the H$\alpha$ and H$\beta$ lines, which is relatively free from line emission, and (2) the S/N is typically much better than in the redder bands. We note that one data point of AT~2018dyb with $P_{V} = 0.00\pm 34.63$\% at +180.3 days \citep{2022Leloudas} is not shown in Fig.~\ref{fig:mainplot} due to the large error bar. In the inset we show the relativistic TDEs Swift J164449.3+573451 and Swift J2058+0516, and AT~2020mot, as the only optical TDE demonstrating high polarisation values. We leave the discussion of these events to Sect.~\ref{sec:ComparisonD18}.

For the non-relativistic TDEs, we can draw the following conclusions: firstly, most polarised (15/16) TDEs have maximum $V$-band polarisation levels in the range 1--2\% (with the 16th, 50th, and 84th percentiles being 0.94, 1.48, and 2.20\%; this includes marginal detections, but excludes the data point of AT~2020mot, which we consider an outlier). An exception is AT~2019dsg, which has $P_{\text{max}}\approx$ 6\% during the first epoch (2019-05-17), although this data point should be treated with caution 
\citep[see the discussion in][]{2022Leloudas}.

Secondly, for those TDEs that have multiple epochs of polarimetry (10/16), 5/10 show a decrease in polarisation after maximum light (given the observed cadence). 2/10 do not show significant changes (AT~2021blz, AT~2023mhs), and 3/10 show a rise after peak, namely AT~2019qiz, AT~2020mot\footnote{Not shown in Fig.~\ref{fig:mainplot}: AT~2020mot has multi-epoch $R$-band polarimetry, but only a single epoch in the $V$-band \citep{2023Liodakis}. The multi-epoch polarimetry is discussed in Sect. \ref{sec:ComparisonD18}.}, and AT~2023clx. However, in the case of the latter, only the first epoch is intrinsic to the TDE -- we will elaborate on this in Sect.\ref{sec:IR}). AT~2019qiz does show a significant (yielding $\chi^2$/dof~=10, $p$=0.002 when testing for variability with a $\chi^2$ test) 1\% pt. increase in polarisation degree from +0 to +29 days. Thirdly, in those ten TDEs, significant changes in the $V$-band polarisation ($\geq$1\% pt.) occur on timescales of 21 -- 50 days, up to 70 days post-peak. The most drastic decrease in polarisation occurs for AT~2019dsg, dropping by at least 5\% pt. within 54 days ($\sim$0.1\% pt. per day), while AT~2018dyb evolves the slowest ($\sim$0.01\% pt. per day). For the others, the average rate of decline in polarisation is $\sim$0.05\% pt. per day. When observed after $\sim$70 days (7/16), 5/7 TDEs have intrinsic polarisation levels consistent with zero. The exceptions are OGLE16aaa and AT~2020mot (as it is noted that the final epoch of AT~2023clx likely probes the host galaxy polarisation; \citealt{2025Uno}).

Within the reprocessing scenario, there are several possible explanations to the decrease in polarisation with time, including (1) the reprocessing layer becoming optically thin to X-rays, and (2) a change in photosphere geometry, where the photosphere transform into a more symmetric system with time.

\indent
For the disruption of a solar-mass star, the time it takes for an electron scattering envelope to become optically thin to X-rays is on the order of a hundred days up to a year, depending on the black hole mass ($10^{5-7}$M$_{\odot}$, with a higher mass yielding shorter timescales; \citealt{2016Metzger}). Hence, if the polarisation is produced by reprocessing in an electron-scattering envelope, the non-detection of polarisation after 70 days can be attributed to the optical depth. Also before that time, the gradual decrease in optical depth may cause a decline in polarisation; this may be applicable to AT~2021blz and AT~2023mhs for example (for which the continuum polarisation stays roughly constant). However, we argue that this scenario cannot hold for all TDEs, and that a change in photosphere geometry is a more favourable explanation. For example, dominant-axis fits to the data of AT~2018dyb \citep{2022Leloudas} and AT~2022dsb (Sect. \ref{sec:specpol}) indicate a transition from an asymmetric photosphere to an axially symmetric or spherical one. In the case of AT~2022dsb, this is supported by the fact that the line widths of its emission lines do not show an obvious decrease between epochs, while they are expected to decrease with time as the envelope becomes optically thin (assuming that electron scattering is responsible for the broadening of these emission lines; e.g., \citealt{2018Roth}). Similarly, in the case of AT~2019dsg, the line widths of the broad H$\beta$ and H$\alpha$ lines were variable in time \citep{2021Cannizzaro} and even increased up to the final epoch of polarimetry. Finally, the decline in polarisation of AT~2020zso happened before peak light, when the envelope mass is still increasing \citep{2016Roth}, and the increase in polarisation seen in AT~2019qiz cannot be explained with a reduced optical depth either.

\indent
Hence, the fact that the polarisation of the TDEs in our sample changes on the order of just two months could signify a sudden change in geometry of their photospheres. If due to the rapid formation and circularisation of an accretion disk, as already suggested by \citet{2022Leloudas} and \citet{2022Patra}, the overall steep drops in polarisation imply a short circularisation timescale. The -- highly uncertain -- circularisation timescale of TDEs is sensitive to various parameters, including the stellar mass and black hole mass, black hole spin, the impact parameter, the eccentricity of the star's orbit, and the radiative cooling efficiency (e.g., \citealt{2015Bonnerot, 2016Hayasaki}), and can vary from months to decades (e.g., \citealt{2021Hayasaki}). Generally, more massive black holes have a higher disk formation efficiency, and prompt disk formation is much less likely to occur for black hole masses of log $M_{\text{BH}}/M_{\odot} < 6$ \citep{2022Wong}. For the TDEs with multi-epoch polarimetry that have black hole mass estimates (AT~2018dyb, AT~2019dsg, AT~2019qiz, AT~2020zso, AT~2022dsb, AT~2023clx), we compared their estimates to the mean of a larger TDE sample, log $M_{\text{BH}}/M_{\odot}\approx$ 6.4 \citep{2022Wong, 2023Mummery, 2023Yao}. 
In these studies, various methods are used to estimate the black hole mass, including the relations between $M_{\text{BH}}$ and the velocity dispersion of the galaxy ($M_{\text{BH}}-\sigma$; e.g., \citealt{2005Ferrarese}), and between $M_{\text{BH}}$ and the total stellar mass of the galaxy ($M_{\text{BH}}-M_{\text{gal}}$; \citealt{2015Reines}). For example, the estimates from the $M_{\text{BH}}-\sigma$ relation are as follows: AT~2018dyb: log $M_{\text{BH}}\sim$6.69 \citep{2022Wong}; AT~2019dsg, AT~2019qiz, AT~2020zso: log $M_{\text{BH}}\sim$ 6.71, 6.35, 6.06, resp. \citep{2023Mummery}; AT~2023clx: log $M_{\text{BH}}\sim$6.49. For  AT~2022dsb, \citet{2024Malyali} inferred log $M_{\text{BH}}\sim$7.3 from the $M_{\text{BH}}-M_{\text{gal}}$ relation. While we do not find a clear trend between the polarisation decline rate and $M_{\text{BH}}$, the $M_{\text{BH}}$ values of these six TDEs are similar or slightly higher than the sample mean (regardless of the estimation method used), so that prompt disk formation is at least not ruled out based on their black hole masses. Finally, if TDEs of the spectroscopic subclass H + He are more likely to be associated with promptly formed accretion disks as opposed to collision shocks \citep{2022Nicholl}, the fact that most TDEs in our sample are of this subclass (Table \ref{tab:multiplewaves}) further motivates this notion.

\subsection{Comparison to models} 

\begin{figure*}[!ht]
\centering
	\includegraphics[width=1.0\textwidth]{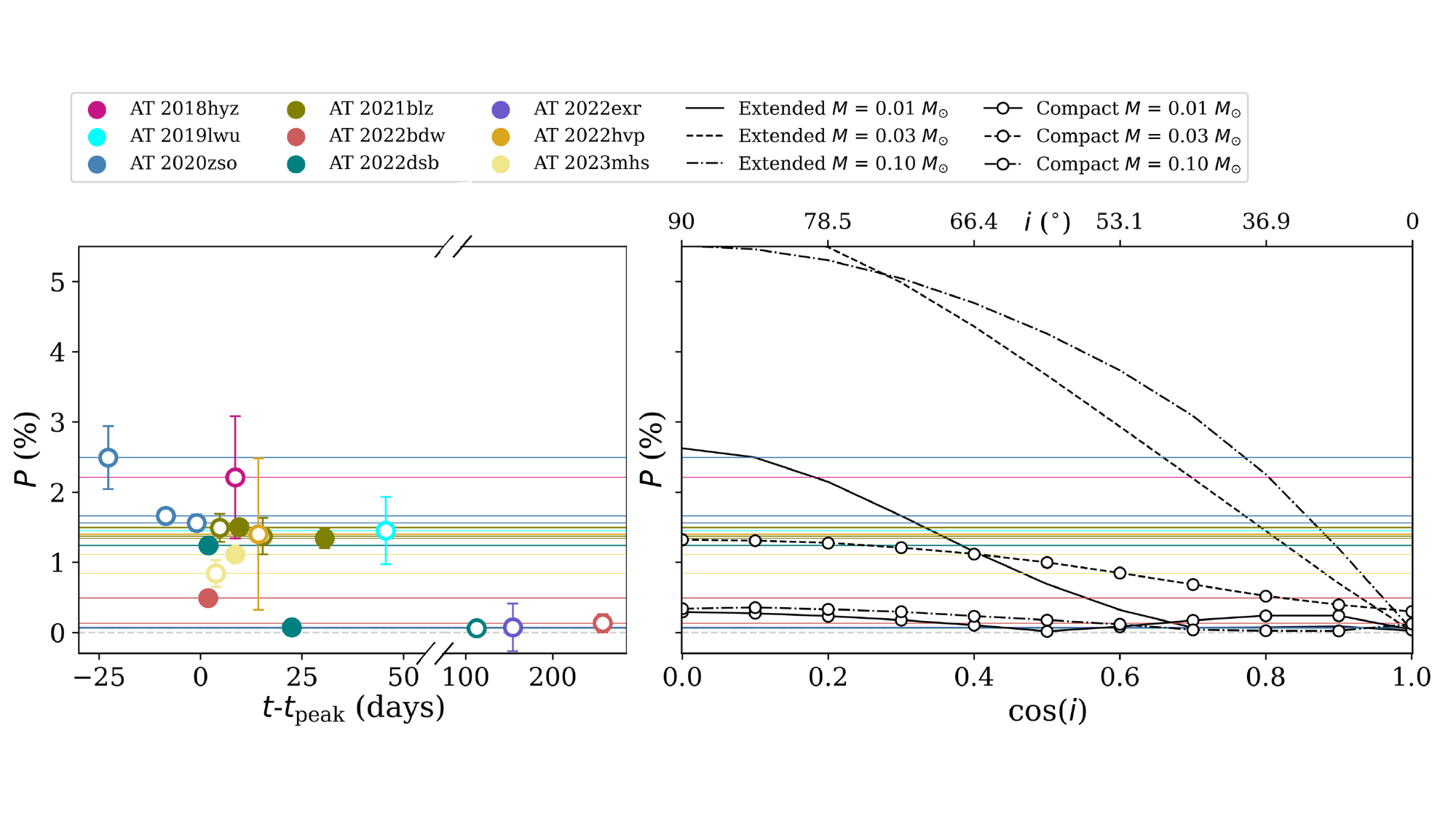}
\caption{Comparison of the measured $V$-band polarisation of TDEs to the predicted polarisation from the super-Eddington disk model \citep{2018Dai} as first produced by \citet{2022Leloudas}. The left panel shows the polarisation as a function of time, similar as in Fig. \ref{fig:mainplot}, now only for the TDEs newly analysed in this work. Horizontal lines indicating the polarisation levels are used to guide the eye to the right panel, which shows the model polarisation as a function of the viewing angle relative to the pole. Six models are included, with three different disk masses ($M$~=~0.01, 0.03, 0.10 M$_{\odot}$) and either an extended disk or a compact disk.}
\label{fig:mainpossis}
\end{figure*}

Potential sources of optical polarisation in tidal disruption events include at least (1) reprocessing via electron scattering in an envelope covering an accretion disk; (2) synchrotron emission, either due to relativistic jets (as seen in gamma-ray bursts) or stream collision shocks; and (3) dust scattering off a disk or torus surrounding the black hole, as seen in active galactic nuclei (AGN). Here we discuss two reprocessing models for which there are polarisation predictions.

\subsubsection{Predictions from reprocessing models}

The super-Eddington accretion disk model proposed by \citet{2018Dai} (hereafter called the \texttt{D18} model) predicts the early formation of a geometrically thick disk in the fallback phase of a TDE, where the fallback rate exceeds the Eddington accretion rate, and the subsequent launch of outflows and relativistic jets. In this scenario, X-ray photons generated in the inner disk are reprocessed in the optically thick outflows, and a viewing angle dependence of the X-ray-to-optical flux ratio is predicted (with this ratio being higher if the line of sight is along the pole, where X-rays can escape freely). This in turn implies a viewing angle dependence of the optical polarisation (which should then be highest along the plane of the disk), as well as an anti-correlation between the polarisation and X-ray flux in the \texttt{D18} model. While the X-rays are reprocessed at early times in an aspherical envelope, producing the optical polarisation, the X-ray flux (polarisation) is expected to increase (decrease) with time as the envelope becomes optically thin later on.

\citet{2022Leloudas} used the \texttt{D18} model as input for the radiative transfer code \texttt{POSSIS} \citep{2019bBulla}. They made predictions for the continuum polarisation whilst varying the inclination angle, disk mass (varied between 0.01 and 0.1 M$_{\odot}$), and disk compactness (i.e., mass density distribution). It was found that a maximum polarisation of $\sim$6\% can be obtained at an edge-on viewing angle, while lower polarisation levels can be achieved with higher inclinations, lower disk masses, or more compact disks. In addition, \citet{2022Leloudas} note that the observed decrease in polarisation levels can be explained by a reduced disk mass as the mass fallback and accretion rates drop with time.
\\ \\
The collision-induced outflow model (CIO model from here on) proposed by \citet{2019LuBonnerot} predicts that debris may become unbound in the self-intersection shock, and fall inward onto the black hole. The accretion flow thus formed between the self-intersection point and the black hole subsequently launches an (aspherical) CIO, which reprocesses X-ray/EUV photons generated in the inner accretion disk at optical wavelengths. This is in contrast to other stream collision models, in which the optical emission is assumed to be produced in the shock itself due to orbital energy dissipation (e.g., \citealt{2015Piran}). Hence, the source of the optical polarisation is reprocessing via electron scattering, similarly to the \texttt{D18} model. As pointed out by \citet{2019LuBonnerot}, their model differs from the one by \citet{2018Dai} in the sense that the X-ray behaviour is not tied to the viewing angle with respect to the symmetry axis of the accretion disk, but to the viewing angle with respect to the outflow direction. Optically bright TDEs may still exhibit strong X-ray emission if the inner accretion disk is not veiled by the stream, for example.

\citet{2023Charalampopoulos} used the CIO model by \citet{2019LuBonnerot} as input for \texttt{POSSIS}. By considering different mass fallback rates and geometrical parameters, they obtain a maximum polarisation of $\sim$9\%. This was obtained for a lower mass fallback rate and at intermediate viewing angles, while the polarisation peaks at equatorial angles for higher mass fallback rates. In addition, depending on the viewing angle, either an increase or decrease in polarisation with time is predicted. \citet{2022Charalampopoulos} noted that in the majority of their simulations with a high peak fallback rate ($\sim$3 M$_{\odot}$/yr), the polarisation level remained below 1\% (and most often below 0.5\%) for any inclination angle. Following the findings by \citet{2020Law-Smith} that most full disruptions of $\geq$~1 M$_{\odot}$ stars should have peak fallback rates $\dot{M_{p}} \geq$~3 M$_{\odot}$/yr, \citet{2022Charalampopoulos} predicted that most TDEs with M$_{*}\geq$~1 M$_{\odot}$ should then have continuum polarisation degrees of $P\lesssim$1\%.

\subsubsection{Comparison to the \texttt{D18} model}
\label{sec:ComparisonD18}

Here we focus mainly on the comparison between the polarisation data and the \texttt{D18} model. In Figure \ref{fig:mainpossis} we compare the polarisation data of our sample to the same models used by \citet{2022Leloudas}. Due to the low-to-moderate polarisation levels of these TDEs, they can easily be accommodated by these models for a range of viewing angles and disk masses, with the lowest disk mass requiring a viewing angle closer to the equatorial plane ($i\lesssim 60^{\circ}$) and higher masses allowing for inclinations closer to (but never along) the pole ($60^{\circ}\lesssim i \lesssim 25^{\circ}$). When considering not just the continuum polarisation but also the polarisation features in the spectra, we find that evidence for depolarisation of broadened emission lines in 7/9 TDEs that have spectropolarimetry data (namely in AT~2018dyb, AT~2019azh, AT~2019dsg, AT~2019qiz, AT~2021blz, AT~2022dsb, and AT~2023clx), further supporting an electron scattering origin \citep{2018Roth, 2022Parkinson}. 

The majority of TDEs in the sample are polarised and have continuum polarisation levels above 1\% and below 6\% (with a median of 1.24\%). In addition, with the exception of AT~2019qiz, AT~2020mot, and AT~2023clx, none of the TDEs with multi-epoch polarimetry observations show a significant increase in their polarisation degree after optical peak light. The increase in polarisation in the case of AT~2019qiz, can, however, be accommodated by various reprocessing scenarios \citep{2022Patra,2023Charalampopoulos}. \citet{2022Patra} explain the null polarisation at peak light with the recession of the spherical reprocessing layer, which later on reveals an asymmetric interior (formed through anisotropic outflows).

\begin{figure}
	\includegraphics[width=\columnwidth]{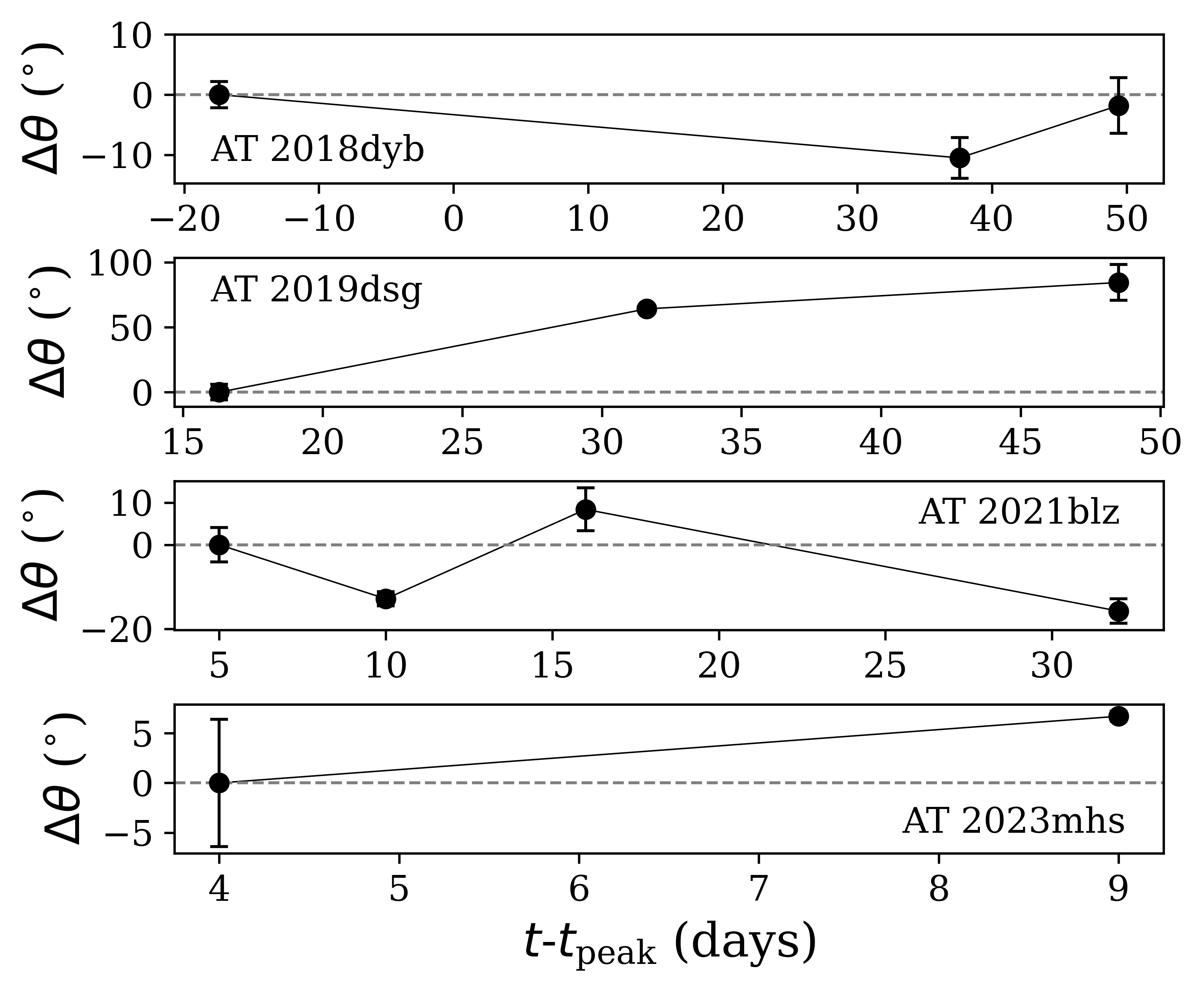}
\caption{Evolution of the $V$-band polarisation angle of AT~2018dyb, AT~2019dsg, AT~2021blz, and AT~2023mhs in time. The change in polarisation angle $\Delta\theta$ is with respect to the first epoch as reference.}
\label{fig:angles}
\end{figure}

Finally, the higher polarisation levels of Swift J164449.3+573451, Swift J2058+0516, and AT~2020mot cannot be accommodated with the reprocessing model. \citet{2023Liodakis} attributed the high polarisation degree of AT~2020mot to synchrotron radiation produced in stream-stream collision shocks. As there are currently no accurate predictions for the optical polarisation resulting from such collisions \citep{2024Jankovic, 2024Koljonen}, it is not possible to further investigate this scenario at this stage. Nevertheless, other studies have attributed a variable polarisation angle ($\theta$) to stream shocks: such variability was found in e.g., AT~2020mot \citep{2023Liodakis} and AT~2023clx \citep{2025Uno,2025Koljonen}. \citet{2025Jordana-Mitjans} found that $\theta$ is variable in two Bowen flares, whereas it is stable in the three classical TDEs they studied. Here, the variability was attributed to a clumpy medium near the black hole, and to a light echo (in the case of AT~2020afhd), similar as was observed in AT~2023clx \citep{2025Uno}. In the \textsc{POSSIS} polarisation predictions for the \texttt{D18} model, we are considering a static configuration, namely that of an axially symmetric density distribution. In order to test the variability of the polarisation angle, time-dependent polarisation predictions are needed, which are currently lacking. Assuming a TDE maintains the same axial symmetry over multiple epochs, we expect $\theta$ to be stable. However, we have already encountered cases where the polarisation points towards a deviation from axial symmetry (e.g., AT~2022dsb and AT~2018dyb). If the accretion disk has already formed at maximum light, we may at least expect $\theta$ to be relatively stable after the peak. We can thus focus on the TDEs in our sample that have multiple epochs of polarimetry after peak light (and on those epochs when the polarisation degree is non-zero). Aside from  AT~2020mot and AT~2023clx, there are four such TDEs (Fig~\ref{fig:angles}). The $V$-band polarisation angles of AT~2018dyb and AT~2023mhs remain stable. In the case of AT~2019dsg, the first detection should be treated with caution \citep{2022Leloudas}, and the third one is marginal, but they point towards a possible variability in $\theta$ nonetheless. Between the first and third epoch show in Fig.~\ref{fig:angles}, the polarisation angle changes by $\Delta\theta =85\pm 15^{\circ}$, which is quite similar to the change in polarisation angle found in AT~2023clx ($\Delta\theta =90\pm 9.7^{\circ}$; \citealt{2025Uno}) and AT~2020afhd ($\Delta\theta =83\pm 8^{\circ}$; \citealt{2025Jordana-Mitjans}). Interestingly, AT~2019dsg also has a mid-infrared excess (see Sect.~\ref{sec:IR}), and its host possibly harbours an AGN, given the narrow emission lines in the galaxy spectrum (although these may arise from a region of star formation; \citealt{2021Cannizzaro}). These hint at a common origin of the change in $\theta$, namely a light echo, between AT~2019dsg,  AT~2020afhd, and AT~2023clx. The polarisation angle of AT~2021blz fluctuates, although this is a more moderate change in $\theta$ than was observed in AT~2020mot and AT~2023clx \citep{2023Liodakis,2025Koljonen}. It could signify a change in geometry, despite its stable polarisation degree. Determining the cause of the variability in $\theta$ is currently challenging, however, due to the lack of time-dependent predictions for the polarisation degree and angle in various scenarios (both reprocessing models and stream-shock models).

\subsection{Unification through multi-wavelength observations}
\label{sec:Discussion-multi-wavelength}

\begin{table*}[!hbt]
\centering
\caption{Multi-wavelength properties of the TDEs in our sample.}
    \begin{tabular}{lllccll} \hline \hline
     TDE & Class & X-ray & Infrared & Radio & Notes & Ref. \\ \hline
     OGLE16aaa & H + He &  \checkmark (delayed) & \xmark & ? & $\bullet$ X-ray bright while polarised & [1][*] \\
     Sw 1644+57 & & \checkmark (early) & \xmark & \checkmark & $\bullet$ Relativistic / jetted TDE & [2][*][3] \\
     Sw 2058+05 & &  \checkmark (early)& \xmark & \checkmark & $\bullet$ Relativistic / jetted TDE & [4][*][5] \\
     AT~2018dyb & H + He&  \xmark & \checkmark & \checkmark & $\bullet$ Polarisation decreases with time, & [6][7,8][9] \\
     & & & & & \textcolor{white}{$\bullet$} line depolarisation &  \\
     AT~2018hyz & H + He & \checkmark (early) & \checkmark & \checkmark & $\bullet$ Double-peaked emission lines & [10][7][11] \\
     AT~2019azh & H + He &  \checkmark (delayed) & \checkmark & \checkmark & $\bullet$ Line depolarisation & [12][7,8][13,14]\\
     AT~2019dsg & H + He & \checkmark (delayed) & \checkmark & \checkmark & $\bullet$ Polarisation decreases with time,  & [15][8,16] \\
     & & & & & \textcolor{white}{$\bullet$} line depolarisation & [17,18]  \\
     AT~2019lwu & H & \xmark & \xmark & \xmark &  & [\dag][*][13] \\
     AT~2019qiz & H + He & \checkmark (delayed) & \checkmark & \checkmark & $\bullet$ Polarisation increases with time, & [19][16][13] \\
     & & & & & \textcolor{white}{$\bullet$} line depolarisation &  \\
     AT~2020mot & H + He & \xmark & \checkmark & \xmark & $\bullet$ $\sim$25\% polarisation & [20][21][13] \\
     AT~2020zso & H + He & \xmark & \xmark & \xmark & $\bullet$ Polarisation decreases with time,  & [22][*][13] \\
     & & & & & \textcolor{white}{$\bullet$} double-peaked emission lines &  \\
     AT~2021blz & H + He & \checkmark (delayed) & \xmark & \xmark & $\bullet$ Roughly constant polarisation, & [23][*][13] \\
     & & & & & \textcolor{white}{$\bullet$} line depolarisation  &  \\
     AT~2022cmc &  & \checkmark (early) & \xmark & \checkmark & $\bullet$ Relativistic / jetted TDE & [24,25][*] \\
     AT~2022bdw & H + He & \xmark & \xmark & \xmark & $\bullet$ Low polarisation at peak light & [\dag][*][13] \\
     AT~2022dsb & H + He & \checkmark (early) & \xmark & \xmark & $\bullet$ Polarisation decreases with time, & [26][*][13] \\
     & & & & & \textcolor{white}{$\bullet$} line depolarisation &  \\
     AT~2022exr & H & \checkmark (delayed) & \xmark & ? & $\bullet$ Double-peaked optical light curve & [27][*][13] \\
     AT~2022hvp & He  & \xmark & \xmark & ? &  & [\dag][*] \\
     AT~2023clx & H + He & \checkmark (delayed) & \xmark & \checkmark & $\bullet$ Line depolarisation, dust  & [28][*][29] \\
     & & & & & \textcolor{white}{$\bullet$} polarisation via light echo &  \\
     AT~2023mhs & H & \checkmark (delayed) & \xmark & ? & $\bullet$ Roughly constant polarisation, & [\dag][*] \\
     & & & & & \textcolor{white}{$\bullet$} wavelength-dependent continuum &  \\
    \end{tabular}
\tablefoot{The second column provides the spectroscopic subtype. In columns 3-5, a tick mark indicates the TDEs in our sample were detected in X-rays, mid-infrared, or radio, whereas a cross indicates non-detections. The information in this table is based on publicly available data, provided by the references in the final column. In some cases, it is unknown whether there are any observations of a TDE, in which case we put down a question mark. For the X-ray detections, we specify whether the X-rays peaked before or during the optical peak ("early") or after ("delayed"). Swift J164449.3+573451 and Swift J2058+0516 have been abbreviated to Sw 1644+57 and Sw 2058+05.
\\ 
\tiny
\textbf{References}. [1]\citet{2020Kajava}, [2] \citet{2016Mangano}, [3] \citet{2021bCendes}, [4] \citet{2012Cenko}, [5] \citet{2015Pasham}, [6] \citet{2020Holoien}, [7] \citet{2021Jiang}, [8] \citet{2022Leloudas}, [9] \citet{2024Cendes}, [10] \citet{2020Gomez}, [11] \citet{2022Cendes}, [12] \citet{2020Hinkle}, [13] \citet{2024Anumarlapudi}, [14] \citet{2022Goodwin}, [15] \citet{2021Cannizzaro}, [16] \citet{2024VanVelzen}, [17] \citet{2021Cendes}, [18] \citet{2021Stein}, [19] \citet{2024Nicholl}, [20] \citet{2023Liodakis}, [21] \citet{2024Newsome}, [22] \citet{2022Wevers}, [23] \citet{2021ATelLiu}, [24]  \citet{2022Andreoni}, [25] \citet{2023Pasham}, [26] \citet{2024Malyali}, [27] \citet{2022ATelGuolo}, [28] \citet{2023Zhu}, [29] \citet{2023ATelSfaradi}.
\\
\dag X-ray observations analysed in this work; see \ref{tab:X-rays}.
\\
* NEOWISE IR light curve inspected in this work.}
\label{tab:multiplewaves}
\end{table*}

We now consider whether implications for the emission mechanism and geometry derived from the polarisation measurements in the optical are in line with multi-wavelength observations of TDEs, focussing especially on X-rays. Due to the diversity within our sample, we summarise the multi-wavelength properties of the sample in this section, from which we draw more general conclusions. In Section \ref{sec:Discussion-Indiv}, we provide more details for individual events.

\subsubsection{X-rays}

\begin{figure*}
\centering
	\includegraphics[width=0.9\textwidth]{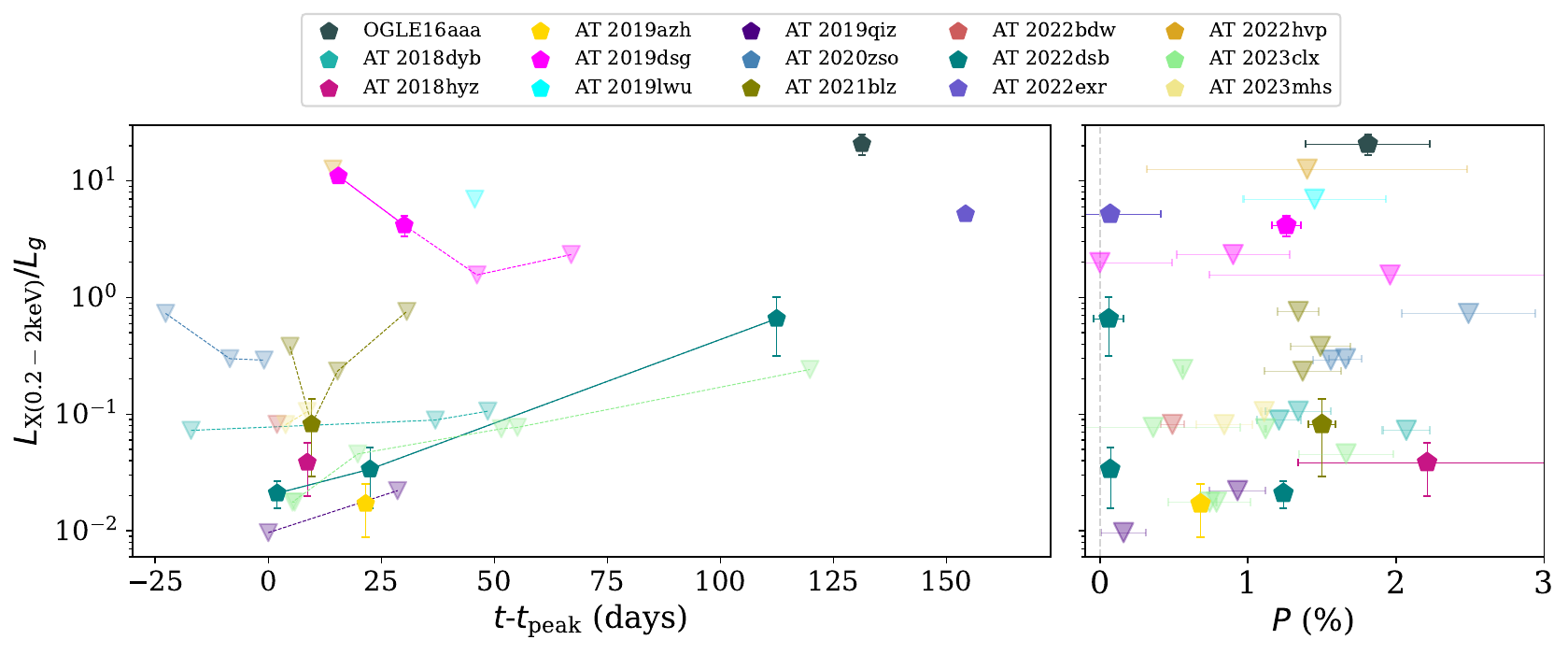}
\caption{The left-hand panel shows the ratio of the 0.2--2 keV X-ray luminosity over the optical $g$-band luminosity, at the times the polarisation observations of the TDEs were made. Inverted triangles denote (X-ray) upper limits, whereas pentagons mark detections. The right-hand panel shows the ratio $L_{\text{0.2-2 keV}}/L_{g}$ plotted against the $V$-band polarisation.}
\label{fig:LXLO}
\end{figure*}

Table \ref{tab:multiplewaves} summarises the multi-wavelength observables of the sample; we loosely define "early" versus "delayed" X-rays as those peaking before or during the optical peak, and those peaking when the optical light curve was already fading. We note that 8/13 TDEs with X-ray detections show a delayed X-ray brightening with respect to the optical peak. 

\begin{figure}
	\includegraphics[width=\columnwidth]{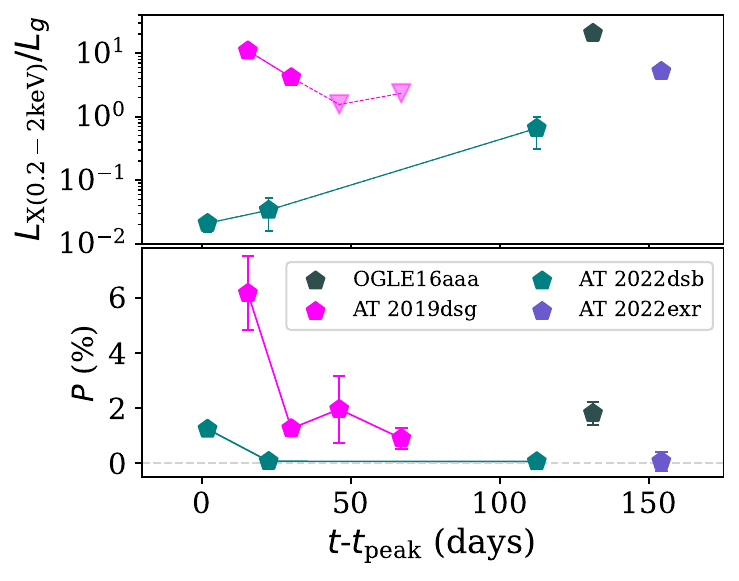}
\caption{Co-evolution of the ratio $L_{\text{0.2-2 keV}}/L_{g}$ and $V$-band polarisation as a function of time since peak light for a selection of TDEs. This selection includes those TDEs that have $L_{\text{0.2-2 keV}}/L_{g}\geq$1 and those TDEs that have multi-epoch X-ray detections.}
\label{fig:LXLO2}
\end{figure}

In the \texttt{D18} model discussed in the previous section, X-rays are only expected to appear at low inclinations or at late times, when the reprocessing envelope becomes transparent. In both cases, we expect a negative correlation between the X-ray-to-optical brightness and the optical polarisation, if the latter is due to reprocessing. On the other hand, in the CIO model of \citet{2019LuBonnerot}, optically bright (and polarised) TDEs may be either X-ray faint or X-ray bright, depending on whether the inner accretion disk is veiled or not. 
\\
\indent
We consider how the X-ray-to-optical ratio $L_{\text{X}}/L_{g}$ evolves with time, and how it behaves relative to the continuum polarisation $P_V$, in Fig. \ref{fig:LXLO}. Here, $L_{g}$ is the $g$-band optical luminosity, and $L_{\text{X}}$ is the X-ray luminosity in the 0.2--2 keV range, which we have collected from the literature and converted to the right energy range if needed, using the spectral model provided in that work (we refer to Appendix \ref{sec:AppendixD} for the references). If necessary, the detected X-ray luminosity was interpolated linearly to the time of polarimetry. For TDEs whose X-ray data were not published before, we used our own reductions (see again Appendix \ref{sec:AppendixD}) In most cases, these are upper limits, and they were converted to the 0.2--2 keV range assuming a $kT$ = 50 eV blackbody spectrum. We note that for AT~2021blz, we stack all Swift/XRT observations spanning 3-4 days over the polarisation measurements at +16 and +32 days (rather than stacking all available observations). Regarding the optical luminosity, if no $g$-band photometry is available at the time of polarimetry, we interpolated from the available bands to the $g$-band using a blackbody fit.
\\
\indent
We focus on the non-relativistic TDEs and exclude AT~2020mot. Out of the remaining TDEs, 7/15 emitted in X-rays at the time the polarimetry observations were made, although the majority of these were faint in X-rays ($L_{\text{X}}/L_{g}\lesssim$1). Figure \ref{fig:LXLO} shows that only three TDEs have $L_{\text{X}}/L_{g}$>1; of these, AT~2022exr was unpolarised at the time of the X-ray detection. Outliers are OGLE16aaa and AT~2019dsg, which were X-ray bright yet at the same time polarised. These are highlighted in the Fig. \ref{fig:LXLO2}. Considering the co-evolution of $L_{\text{X}}/L_{g}$ and $P_V$, AT~2019dsg shows a simultaneous decrease in 
$L_{\text{X}}/L_{g}$ and $P_V$, as opposed to the expected contra-evolution of these two quantities (which is shown in AT~2022dsb, for example, where the polarisation decreases as $L_{\text{X}}/L_{g}$ rises). 
In the end, this leaves 2/15 outliers in a sample of optically selected TDEs that are otherwise consistent with \texttt{D18} model when it comes to their X-ray behaviour.

\subsubsection{Infrared}
\label{sec:IR}

Infrared emission associated with TDEs can be attributed to light echoes off a dusty environment (e.g., \citealt{2024Masterson}). Constraining the dust content of the host galaxy is especially important in polarisation studies, as dust scattering can be another contributor to the optical polarisation. In AT~2023clx, \citet{2025Uno} identified two polarisation components: a $\sim$1\% level associated with the TDE, and a second component that showed a decrease from $\sim$2\% in the blue to $\sim$0\% in the red. Between two epochs of observations, they found a 90$^{\circ}$ flip in the polarisation angle (something not yet observed in other TDEs). Line depolarisation in the spectrum at +5 days, the wavelength-dependent continuum of the spectrum at +52 days, and the change in polarisation angle between the two epochs, led the authors to attribute the two epochs to different sources of polarisation. One is due to electron scattering and intrinsic to the TDE (+5 days), and one is due to dust scattering (+52 days). Specifically, they linked the second polarisation component to a light echo off a pre-existing dust plane or torus, aligned perpendicularly to the outflow (as evidenced by the 90$^{\circ}$ flip). Indeed, \citet{2024Charalampopoulos} found evidence for a dust echo based on a near-infrared excess about two months after the optical peak.

In some AGN, an increase in polarisation towards shorter wavelengths provides evidence for dust scattering off a dusty torus surrounding the supermassive black hole, owing to the larger scattering cross section at these wavelengths. However, \citet{2007Goosmann} found that dust scattering polarisation spectra are mostly wavelength-independent at optical wavelengths ($\gtrsim$2500\,Å), and that a wavelength-independent polarisation spectrum is thus not unique to electron scattering, making it difficult to disentangle contributions from dust scattering and electron scattering in optical TDEs. 
\\
\indent
Although it is thus difficult to rule out dust contributions to the optical polarisation of the TDEs in our sample, their infrared emission can give some insights into the presence of dust in their environments. Similarly to \citet{2022Leloudas}, we searched the NEOWISE \citep{2011Mainzer, 2014Mainzer} Reactivation Database for $W1-$ and $W2-$band (3.4, 4.6 $\mu$m) observations of the TDEs in our sample, and find evidence for a mid-infrared signal in AT~2018dyb, AT~2018hyz, AT~2019azh, AT~2019dsg, AT~2019qiz, and AT~2020mot, as already found by e.g., \citet{2021Jiang, 2022Leloudas, 2024VanVelzen}, and \citet{2024Newsome}. For most of these, the dust covering fractions of their host galaxies were found to be low \citep{2021Jiang,2022Leloudas}, and the null polarisation measurement of AT~2019qiz at peak light suggests that there was no contribution to the polarisation from host dust. 
\\
\indent
In addition, we have considered the 4000\,Å break index $D_{4000}$ as a probe of dust in the optical host galaxy spectra. The index $D_{4000}$ is defined as the ratio of the average flux density in the 4050--4250\,Å range over the average flux density in the 3750--3950\,Å range, where $D_{4000}\gtrsim 1.55$ is an indicator of a passive host galaxy and thus likely dust-free environment \citep{1983Bruzual, 2003Kauffmann}. Of the newly analysed TDEs in this work, AT~2020zso and AT~2021blz have indices 1.55 and 1.54. While AT~2022bdw and AT~2022dsb have lower indices, their late-time polarimetry measurements after the ISP corrections (thus expected to be governed by the host contribution only) were consistent with zero polarisation in all filter bands. AT~2023mhs has the lowest index (1.24), and in this case it is thus particularly difficult to rule out any host dust contributions to the polarisation. The flux spectra of AT~2023mhs lack evidence for AGN activity \citep{2023TNSWise}, and its infrared light curve does not have a dust echo. 
\\
\indent
Curiously, AT~2023mhs does show a slight decrease in the continuum polarisation with wavelength, while the electron scattering opacity is wavelength-independent. A wavelength dependence of the continuum polarisation may be introduced due to depolarisation by bound-free and free-free atomic transitions (whose contributions increase with wavelength; e.g., \citealt{2016Roth}). \citet{2022Leloudas} investigated this possibility for AT~2018dyb by fitting polarisation models including contributions from electron scattering plus absorptions and from electron scattering only. They found that scenarios with moderate to high absorption opacities could not be ruled out, although this is in part due to the low degrees of freedom in these fits. Possibly, such a scenario could explain the moderate drop in polarisation in the red end of the spectrum of AT~2023mhs.

\subsubsection{Radio}

9/19 TDEs are known to have radio emission, implying the presence of a relativistic jet (as in AT~2022cmc, Swift J164449.3+573451 and Swift J2058+0516) or relativistic outflows \citep{2024Cendes}. AT~2020zso, AT~2020mot, AT~2022bdw, and AT~2022dsb were detected in radio, but this emission likely stems from the underlying AGN or host and not from the transient \citep{2024Anumarlapudi}. A transient radio flare was detected in AT~2018dyb, AT~2018hyz, AT~2019azh, AT~2019dsg, AT~2019qiz, and AT~2023clx; in these cases, contributions from synchrotron emission to the optical polarisation can be ruled out based on the radio-faintness at the time of polarimetry, or a delayed launch time of the flare with respect to the optical discovery \citep{2022Leloudas,2022Cendes,2024Cendes,2024Anumarlapudi,2025Uno}. 

\section{Discussion: Focus on individual events}
\label{sec:Discussion-Indiv}

In this section, we provide more details on the newly analysed TDEs in context of the unification scenario, building on similarities between events.

\subsection{Double-peaked emission lines: AT~2018hyz and AT~2020zso}

We first examine AT~2018hyz and AT~2020zso, two TDEs with some of the highest polarisation levels in our sample, which share a similarity in terms of their optical spectra. Both events are characterised by asymmetric, double-peaked Balmer lines in their early spectra \citep{2020Gomez, 2020Short, 2020Hung, 2022Wevers}, indicating that (1) an accretion disk had already formed at early times, and (2) this disk cannot have been observed face-on. In the case of AT~2018hyz, several inclination estimates were made assuming either a circular accretion disk (30.5$^{\circ}$; \citealt{2020Short}), a slightly eccentric disk (57$^{\circ}$; \citealt{2020Hung}), or by considering the radio data; \citet{2023Sfaradi} inferred a viewing angle of $\sim$42$^{\circ}$ based on the detection of a radio flare over two years after its discovery, possibly caused by an off-axis relativistic jet. To match these inclination estimates, a disk mass of 0.03 M$_{\odot}$ or 0.10 M$_{\odot}$ is required to explain the polarisation of AT~2018hyz (see Figure \ref{fig:mainpossis}).
We conclude that AT~2018hyz was likely observed at an intermediate-to-high inclination, which could explain the faint X-rays \citep{2020Hung} in the unification scenario. It should be noted, however, that this is only a 2.5$\sigma$ detection.
\\
\indent
AT~2020zso is the only TDE with multiple epochs prior to peak light. Together with the presence of the high-ionisation lines and complete absence of X-ray emission, \citet{2022Wevers} concluded that this TDE fits well in the unification model. Through modelling, they found that the highly eccentric ($e$ = 0.97) accretion disk likely formed early (near peak light) and that it was observed at an inclination of 85$^{\circ}$. When comparing the value of $P_V = 2.49\pm 0.45$\% of AT~2020zso during its first epoch to the models in Fig. \ref{fig:mainpossis}, in particular the 0.01 M$_{\odot}$ extended-disk model, this value implies an inclination of $\sim$80$^{\circ}$, consistent with the one derived by \cite{2022Wevers}.

Interestingly, both the change in polarisation as well as the evolution of the flux spectra reveal a change in geometry. The timeline according to \citet{2022Wevers} is as follows: at -14 days prior to peak light\footnote{We note that our inferred peak time from the optical light curves (MJD 59195) differs from that of \citet{2022Wevers} (MJD 59184), which is that of the bolometric light curve. For ease of comparison, all phases in this paragraph refer to MJD 59184.}, which corresponds to our first epoch of polarimetry, X-rays are likely being reprocessed (given the presence of the Bowen lines at this time), while the accretion disk is still forming. This is consistent with the high polarisation measured at this time. After this, the polarisation experiences a steep drop, suggesting a change in geometry during this period. Based on consistent spectral fits at --3 days and +10 days, \citet{2022Wevers} note that the eccentricity of the disk and inclination remain constant at least during this period. Hence, it is possible that between --14 and --3 days, the change in polarisation traced the assembly of the accretion disk (whose geometry influences the surrounding reprocessing envelope) before reaching a stable configuration. A caveat is that the presence of the clear, double-peaked structure in the flux spectrum at +10 days suggests the photosphere had become optically thin. At this phase AT~2020zso was still quite polarised, and this is hard to explain in the reprocessing scenario if the photosphere was indeed already optically thin by this time. Unfortunately, there are no further post-peak epochs of polarimetry for this TDE, so we cannot determine at what time the polarisation would drop to zero. 

\subsection{Line depolarisation: the slow and fast AT~2021blz and AT~2022dsb}
In many aspects, the polarisation properties of AT~2021blz and AT~2022dsb (Figs. \ref{fig:specpol_AT2021blz}, \ref{fig:specpol_AT2022dsb}, \ref{fig:specpolpca}) are similar to those of AT~2018dyb and AT~2019dsg \citep{2022Leloudas}; all exhibited Bowen lines in their spectra, and depolarised broad lines, a signature of electron scattering. They show a decrease in polarisation with time. In addition, the multi-epoch spectropolarimetry points towards strong deviations from axial symmetry at early times in case of AT~2018dyb and AT~2022dsb, which later transitioned to a more symmetric system: \citet{2022Leloudas} hypothesised that this evolution traces the formation and circularisation of the accretion disk. 

Focussing on the evolution of these four TDEs, the polarisation of AT~2022dsb and AT~2019dsg evolved significantly faster than AT~2018dyb and AT~2021blz. The former two also have multiple X-ray detections, allowing for a more detailed look into the co-evolution of their polarisation and X-ray emission.

AT~2022dsb evolved quite rapidly in X-rays, exhibiting early (--14 days) soft X-ray emission which subsequently faded as the optical light curve rose \citep{2024Malyali}. Subsequently, the X-ray emission remained more or less constant while the optical emission, and also the polarisation, faded. This yields a contra-evolution between the X-ray-to-optical ratio and the polarisation, as highlighted in Fig. \ref{fig:LXLO2}, as may be expected in the \texttt{D18} unification scenario. The increase in the X-ray-to-optical ratio was preceded by a steep decrease based on the pre-peak eROSITA detection \citep{2024Malyali}. Unfortunately, no contemporaneous polarimetry was obtained to see whether the apparent co-evolution of $L_{\text{X}}/L_{\text{opt}}$ and $P_{V}$ extends to the pre-peak phase. 

If AT~2022dsb was observed face-on or at an intermediate viewing angle, as implied by the pre-peak X-ray detection, a veil of debris formed shortly after circularisation (as suggested by \citealt{2024Malyali}) could explain the strong asymmetry of the surrounding material as inferred from the polarisation at +2 days. During the second epoch (+23 days), the X-ray emission was still detectable but faint. If most X-rays are still being reprocessed at this point, the low polarisation degree at this epoch suggests the envelope has become nearly spherical over the course of three weeks. Possibly, a strong spherical but dense outflow was present from the start, and was able to restore symmetry briefly after obscuration. 
 
AT~2019dsg showed an even steeper rate of change in polarisation, but the X-ray emission evolved slower than in AT~2022dsb. It was first observed at +20 days, a delayed detection which could be due to the clearing of obscuring material present at initial times \citep{2021Cannizzaro}. Hence, a similar mechanism that causes the obscuration of the X-rays might be at play in both TDEs, just occurring at different phases. 

AT~2018dyb and AT~2021blz evolve similarly, with the polarisation dropping slowly and remaining roughly constant (at $\gtrsim$1\%) for several weeks. AT~2018dyb remained undetected in X-rays until late times, consistent with reprocessing of X-rays in a dense wind emitted from a disk viewed nearly edge-on, as concluded by \citet{2019Leloudas, 2020Holoien, 2022Leloudas}. AT~2021blz has one X-ray detection at +10 days \citep{2021ATelLiu}, but remained undetected in X-rays at +16 and +32 days (see Fig. \ref{fig:LXLO}). The polarisation $P_{V} = 1.5$\% at +10 days can be accommodated by the \texttt{POSSIS} models for inclinations $\gtrsim 30^{\circ}$, depending on the disk mass (see Fig. \ref{fig:mainpossis}). Combined with the weak X-ray emission, this implies an intermediate viewing angle for AT~2021blz, making it fit well into the unification model involving a reprocessing scenario.

\subsection{Low polarisation at peak: AT~2022bdw and AT~2019qiz}

 AT~2022bdw has a relatively low polarisation close to peak light, which could imply that it was observed almost face-on, along the pole of the accretion disk. However, this may be at odds with the lack of observed X-rays, and the presence of the high-ionisation N~III lines makes an edge-on view more likely \citep{2019Leloudas}. Combining all Swift observations of AT~2022bdw (starting from one day after peak up to $\sim$ 100 days), we obtain $L_{X}/L_{g}\lesssim 0.08$. In comparison to AT~2019qiz, which is another TDE with low polarisation at peak and which has $L_{X}/L_{g}\lesssim 0.01$, this upper limit is less constraining. This leaves open the possibility that AT~2022bdw was observed at a low to intermediate inclination, and it is known that the CIO model (which was also applied to AT~2019qiz) can produce zero or very low polarisation values at such viewing angles \citep{2023Charalampopoulos}. On the other hand, for example the 0.01, 0.1 M$_{\odot}$ compact-disk models (Fig.~\ref{fig:mainpossis}) also allow for a low polarisation at higher inclinations. Inclination estimates for AT~2022bdw are therefore inconclusive.

\subsection{Late-time X-ray brightening: AT~2022exr and AT~2023mhs}

We find that a post-peak emergence of X-rays is not uncommon in our sample of TDEs, occurring either on a timescale of a few days - weeks after optical peak (AT~2019dsg and AT~2021blz) or on a timescale of $\gtrsim$ 100 days (OGLE16aaa, AT~2019azh, AT~2019qiz, AT~2022exr, AT~2023clx, and AT~2023mhs). 

In the case of AT~2022exr, we note that its optical light curve has two peaks; similar double-peaked optical light curves have been observed for a handful of TDEs now (e.g., AT~2019avd; \citet{2021Malyali, 2022Chen}, AT~2019aalc; \citealt{2024Veres}, AT~2020vdq; \citealt{2025Somalwar}, AT~2022dbl; \citealt{2024Lin,2025Makrygianni}). These were identified as (candidate) partial TDEs, producing two or more flares, often separated by several years. 
AT~2022exr exhibited a delayed X-ray brightening $\sim$140 days and $\sim$25 days after the first and second peak, respectively \citep{2022ATelGuolo}. The null polarisation measurement is compatible with the presence of X-rays in the unification scenario, as the broadband polarimetry was obtained after the X-rays emerged several weeks earlier (Fig.\ref{fig:LXLO2}). 

Similarly, we find that AT~2023mhs was X-ray faint at early times, at the time this TDE was polarised, and the X-ray detection occurred almost a year later (see Appendix \ref{sec:AppendixD}). This is in contrast to OGLE16aaa, whose X-ray flux quickly increased after the first detection at +141 days \citep{2020Shu}, and which was still polarised at +153 days \citep{2018Higgins}. For this source, it has been debated whether the late-time brightening is due to a delayed accretion onset or reprocessing of X-rays \citep{2020Kajava, 2020Shu}. The rather large uncertainty on the polarisation does not allow us to draw strong conclusions about its origin, but it serves as a motivation for obtaining polarimetry measurements even at late times, when possible.

\section{Conclusion}
\label{sec:Conclusion}
We have added nine tidal disruption events (TDE) to an existing sample of TDEs with optical linear broadband polarimetry observations, now counting sixteen optical events in total (when excluding relativistic TDEs). Polarimetry can give insights into the photosphere geometry and (yet uncertain) emission mechanism responsible for the optical/UV radiation in TDEs, and together with multi-wavelength observations, these may be tied together under the unification scenario proposed by \citet{2018Dai}. We have analysed the new broadband- and spectropolarimetric data in detail, and discussed the polarisation properties and multi-wavelength data (X-rays, infrared, radio) of the full sample in context of TDE unification scenario. We draw the following conclusions:
\\ \\
(1) The majority of optical (non-relativistic) TDEs (15/16) are polarised at some point during their evolution, indicating an aspherical geometry and, in some cases, a deviation from axial symmetry as well. 15/16 events have continuum polarisation levels $\lesssim$6\%, which mostly lie in the range $\sim$1--2\%. For these 15 TDEs, the distribution of the maximum $V$-band polarisation has 16th, 50th, and 84th percentile values of 0.94, 1.48, and 2.20\%, respectively (including marginal detections). This is consistent with a reprocessing scenario in which an envelope is formed through outflows launched from a super-Eddington accretion disk, as in the model of \citet{2018Dai} \citep{2022Leloudas}. 
\\ \\ 
(2) Of those optical TDEs that have multiple epochs of polarimetry (10/16), 5/10 show a decrease in polarisation after maximum light, while 2/10 do not show significant changes, and 3/10 show a rise after peak. Significant changes in polarisation occur up to 70 days post-peak, with the exception of AT~2020mot. When observed after $\sim$70 days (7/16), 5/7 TDEs have intrinsic polarisation levels consistent with zero. This corresponds to the timescale on which the ejecta are expected to become optically thin to X-rays, and is compatible with a reprocessing scenario for the optical emission and a rapid formation of an accretion disk. Obscuration by debris may further contribute to the diversity seen in our sample, which can explain the polarisation behaviour in certain events.
\\ \\
(3) Of those TDEs that have spectropolarimetric data, 7/9 events have depolarised features of the same width as the corresponding broadened spectral features, which in TDEs is attributed to electron scattering. These depolarised emission lines are thus signatures of reprocessing in an electron scattering envelope around the black hole. Further constraints on the geometry and inclination of the emission line regions can be obtained from the flux spectra, in which the absence or presence of certain lines, as well as their shapes, favour face-on or edge-on inclinations.
\\ \\
(4) The three TDEs with polarisation levels $\gtrsim$7\% (AT~2020mot, and the two relativistic TDEs Swift J164449.3+573451 and Swift J2058+0516) are most likely to produce optical radiation through synchrotron emission, either generated at the shock front of a relativistic jet, or in shocks due to self-intersection of the stellar debris stream. Of the latter scenario, model predictions are currently lacking.
\\ \\
(5) When combining the optical polarisation measurements with multi-wavelength observations (especially in X-rays), we find that 14/16 TDEs can be accommodated by the unification model, e.g., constraints on the inclination from polarimetry are consistent with either detections or non-detections of X-rays that would imply a face-on or edge-on viewing angle. 
\\ \\
(6) However, TDEs that share similar behaviour in X-rays are not necessarily similar in terms of their optical polarisation. This can be due to differences in inclination and phase with respect to peak light, but also due to differences in intrinsic geometry: e.g., the shape of the disk (AT~2020zso), the symmetry of outflows (AT~2019qiz), the distribution of obscuring material if present (AT~2019dsg, AT~2022dsb), and so on. The observed diversity in polarisation properties indicates that the unification scenario may not account for the full range of behaviours seen in TDEs.
\\ \\
For future events, we emphasise the importance of obtaining multi-epoch polarimetry alongside X-ray observations. Combined with models that incorporate an explicit time dependence of the polarisation, or alternatives to the reprocessing models considered here, these would improve our understanding of the observed diversity.

\section*{Data Availability}
All VLT/FORS2 and NOT/ALFOSC polarimetric observations in this work are publicly available on the ESO Science Archive and NOT data archive. X-ray observations are publicly available from the archives of the facilities mentioned in the text. 

\begin{acknowledgements}
We would like to thank the anonymous referee for providing helpful suggestions that improved our manuscript. We thank Kohki Uno for sharing data and polarimetry values for AT~2023clx and Klaas Wiersema for checking the polarimetry of OGLE16aaa.
GL was supported by a research grant (VIL60862) from VILLUM FONDEN. MP acknowledges support from a UK Research and Innovation Fellowship (MR/T020784/1). MB acknowledges the Department of Physics and Earth Science of the University of Ferrara for the financial support through the FIRD 2024 grant. LD acknowledges the support from the National Natural Science Foundation of China and the Hong Kong Research Grants Council (N\_HKU782/23, HKU 17314822, 17305124). TEMB is funded by Horizon Europe ERC grant no. 101125877. CPG acknowledges financial support from the Secretary of Universities and Research (Government of Catalonia) and by the Horizon 2020 Research and Innovation Programme of the European Union under the Marie Sk\l{}odowska-Curie and the Beatriu de Pin\'os 2021 BP 00168 programme, from the Spanish Ministerio de Ciencia e Innovaci\'on (MCIN) and the Agencia Estatal de Investigaci\'on (AEI) 10.13039/501100011033 under the PID2023-151307NB-I00 SNNEXT project, from Centro Superior de Investigaciones Cient\'ificas (CSIC) under the PIE project 20215AT016 and the program Unidad de Excelencia Mar\'ia de Maeztu CEX2020-001058-M, and from the Departament de Recerca i Universitats de la Generalitat de Catalunya through the 2021-SGR-01270 grant. MN is supported by the European Research Council (ERC) under the European Union’s Horizon 2020 research and innovation programme (grant agreement No.~948381).
Based on observations collected at the European Southern Observatory under ESO programme(s) 199.D-0143(R), 0103.D-0350(A), 106.214S.001, 106.216C.008, 106.216C.009, 106.216C.011, 108.222Q.001, 109.23FR.001, 111.24ME.001, 112.25JQ.001, 112.25JQ.005, and 112.25JQ.008.
Based on observations made with the Nordic Optical Telescope, owned in collaboration by the University of Turku and Aarhus University, and operated jointly by Aarhus University, the University of Turku and the University of Oslo, representing Denmark, Finland and Norway, the University of Iceland and Stockholm University at the Observatorio del Roque de los Muchachos, La Palma, Spain, of the Instituto de Astrofisica de Canarias. The data presented here were obtained with ALFOSC, which is provided by the Instituto de Astrofisica de Andalucia (IAA) under a joint agreement with the University of Copenhagen and NOT.
The ZTF forced-photometry service was funded under the Heising-Simons Foundation grant \#12540303 (PI: Graham). This work has made use of data from the Asteroid Terrestrial-impact Last Alert System (ATLAS) project. The Asteroid Terrestrial-impact Last Alert System (ATLAS) project is primarily funded to search for near earth asteroids through NASA grants NN12AR55G, 80NSSC18K0284, and 80NSSC18K1575; byproducts of the NEO search include images and catalogs from the survey area. This work was partially funded by Kepler/K2 grant J1944/80NSSC19K0112 and HST GO-15889, and STFC grants ST/T000198/1 and ST/S006109/1. The ATLAS science products have been made possible through the contributions of the University of Hawaii Institute for Astronomy, the Queen’s University Belfast, the Space Telescope Science Institute, the South African Astronomical Observatory, and The Millennium Institute of Astrophysics (MAS), Chile.
This research has made use of the SVO Filter Profile Service (\url{http://svo2.cab.inta-csic.es/theory/fps/}) supported from the Spanish MINECO through grant AYA2017-84089.
This research has made use of data and/or software provided by the High Energy Astrophysics Science Archive Research Center (HEASARC), which is a service of the Astrophysics Science Division at NASA/GSFC. 
This work made use of data supplied by the UK Swift Science Data Centre at the University of Leicester, and data obtained through the High Energy Astrophysics Science Archive Research Center online service, provided by the NASA/Goddard Space Flight Center. This research has made use of data obtained from the Chandra Data Archive provided by the Chandra X-ray Center (CXC). Based on observations obtained with XMM-Newton, an ESA science mission with instruments and contributions directly funded by ESA Member States and NASA
This publication makes use of data products from the Near-Earth Object Wide-field Infrared Survey Explorer (NEOWISE), which is a joint project of the Jet Propulsion Laboratory/California Institute of Technology and the University of Arizona. NEOWISE is funded by the National Aeronautics and Space Administration.
This research has made use of the NASA/IPAC Extragalactic Database (NED) which is operated by the Jet Propulsion Laboratory, California Institute of Technology, under contract with the National Aeronautics and Space Administration.
This work has made use of data from the European Space Agency (ESA) mission
{\it Gaia} (\url{https://www.cosmos.esa.int/gaia}), processed by the {\it Gaia}
Data Processing and Analysis Consortium (DPAC,
\url{https://www.cosmos.esa.int/web/gaia/dpac/consortium}). Funding for the DPAC has been provided by national institutions, in particular the institutions
participating in the {\it Gaia} Multilateral Agreement.
\end{acknowledgements}

\bibliographystyle{aa} 
\bibliography{references} 

\begin{appendix}
\nolinenumbers

\section{Wavelet transforms of TDE spectra}
\label{sec:AppendixB}
\begin{figure}
	\includegraphics[width=\columnwidth]{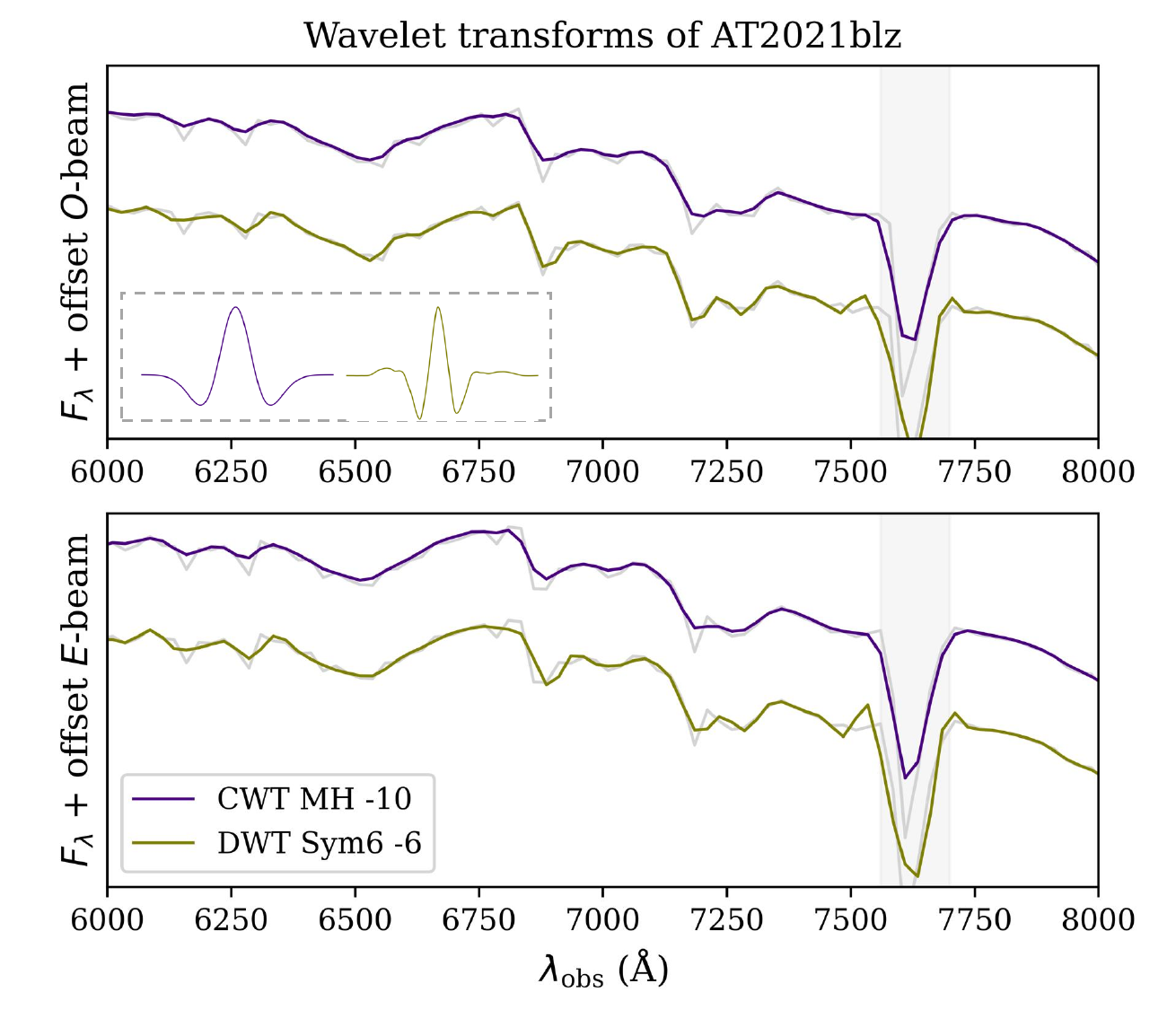}
\caption{Comparison of the continuous wavelet transform (CWT) versus the discrete wavelet transform (DWT), in this example applied to the ordinary beam (upper panel) and extraordinary beam (lower panel) of one spectrum of AT~2021blz. The original spectra (binned by 25\,Å) are shown in grey, and the grey shaded region marks the location of one telluric absorption line. We compare the reconstructed spectra de-noised via a CWT (blue) with a Mexican Hat wavelet, where the detail coefficients of the first ten levels were removed, to the reconstructed spectra de-noised via DWT (green) with the sixth-order Symlet wavelet, where the coefficients of the first six levels were removed. The inset in the upper panel illustrates the shapes of these wavelets, with the Mexican Hat wavelet on the left in blue, and the Symlet wavelet on the right in green.}
\label{fig:WT}
\end{figure}

Wavelet transforms have been used to remove noise from spectra of e.g., supernovae \citep{2010Wagers,2019Cikota}. A wavelet transform decomposes a signal into a set of so-called wavelets of finite extent, which are versions of the same mother wavelet but rescaled and translated (e.g., \citealt{1992Daubechies}, \citealt{2011GaoYan}). In the case of a continuous wavelet transform (CWT), the scale parameter and translation parameter vary continuously, while in the case of a discrete wavelet transform (DWT) they are discretised. The deconstructed signal is then expressed in terms of the base wavelets and wavelet coefficients (consisting of a set of approximation coefficients and a set of detail coefficients, corresponding to the low- and high-frequency components in the signal, respectively). By omission of the detail coefficients at the smallest scales before reconstructing the signal, the narrow, high-frequency features in the original spectrum, assumed to be noise, are then removed.
\\
\indent
We have checked the performance of either the CWT or DWT applied to the ordinary- and extraordinary-beam spectra of four TDEs when it comes to removal of noise from these spectra. For the CWT, we use the Python package \texttt{PyCWT} \citep{pycwt}, and for the DWT we use the \texttt{PyWavelets} \citep{2019Lee}. Specifically for the DWT, we use the multilevel decomposition (\cite{1989Mallat}; see also \citealt{1997FliggeSolanki}), which further decomposes the approximation coefficients and detail coefficients at different levels, until the desired scale at each level is reached. 
The choice of the mother wavelet depends on the expected shape (width, symmetry, and so on) of the features in the signal. Emission lines in TDEs are typically broad and overall symmetric, although some may have extended red wings (e.g., \citealt{2016Roth}).
We have opted for the orthogonal and near-symmetric Symlet wavelet of base six (Sym6) when applying the DWT with \texttt{PyWavelets}, and for the Mexican Hat wavelet implemented in \texttt{PyCWT} when applying the CWT. For the DWT, we omit the detail coefficients up to the sixth level (and refer to this configuration as "\texttt{DWT Sym6 --6}"), while for the CWT we set the coefficients of the lowest ten scales to zero (and refer to it as "\texttt{CWT MH --10}"). These choices are the result of inspecting the performance of the wavelet transform by eye, and they are admittedly subjective. Here we only motivate our choice of wavelet by means of an illustration; see Fig. \ref{fig:WT}. This figure shows that the \texttt{CWT MH --10} configuration captures both broad TDE emission lines and narrow telluric absorption features well, and it was found to remove an acceptable amount of noise without introducing or removing real features in either the flux spectra or the resulting polarisation spectra. Although the \texttt{DWT Sym6 --6} reconstruction looks similar, this type of wavelet transform introduces peaks near the telluric absorption feature, an issue that we also encounter in the other TDEs that have narrow lines (host emission lines or tellurics). This is likely due to the fact that this wavelet form has stronger negative peaks adjacent to the main positive peak compared to the Mexican Hat wavelet, as illustrated in the inset of Fig.~\ref{fig:WT}. Also the slightly asymmetric nature of this wavelet form is apparent. In addition, in the telluric absorption region within the broad H$\alpha$ emission line ($\sim$6800--6900\,Å), this wavelet form is more sensitive to small fluctuations between the spectra of the ordinary and extraordinary beam, which could create artificial polarisation peaks. For this reason, we have opted for the \texttt{CWT MH --10} wavelet transform in our analysis in the main text.

\section{Corrections for the interstellar polarisation and host contribution}
\label{sec:AppendixA}

\subsection{ISP corrections}
Table \ref{tab:ISP} lists the ISP estimates for those TDEs in our sample that have broadband polarimetry data, as well as the fit parameters related to the fits of our broadband ISP measurements to the Serkowski law $P(\lambda) = P_{\text{max}}\times\text{exp}\left({-K\text{ln}^2(\lambda_{\text{max}}/\lambda)}\right)$, with the parameter $K$ either kept as a free parameter, fixed to $K=1.15$, or assumed to follow the \citet{1992Whittet} relation $K =(0.10 \pm 0.05) + (1.86\pm 0.09)\lambda_{\text{max}}$, depending on the available number of data points (we use the former for AT~2022bdw, as it has broadband polarimetry in four bands, and the latter for the others).
Based on a sample of 364 stars, \cite{1975Serkowski} derived an upper limit on the Galactic ISP of $P_{\text{max}} < 9\times E(B-V)$. Using the $E(B-V)$ extinction in the direction of the TDEs in our sample \citep{2011Schlafly}, we verify that our $P_{\text{max}}$ values inferred from the Serkowski fits do not exceed this upper limit. However, \cite{1975Serkowski} found that for some stars, this upper limit might be as high as 12\% per magnitude, and more recent measurements of the Galactic ISP indicate an upper limit of $P_{\text{max}} < 13\times E(B-V)$ \citep{2019Panopoulou, 2021Hensley}. 
\\ 
\indent
In addition, we verify our derived values of the ISP with the values from the Heiles catalogue \citep{2000Heiles} within a radius of $1^{\circ}-4^{\circ}$ from each target (requiring at least three stars to be found in the catalogue). We provide the average polarisation of the found stars, their distance range (obtained by cross-matching the catalogue with Gaia DR2), and the polarisation of the most distant star. We find that in most cases, the Heiles catalogue stars are too nearby to probe the Galactic ISP along the line of sight, but we consider these to be a lower limit on the ISP in the (approximate) direction of the target.

\begin{table*}
\centering
\caption{Measurements and limits on the interstellar polarisation.}
    \begin{tabular}{c|c|ccc|ccc} \hline
     TDE & $P_{\text{max}}$ (\%) & $P_{\text{max}}$ (\%) & $P_{\text{max}}$ (\%) & Distance & $P_{\text{avg}}$ (\%) &  $P_{\text{dist}}$ (\%) &  Distance \\
      & (extinction) & (measured) & (Serkowski) & (kpc) & (Heiles) & (Heiles) & (kpc) \\ \hline \hline

     AT~2018hyz & 0.39 & ... & ...  & ... & 0.15$\pm$0.04 & 0.18$\pm$0.08 & 0.41-0.45 \\ \hline 
     AT~2019lwu & 0.429 & 0.22$\pm$0.03 (B) & ... & 0.5--4.3 & 0.05$\pm$0.01 & 0.060$\pm$0.035 & 0.05--0.18\\ \hline
     AT~2020zso & 0.780 & 0.74$\pm$0.02 (B) & 0.73$\pm$0.03 & 0.4--0.80 & 0.33$\pm$0.01 & 0.260$\pm$0.038 & 0.10--0.27 \\ 
      &  & 0.72$\pm$0.03 (B) & 0.75$\pm$0.31 & & & & \\
      &  & 0.77$\pm$0.+03 (B) & 0.72$\pm$0.19 & & & & \\ \hline
     AT~2021blz & 0.234 & 0.14$\pm$0.04 (B) & 0.11$\pm$0.13 & 0.9--2.5 & 0.065$\pm$0.035 & 0.020$\pm$0.035 & 0.18--0.19\\
      &  & 0.12$\pm$0.03(B) & 0.04$\pm$0.25 & & & & \\ \hline
     AT~2022bdw & 0.559 & 0.23$\pm$0.04 (B) & 0.21$\pm$0.04 & 1.3--1.5 & 0.06$\pm$0.02 & 0.071$\pm$0.044 & 0.15--3.27 \\ \hline
     AT~2022dsb & 2.665 & 0.18$\pm$0.01 (R) & 0.17$\pm$0.01 & 1.3--4.3 & 0.27$\pm$0.01 & 0.650$\pm$0.200 & 0.11--0.17 \\ \hline
     AT~2022exr & 0.754 & 0.05$\pm$0.02 (V) & ... & 0.5--3.6 & 0.11$\pm$0.04 & 0.12$\pm$0.04 & 0.02-0.25 \\ \hline
     AT~2022hvp & 0.91 & ... & ... & ... & ... & ... & ... \\ \hline   
     AT~2023mhs & 0.260 & ... & ...  & ... & 0.07$\pm$0.02 & 0.260$\pm$0.070 & 0.10--0.53 \\ \hline
    \end{tabular}
\tablefoot{Columns: (1) TDE name; (2) $P_{\text{max}}$ estimated from the extinction in the direction of the TDE, assuming $P_{\text{max}}= 13\times E(B-V)$; (3) maximum measured polarisation of field stars, with the FORS2 broadband filter in which it is maximum in parentheses; (4) $P_{\text{max}}$ estimated from a Serkowski fit to the multi-band polarimetry of field stars; (5) range of distances from Earth in which the used field stars lie; (6) average polarisation of stars from the Heiles catalogue; (7) polarisation of the most distant star from the Heiles catalogue; (8) range of distances from Earth in which the stars from the Heiles catalogue lie. The second epoch of AT~2020zso has been omitted from the table, as it is unreliable. AT~2018hyz, AT~2022hvp, and AT~2023mhs do not have suitable field stars to estimate the ISP, and for AT~2022hvp there were no sources in the Heiles catalogue within 4$^{\circ}$.}
\label{tab:ISP}
\end{table*}

\subsection{Host corrections}

\begin{figure}
	\includegraphics[width=\columnwidth]{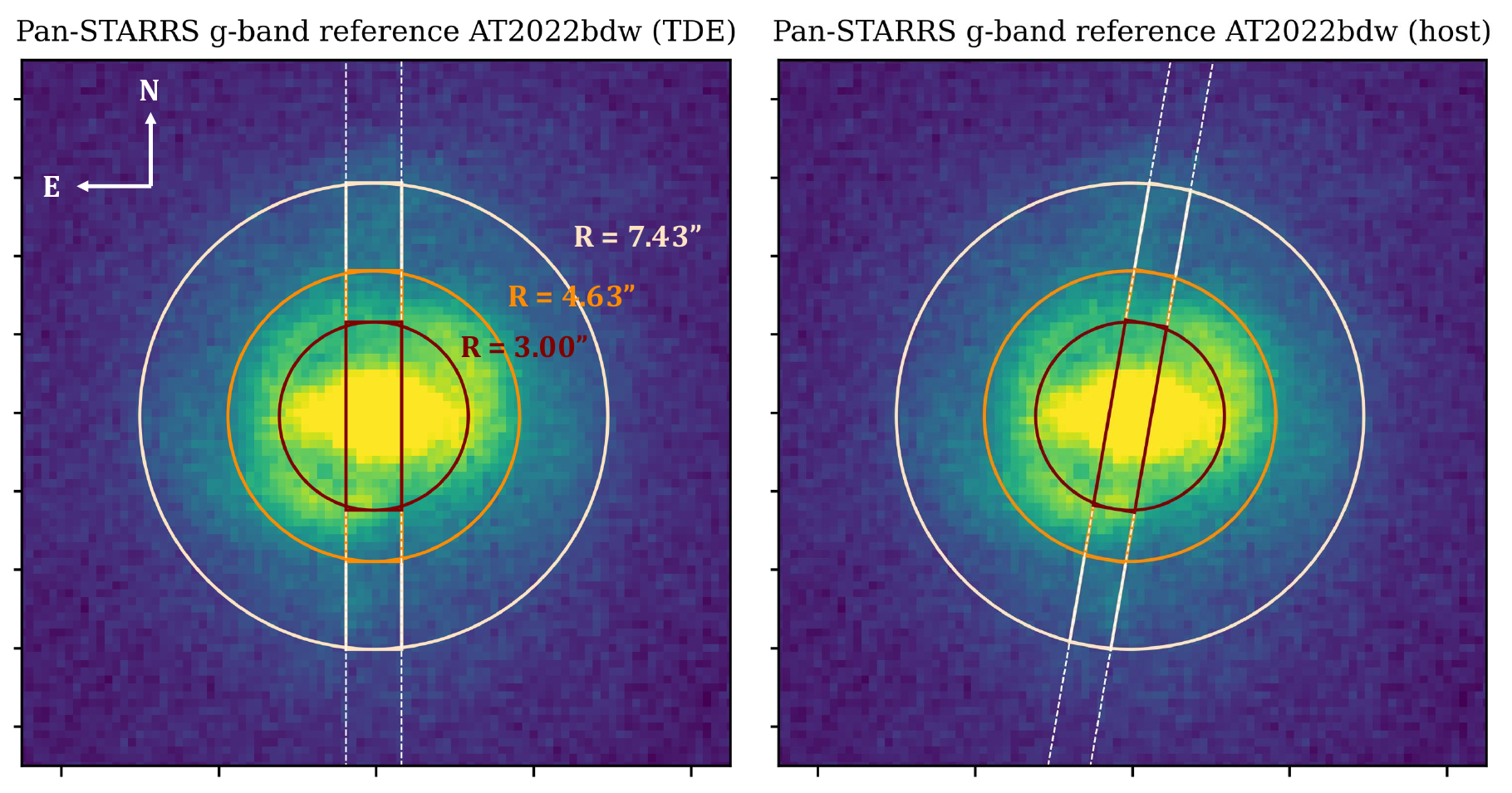}
	\includegraphics[width=\columnwidth]{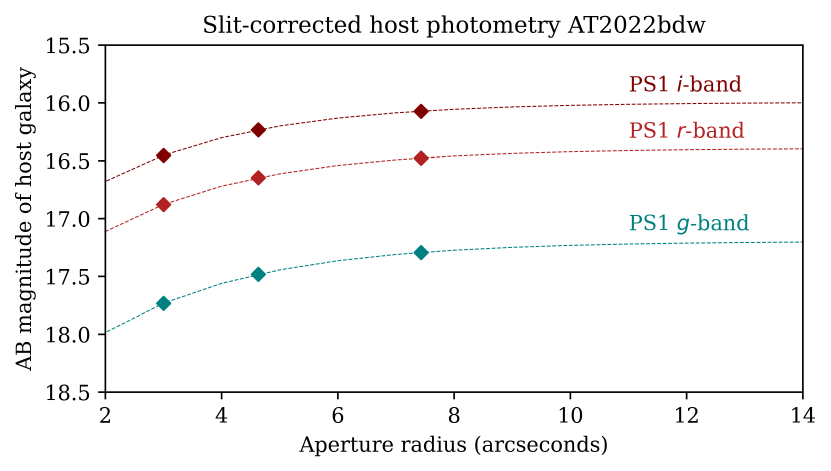}
\caption{Example of slit-loss correction applied to the host galaxy of AT~2022bdw. The top panels show the $g$-band PS1 stack image cutout of the galaxy, and overlaid are the rectangular apertures with the same slit position as at the time the TDE spectrum was taken (left) and as at the time the host spectrum was taken (right). Also plotted are the three circular apertures of radii R5, R6, and R7 that were used to calculate a slit ratio: for example, for R5 = 3.00", the slit ratio would be the number of counts in the dark red rectangular aperture divided by the number of counts in the dark red circular aperture. In the bottom panel, we show how the slit-corrected PS1 $g,r,i$ band magnitudes of the host of this TDE increase with circular aperture size, until they reach a plateau as indicated by the dashed lines.}
\label{fig:PS}
\end{figure}

We aim to obtain a reliable slit-loss-corrected apparent magnitude $m_{\text{host}}$ of each host galaxy, i.e., an apparent magnitude which captures the fraction of host galaxy flux let through the spectrograph slit when the TDE was observed. We start out with the Pan-STARRS (The Panoramic Survey Telescope and Rapid Response System; \citealt{2010Kaiser}) PS1 StackApFlx table\footnote{\url{https://outerspace.stsci.edu/display/PANSTARRS/PS1+StackApFlx+table+fields\#}}. This table contains fixed-aperture fluxes (in Jy) corresponding to aperture radii of 3.00”, 4.63”, and 7.43”, which are equivalent to the R5, R6 and R7 circular apertures used in the Sloan Digital Sky Survey (SDSS; \citealt{2020Magnier}). We use the PS1 Image Cutout Service to obtain sky-subtracted stack image cutouts in the PS1 $g,r,i$ bands of each galaxy. Using \textsc{photutils}, we place circular apertures with radii of $R$ = 3.00”, 4.63”, and 7.43” in each image, as well as a rectangular aperture which has the same position and position angle of the slit used when taking the spectrum of the TDE or of the host (information on the slit is obtained from the FITS headers of these spectra). This is illustrated in Fig. \ref{fig:PS}. To account for the differences in seeing between the Pan-STARRS images and the FORS2 / EFOSC2 spectra, we measure the FWHM of field stars in the Pan-STARRS cutouts and the FORS2 and EFOSC2 acquisition images, and we use their ratio to scale the slit width (i.e., the width of the rectangular aperture). For each aperture size, we calculate a slit loss ratio by dividing the number of counts within the rectangular aperture of height 2$\times R$ by the total number of counts within the circular aperture. We multiply the aforementioned fixed-aperture fluxes with the slit ratio, and convert them to AB magnitudes. As illustrated in Fig. \ref{fig:PS}, the apparent magnitude depends on the circular aperture size used, so we fit an exponential function to the three magnitude values to obtain the converged magnitude value, at which the "curve of growth" reaches a plateau. As a final step, we convert the slit-corrected, curve-of-growth-corrected host magnitude $m_{\text{host}}$ from the PS1 $g,r,i$ bands to the ZTF $g,r$ bands and/or ATLAS $c,o$ bands (following the conversion formulae given by \citealt{2018Tonry,2020Medford}), and subsequently combine it with the ZTF / ATLAS PSF photometry of the TDE to obtain the host-contaminated TDE photometry $m_{\text{tot}}$. $m_{\text{tot}}$ and the host photometry $m_{\text{host}}$ can then be used to scale the corresponding spectra, as described in the main text.

For broadband polarimetry, we obtain the curve-of-growth corrected magnitude of the host in the same manner, but we do not apply a slit correction. In case of AT~2023mhs, which has an archival host galaxy spectra in SDSS, we calculate the slit ratio with a circular "slit" with the same diameter as the fibre, 3". 
\\
\indent
The host of AT~2021blz was not observed by Pan-STARRS. We thus used the ATLAS photometry server to obtain the apparent magnitudes of the TDE and host at the time the TDE was ongoing, and those of the host prior to the start of the flare, for which we computed a weighted average of several epochs. We used this photometry to scale the host-contaminated TDE spectrum and the host spectrum. However, the PSF-magnitude of the host obtained this way might be unreliable due to the galaxy's extended nature. As a verification, we convolve the $\alpha(\lambda)$ spectra with ATLAS $c,o$ filter transmission curves. We then perform our own aperture photometry on the ATLAS reduced images and reference images, with which we calculate the ratio of the flux of the target in the reduced image and the flux of the host in the reference image, normalised by the same (average) ratio of several field stars. This yields an independent estimate of $\alpha$ in the ATLAS $c,o$ bands (including uncertainties), and we can compare their values at the time of polarimetry to the synthetic broadband $\alpha(c,o)$ values derived from $\alpha(\lambda)$.
We find a good agreement between the $c,o$ host contribution derived from the scaled spectra on one hand, and the $c,o$ host contribution derived from our aperture photometry on the ATLAS images on the other hand. We thus continued our host correction with $\alpha(\lambda)$ without further attempts at corrections for slit losses.
\\ \\
In what follows, we describe the next steps needed to do the host corrections in Case I -- III as defined in the main text, given that we now have a slit-loss corrected host magnitude $m_{\text{host}}$ and PSF photometry of the TDE $m_{\text{TDE}}(t)$, which we can combine to get the total host-contaminated photometry $m_{\text{tot}}(t)$. 
\\ \\
\textit{Case I}. We obtained the most recent late-time ($\geq$ 2 years after maximum light) spectra, assuming that at this time the TDE has already faded so that these spectra contain purely host light (see Table \ref{tab:obslog}). We refer to these as the host flux spectra ($I_{\text{host}}(\lambda)$). They were obtained with the EFOSC2 spectrograph mounted on the New Technology Telescope (NTT) under the ePESSTO (extended Public ESO Spectroscopic Survey for Transient Objects Survey) programme, and reduced using the ePESSTO pipeline \citep{2015Smartt}. Only in case of AT~2023mhs an archival SDSS spectrum was available, which we used instead. We extract a high S/N flux spectrum from the spectropolarimetry data itself by computing a weighted sum of the fluxes from both the ordinary and extraordinary beam from four HWP angles, and we flux-calibrate this using two archival spectroscopic standard stars (EG 274 and Feige 110) that were observed with the instrumental polarimetry setup in place. We verified this method of flux calibration with an NTT/EFOSC2 taken during the same night as the first epoch of spectropolarimetry of AT~2021blz, and found that despite the inherently imperfect flux calibration, the differences between the two spectra were negligible at wavelengths $\lambda_{\text{obs}}\gtrsim$~3800\,Å (after scaling both with photometry). We use our own extracted spectra as our host-contaminated TDE flux spectra ($I_{\text{tot}}(\lambda) = I_{\text{TDE}}(\lambda) + I_{\text{host}}(\lambda)$), and ignore the part of the spectrum below $\lambda_{\text{obs}}\sim$~3800\,Å.

We then scale the host spectrum $I_{\text{host}}(\lambda)$ with the host photometry $m_{\text{host}}$, and the host-contaminated TDE spectrum $I_{\text{tot}}(\lambda)$ with the host-contaminated TDE photometry, $m_{\text{tot}}(t_{\text{pol}})$. Here, $m_{\text{tot}}(t_{\text{pol}})$ is the light curve $m_{\text{tot}}(t)$ interpolated to the time of polarimetry, $t_{\text{pol}}$. Depending on the sampling of the light curve, this interpolation was done by fitting the multi-colour light curve with an empirical Gaussian Rise, Exponential Decay (GRED) model, or simply with a polynomial function. In each band we then compute a scaling factor to match the flux of the spectrum with the flux inferred from the photometry, and we apply a linear interpolation between scaling factors of different bands (i.e., ZTF $g,r$ and/or ATLAS $c,o$) to obtain a linear function that we scale the total spectrum with ("mangling"). We then take the ratio of the scaled spectra to calculate $\alpha(\lambda)$ directly, which allows us to correct the polarisation spectra as described before.
\\
\textit{Case II}. In this case, we used the spectra that were taken closest in time to the epoch of broadband polarimetry to obtain $I_{\text{tot}}(\lambda)$. The rest of the procedure is the same as in Case I, except that in this case we convolve $\alpha(\lambda)$ with the FORS2 filter transmission curves\footnote{\url{https://www.eso.org/sci/facilities/paranal/instruments/fors/inst/Filters/curves.html}} to obtain the host contribution in the relevant bands, $\alpha(B,V,R,I)$.
\\
\textit{Case III}. In this case, there are no spectra involved. As in Case I, we fit the host-contaminated TDE light curve and interpolate it to the time of polarimetry. Together with the host magnitude $m_\text{host}$, this gives an estimate of the value of $\alpha$ in one or more of the broadband filters of e.g., ZTF ($g,r$) and/or ATLAS ($c,o$). We convert these to the FORS2 bands through a simple linear interpolation or extrapolation of $\alpha(g,c,r,o)$ to obtain $\alpha(B,V,R,I)$.
\\
Our uncertainties on $\alpha$ include measurement uncertainties on the spectra and/or photometry, as well as standard deviations of fit parameters derived from their covariance matrix, and were computed via standard error propagation (under the assumption that all variables are independent and uncorrelated). 

\section{Supplementary figures}
\label{sec:AppendixC}

\begin{figure*}[!h]
\centering
	\includegraphics[width=0.5\columnwidth]{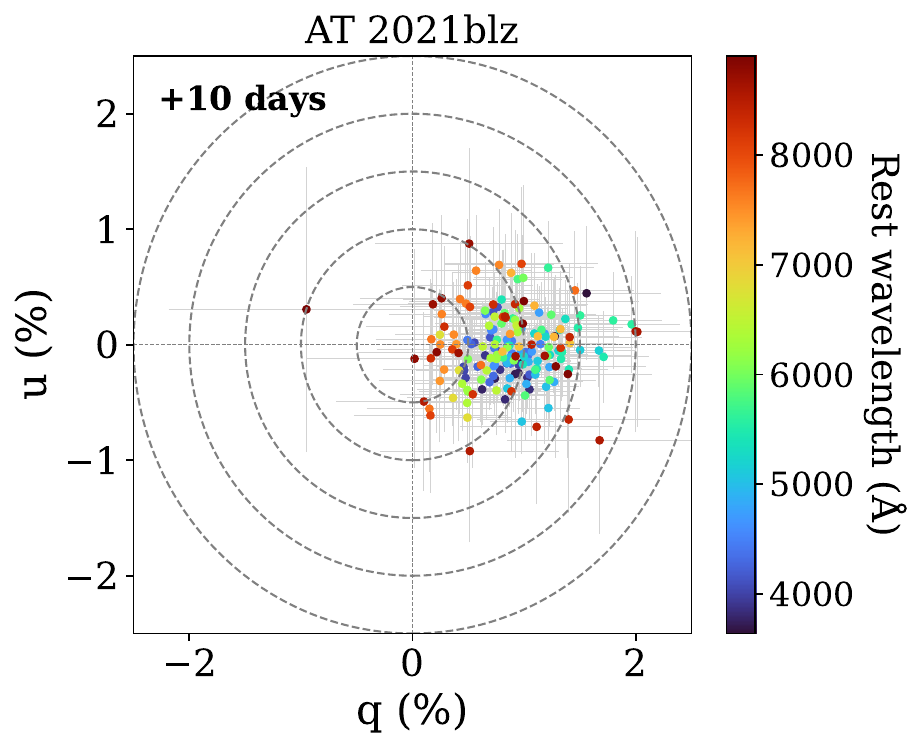}
	\includegraphics[width=0.5\columnwidth]{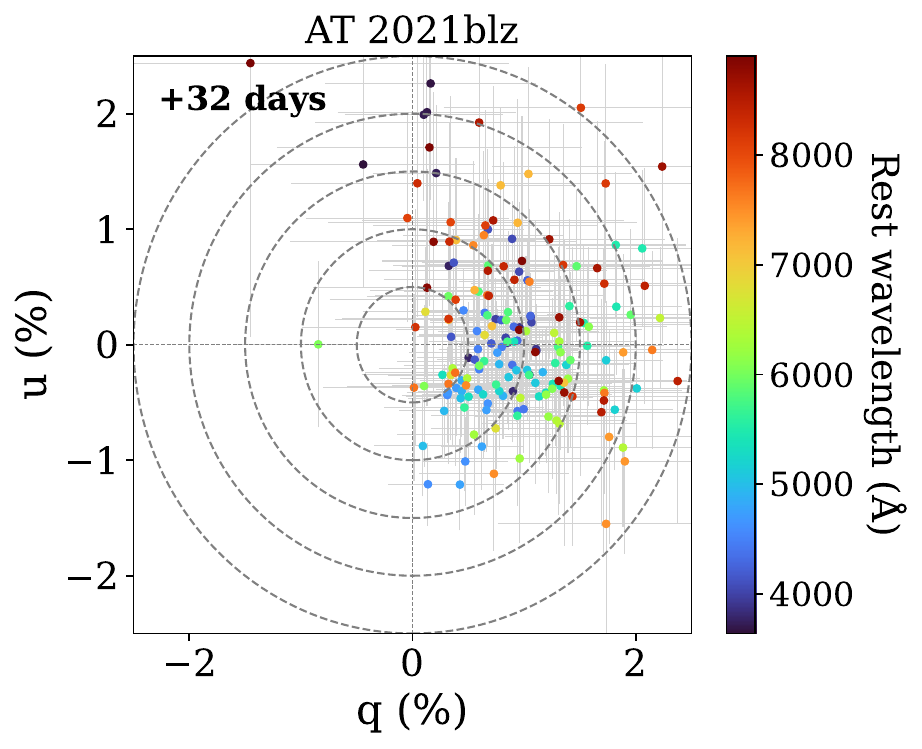}
	\includegraphics[width=0.5\columnwidth]{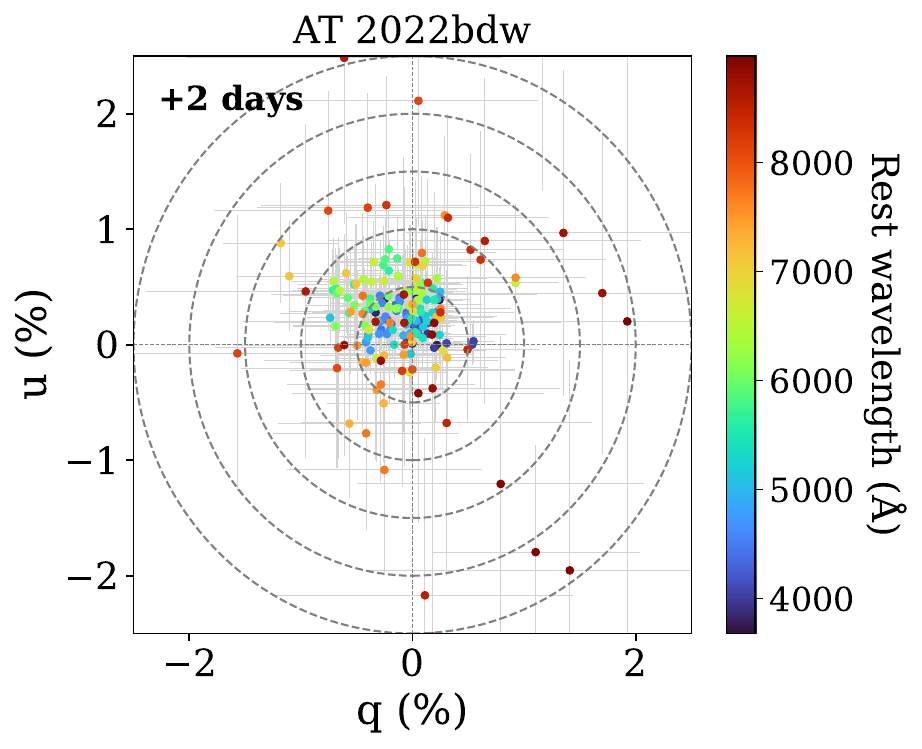}
	\includegraphics[width=0.5\columnwidth]{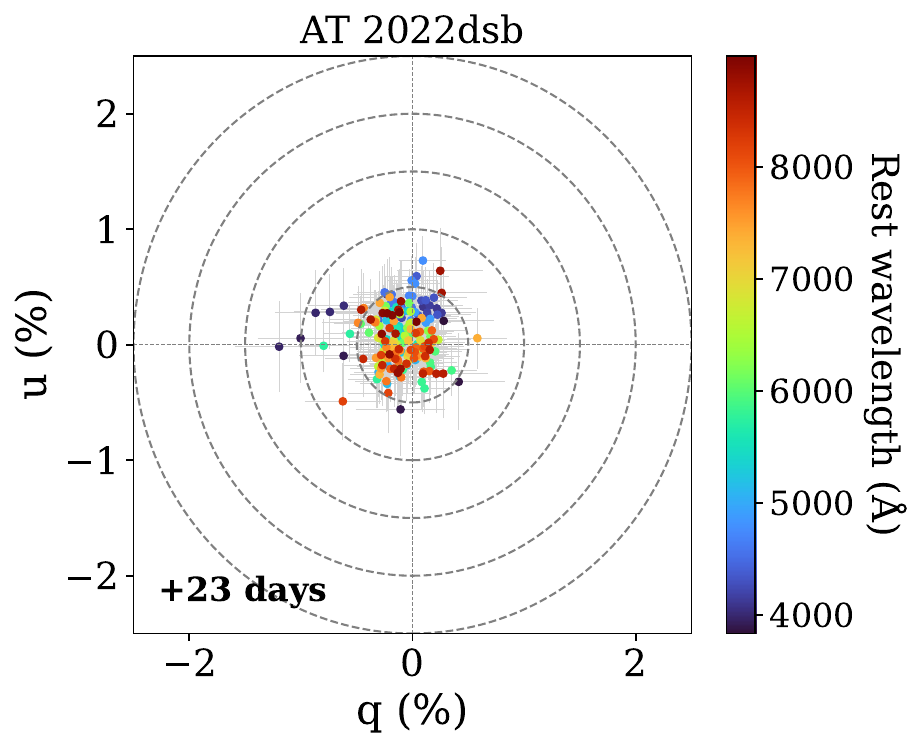}
\caption{Stokes $q,u$ plane of AT~2021blz at +10 and +32 days, AT~2022bdw at +2 days, and AT~2022dsb at +23 days.}
\label{fig:specpolpca_appendix}
\end{figure*}

\begin{figure*}[!h]
\centering
	\includegraphics[width=0.65\columnwidth]{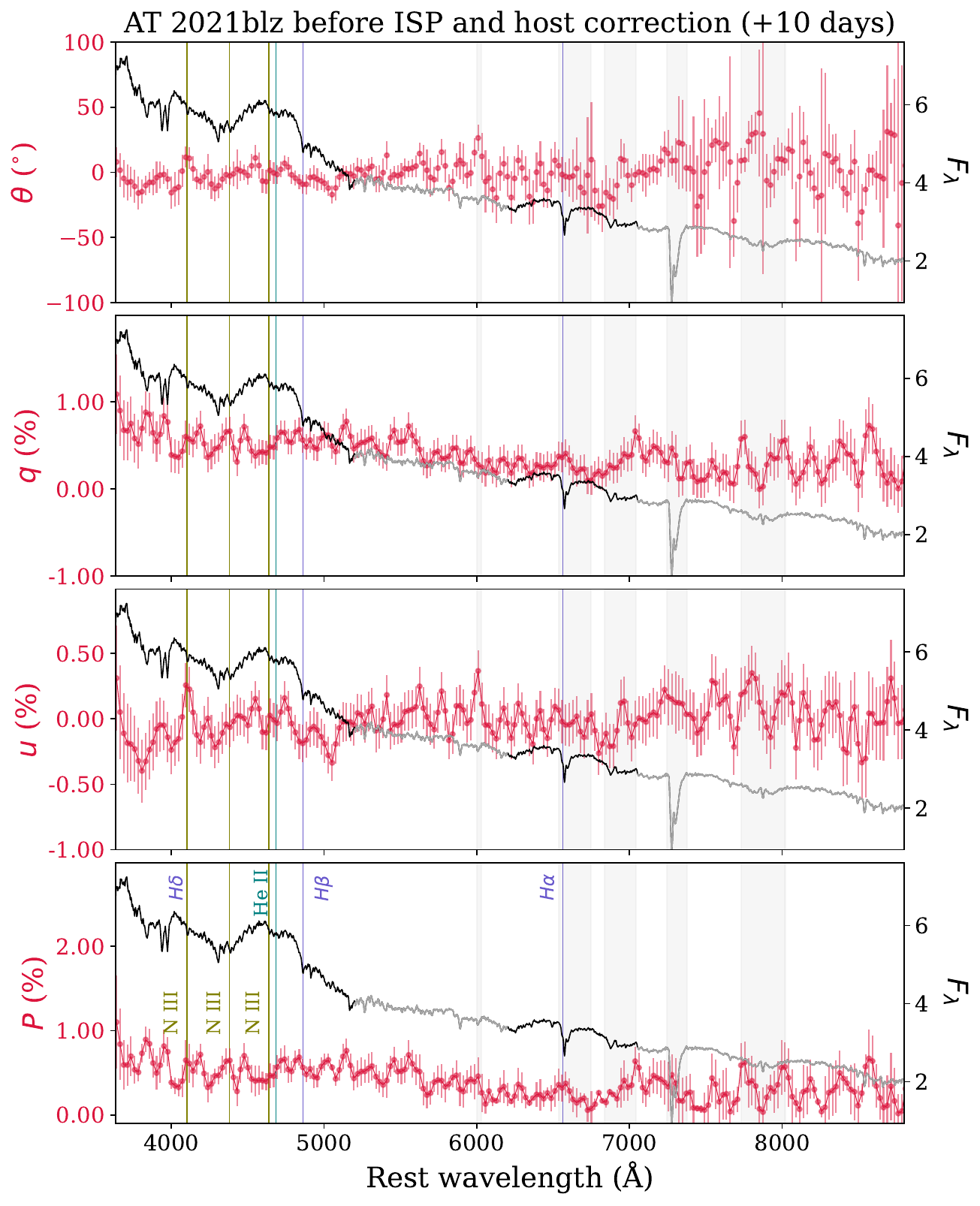}
	\includegraphics[width=0.65\columnwidth]{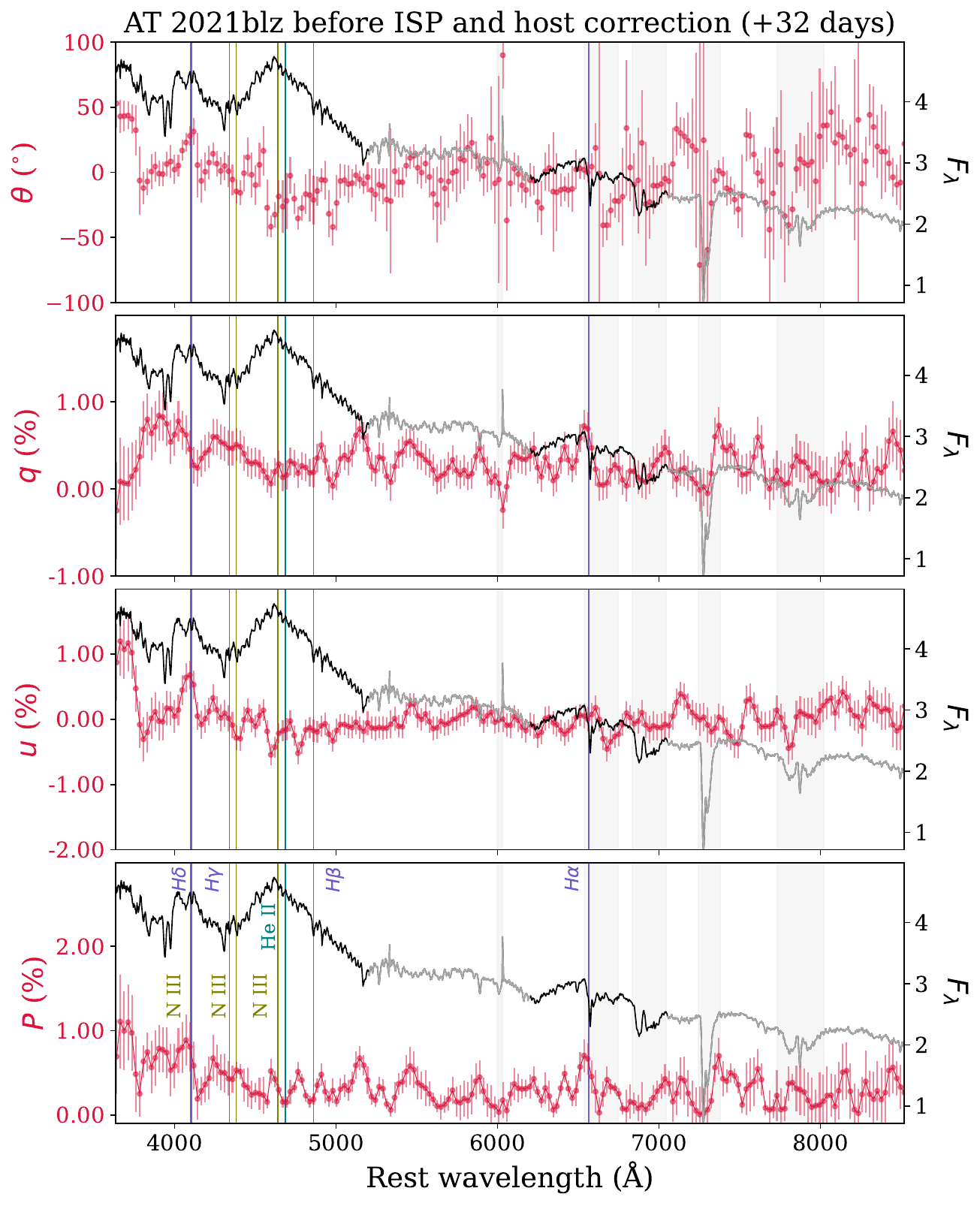}
	\includegraphics[width=0.65\columnwidth]{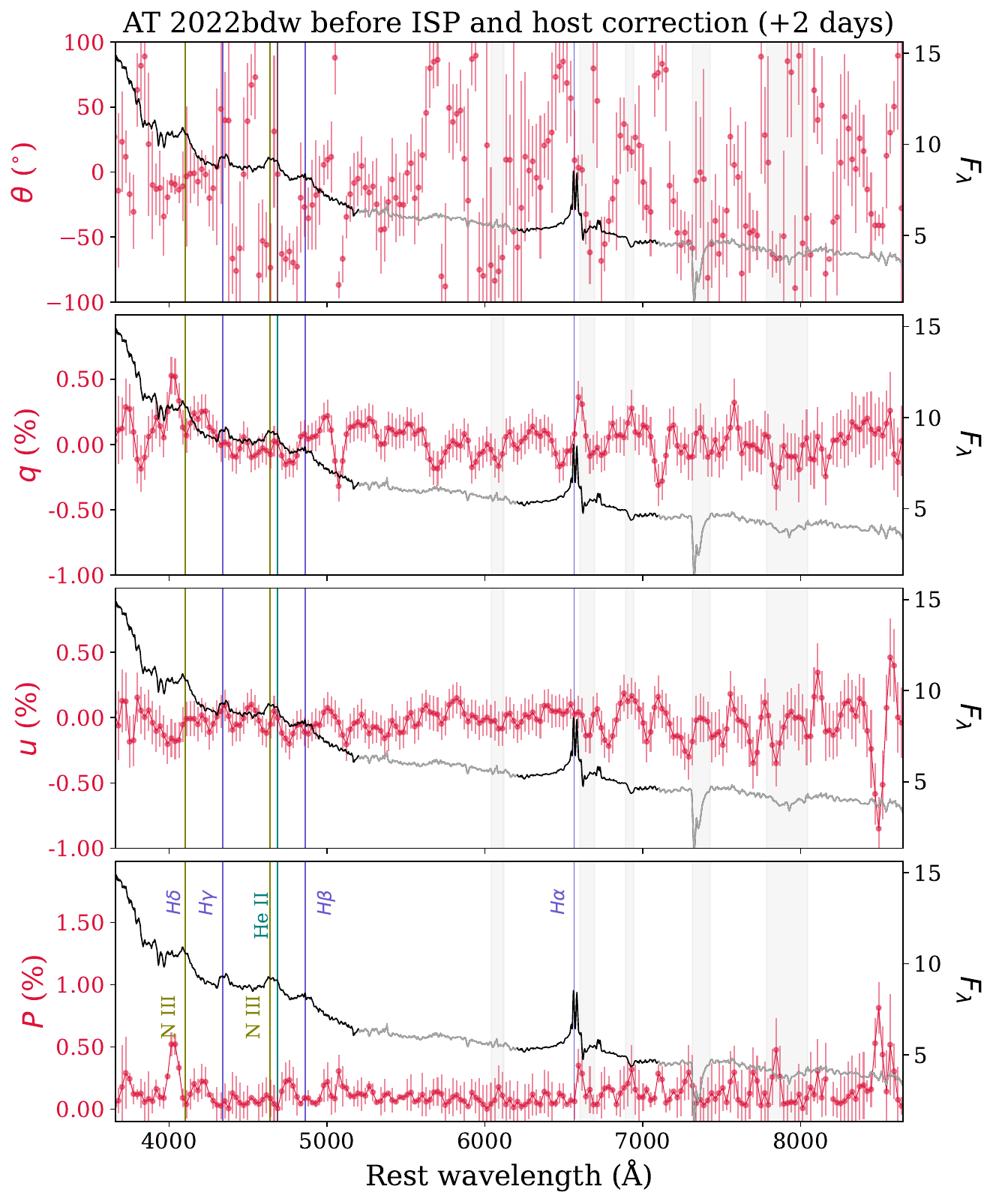}
\caption{Same as Figures \ref{fig:specpol_AT2021blz}, \ref{fig:specpol_AT2022dsb}, \ref{fig:specpol_AT2022bdw}, and \ref{fig:specpol_AT2023mhs} in the main text, but before applying the ISP and host corrections.}
\end{figure*}

\begin{figure*}\ContinuedFloat
\centering
	\includegraphics[width=0.65\columnwidth]{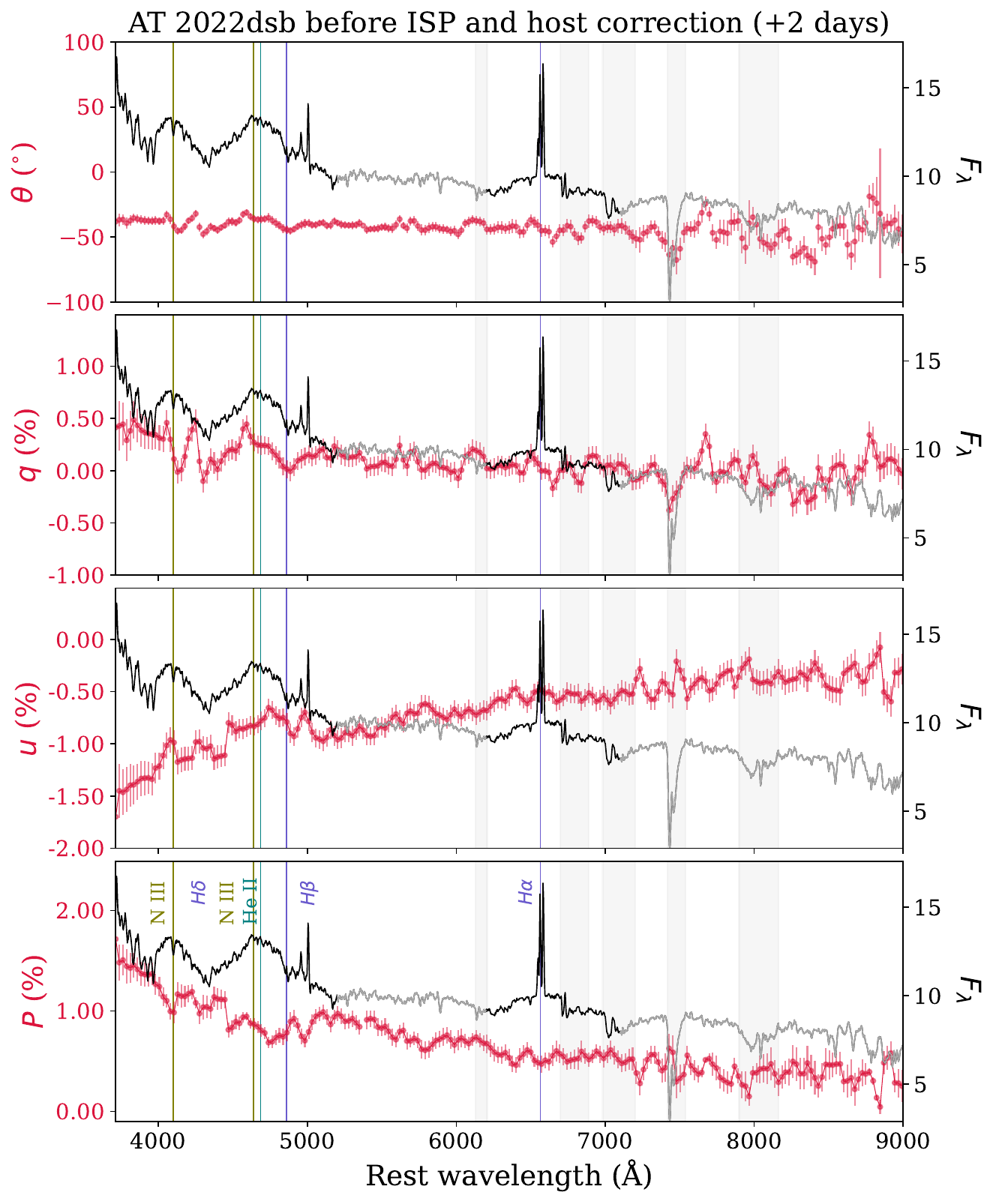}
	\includegraphics[width=0.65\columnwidth]{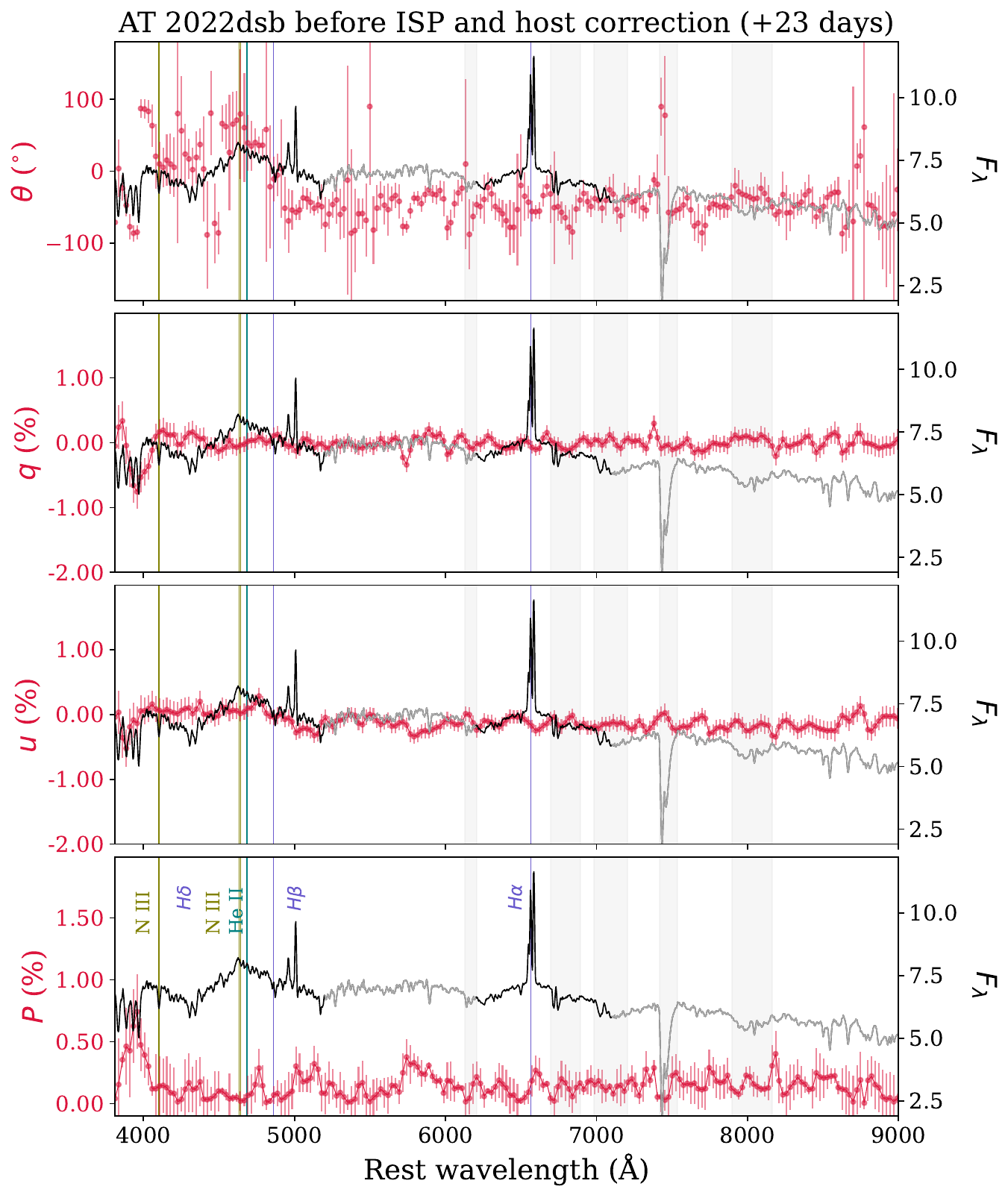}
	\includegraphics[width=0.65\columnwidth]{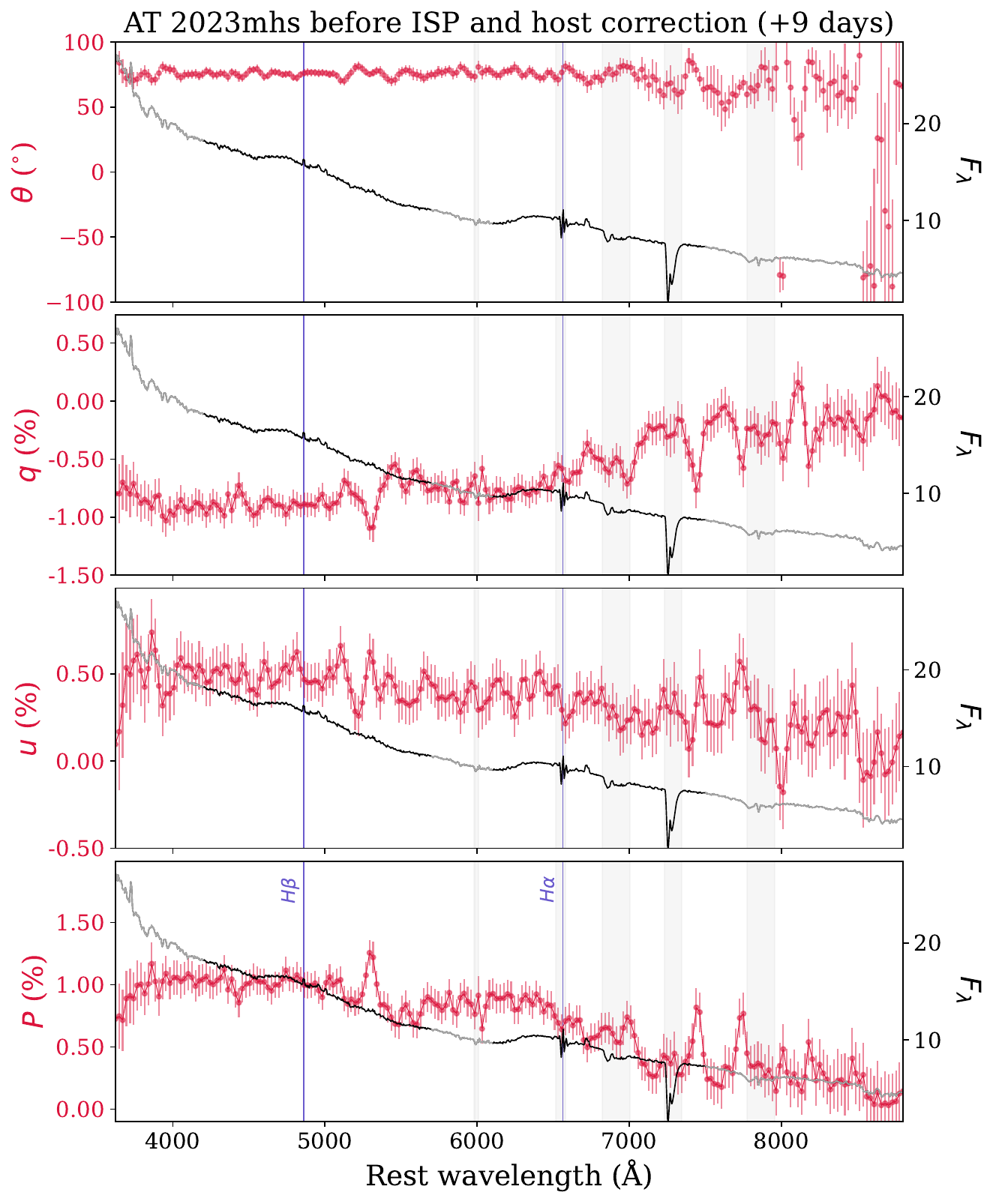}
\caption{continued.}
\label{fig:specpol_before}
\end{figure*}

\newpage
 \section{X-ray data}
\label{sec:AppendixD}

Table \ref{tab:X-rays} lists the X-ray data for all TDEs in our sample, and for different instruments. Note that we only reduce the observation epochs that are not already covered by other publications. For this reason, the table provides the relevant reference and date of the last observation covered in that work, and the data in the table include only new observations analysed by us. Conversions from count rates to fluxes are done with the online \texttt{WebPIMMS}\footnote{\url{https://heasarc.gsfc.nasa.gov/cgi-bin/Tools/w3pimms/w3pimms.pl}} tool, using the hydrogen column density $n_H$ \citep{HI4PI}, the redshift $z$, and adopting spectral fit parameters from our own fits or from the literature (in case of detections) or adopting a blackbody spectrum with typical temperature $kT$ = 50 eV \citep{2021Saxton} (in case of non-detections). The mission-specific data reduction procedures are as follows:
\\ \\
\indent
\textit{Chandra}. We searched for the available public X-ray observations of all targets in our sample using the CIAO software (v. 4.15; \citealt{2006Fruscione}) of the Chandra X-ray Observatory. Chandra/ACIS data were reprocessed using the \texttt{chandra\_repro} task in CIAO v. 4.15. The \texttt{wavdetect} task was used for source detection, \texttt{srcflux} to measure the 0.5--7.0 keV flux, and \texttt{aplimits} to measure flux upper limits. There are observations for three TDEs, AT~2018hyz, AT~2018dyb, and AT~2019qiz (Table \ref{tab:X-rays}). AT~2019qiz has observations with NICER, Swift, and XMM-Newton in addition, part of which were analysed by \citet{2024Nicholl}, who identified quasi-periodic eruptions (QPEs) in the late-time X-ray data of this TDE. While of great scientific interest, a re-analysis of these QPEs is beyond the scope of this work, and we thus omit AT~2019qiz from our analysis. 
\\
\indent
The first epoch of AT~2018hyz was analysed by \citet{2022Cendes}, who measured a 0.3--10 keV flux of 1.78$\times 10^{-14}$ erg cm$^{-2}$ s$^{-1}$ over a thousand days after peak light. We obtained the 0.5--7.0 keV count rates of four consecutive detections another two years later, and converted these to unabsorbed 0.3--10 keV fluxes using the spectral model parameters measured by \citet{2022Cendes}. This yields a time-averaged flux of $F_{\text{0.3-10 keV}}\approx 1.25\times 10^{-14}$ erg cm$^{-2}$ s$^{-1}$, i.e., we do not find evidence for any rebrightening. We found no source at the position of AT~2018dyb (\texttt{wavdetect} did detect a source, but it was 1.37" offset from the optical position of the target \citep{2019Leloudas}, outside the total 95\% confidence positional uncertainty of 0.61"). We place a 90\% detection limit of $3.9\times 10^{-4}$ cts/s given a background rate of $5.1\times 10^{-4}$ cts/s.
\\
\indent
\textit{NICER}. We extracted the cleaned event lists with the \texttt{nicerl2} task of the NICER Data Analysis Software (HEASoft 6.34), using the calibration data release of February 2024. We merged the observations for each TDE with the \texttt{niobsmerge} task, and subsequently ran the \texttt{nicerl3-spect} and \texttt{nicerl3-lc} tasks with default parameters to first generate spectra and 0.4--1.0 keV light curves, for which we adopted the Scorpeon background model and a time bin size of 32s. In case of non-detections, we calculate the upper limit on the source count rate as CR$_{\text{tot}}$ + 3$\sigma_{\text{tot}}$, where CR$_{\text{tot}}$ is the total count rate and $\sigma_{\text{tot}}$ its corresponding error. In the cases of AT~2018hyz, AT~2022dsb, AT~2022exr, and AT~2023mhs additional (non-standard) filtering was applied due to strong flaring in the background and source count rate. We set the cut-off rigidity to >~4 Gev/c and the average overshoot rate per detector to $\leq$ 5 cts/s. A source is detected for AT~2022exr and AT~2023mhs, and we report the background-subtracted peak count rate in the table. Owing the to the scatter in the light curves of AT~2022exr and AT~2023mhs, they were re-binned by 1 ks. For the same reason, we omit a spectral fit and instead rely on the spectral fitting from the XMM-Newton detections (see the next paragraph).
\\
\indent
\textit{XMM-Newton}. We first processed the EPIC-pn data of all TDEs using the standard \texttt{epproc} task of the XMM-Newton Science Analysis System (SAS v. 22.1.0). The \texttt{tabgtigen} task was used to select time intervals devoid of background flaring in the 10--12 keV range (when the background count rate is below 0.4 cts/s), and the \texttt{evselect} task was used to create filtered event lists and images. In doing so, we followed the standard event screening recommended by the SAS analysis threads\footnote{\url{https://www.cosmos.esa.int/web/xmm-newton/sas-threads}}.
We extracted a source spectrum with a 30" aperture centred on the optical position of the TDE, and a background spectrum via a 50" aperture located on the same detector as the source. In the case of detections, the \texttt{evselect} task was again used to create 0.3--1.0 keV light curves (which were corrected for several instrumental effects with the \texttt{epiclccorr} task) and spectra (which were regrouped using the \texttt{specgroup} command to have at least one count per spectral channel, and an oversampling factor of 3). In the case of non-detections, we used the \texttt{eupper} task to obtain 3$\sigma$ upper limits for each observation. 
\\
\indent
We note that we rejected the single observation of AT~2020mot as the full observation was affected by background flaring. We detected a source in the first two XMM-Newton epochs of AT~2022exr, and for the single epoch of AT~2023mhs, for which we then also analysed the available EPIC-MOS data (again following the recommended procedures for MOS). We analysed the spectra separately in XSPEC (v.12.14.1), adopting the \texttt{TBabs*zashift*diskbb} model to perform a spectral fit. We froze the hydrogen column density and redshift. In the case of AT~2022exr, our best-fit ($C$-stat/dof = 27.62/25 -- 34.72/19) inner disk temperature is 86.1 $\pm$0.9 eV and 82.2 $\pm$1.0 eV (90\% confidence interval) for the first and second epoch, respectively. In the case of AT~2023mhs, we obtain $T_{\text{in}} = 115\pm 2.3$ eV ($C$-stat/dof = 14.4/19--48.7/25).
\\
\indent 
\textit{Swift}. We used the online Swift/XRT product building software \citep{2007Evans, 2009Evans} to create 0.3--1.0 keV light curves and images. Since we are considering late-time data, we found mostly non-detections for the TDEs in our sample. We used the XIMAGE \texttt{uplimit} task and an extraction aperture radius of 50" to calculate upper limits from the combined image (calculated with the Bayesian method; \citealt{1991Kraft}). The exception is AT~2022exr, for which again the background-subtracted count rate is listed in the table.

\longtab[1]{
\footnotesize
%\centering
\begin{longtable}{llllllll}
\caption{X-ray detections and upper limits of TDEs from different facilities.}
\label{tab:X-rays}
\\ \hline
TDE & $z$ & $N_{\text{H}}$ & ObsID & Instrument & Phase & Count rate & Reference \\
& & & & & (days) & (cts/s) &  \\ \hline \hline
\endfirsthead
\caption{continued.}\\
\hline
TDE & $z$ & $N_{\text{H}}$ & ObsID & Instrument & Phase & Count rate & Reference \\
& & & & & (days) & (cts/s) &  \\
\hline \hline
\endhead

\endfoot

    OGLE16aaa & 0.1655 & 2.7 & 00097673001 & Swift/XRT & +1481 & < 3.11$\times 10^{-3}$ & \citet{2020Shu} \\ 
    & & & & & & & 2020-02-09 \\ \hline
     AT~2018dyb & 0.0180 & 17.8 & 26789 & Chandra/ACIS & +1823 & < $3.9\times 10^{-4}$ &  \citet{2020Holoien} \\ 
      & & & 7204610101-6 & NICER/XTI & +2149 to +2154 & < 5.3$\times 10^{-1}$ & 2019-10-29 \\
      & & & 00010764001--60, & Swift/XRT & --23 to +2278 & < 6.42$\times 10^{-4}$ &  \\
      & & & 00095143001--14, & & & &  \\ 
      & & & 00096021001, & & & &  \\ 
      & & & 00096579001--2, & & & &  \\ 
      & & & 00097627001, & & & &  \\ 
      & & & 03105209001 & & & &  \\ \hline 
     AT~2018hyz & 0.0457 & 2.6 & 26647, & Chandra/ACIS & +1915 & 5.32$\times 10^{-4}$ & \citet{2022Cendes}  \\
      & & & 29279--81 & & & & \\
      & & & 7204520101--5 & NICER/XTI & +2047 to +2051 &  < 3.8$\times 10^{-1}$ & 2022-03-19 \\ 
      & & & 03111719001--17, & Swift/XRT & +1436 to +2248 & < 1.91$\times 10^{-3}$ & \\ 
      & & & 00096580001--2, & & & & \\ 
      & & & 00097571001--2 & & & & \\ \hline
     AT~2019azh & 0.0220 & 4.2 & 00097565001, & Swift/XRT & +1903 to +2174 & < 1.11$\times 10^{-2}$ &  \citet{2024Guolo}\\ 
      & & & 00097565003 & & & &  2023-04-15  \\
      & & & 842592601 & XMM/EPIC & +412 & < 4.55$\times 10^{-2}$ & \\ \hline 
     AT~2019dsg & 0.0512 & 6.4 & 2200680101-12, & NICER/XTI & +20 to +1911 & < 2.4$\times 10^{-1}$ & \citet{2021Stein} \\
      & & & 7200680101-3 & & & & 2019-10-23 \\
      & & & 00011396035, & Swift/XRT & +182 to +1881 & < 6.40$\times 10^{-4}$ &  \\ 
      & & & 00011396037--44, & & & & \\
      & & & 00096583001--2, & & & & \\
      & & & 00097573002--3, & & & & \\
      & & & 00097573005, & & & & \\ \hline
     AT~2019lwu & 0.1170 & 3.3 & 7204830101--3  & NICER/XTI & +1852 to +1855  & < 2.8$\times 10^{-1}$ & ... \\
      & & & 00011520001, & Swift/XRT & +23 to +50 & < 1.49$\times 10^{-3}$ & \\
      & & & 00011520003, & & & & \\ 
      & & & 00011520005--7 & & & & \\ \hline
     AT~2020mot & 0.070 & 5.7 & 00013647015--16,& Swift/XRT & +586 to +1638 & < 8.57$\times 10^{-4}$ & \citet{2023Liodakis} \\
     & & & 00097579001--3 & & & & 2021-04-04 \\
     & & & 894200101 & XMM/EPIC &  & -- &  \\ \hline
     AT~2020zso & 0.0563 & 5.5 & 00013884001--2, & Swift/XRT & --22 to +1216 & < 8.05$\times 10^{-4}$  &  \citet{2022Wevers} \\
      & & & 00013884004--18, & & & & 2021-05-21\\
      & & & 00097584001 & & & & \\
      & & & 902760201 & XMM/EPIC & +517 & < 1.09$\times 10^{-2}$ & \\ \hline 
     AT~2021blz & 0.0450 & 1.5 & 3201990101 & NICER/XTI & +3 & < 1.2$\times 10^{-1}$ &  \citet{2021ATelLiu} \\
      & & & 00014054001--26, & Swift/XRT & +11 to +1164 & < 5.49$\times 10^{-4}$&  2021-02-12 \\
      & & & 00096588001--3, &  & & & \\
      & & & 00097637001--3 &  & & & \\ \hline
     AT~2022cmc & 1.193 & 0.9 & ... & Swift/XRT &  & ... &  \citet{2024Eftekhari}\\
      & & & &  & & & 2023-03-29 \\ \hline
     AT~2022bdw & 0.0378 & 4.7 & 4202590101-2 & NICER/XTI & --3 to --2  & < 1.9$\times 10^{-1}$  & ... \\ 
      & & & 00015033001--18, & Swift/XRT & +1 to +1123 & < 5.62$\times 10^{-4}$ &  \\
      & & & 00097596001 & & & &  \\ \hline
     AT~2022dsb & 0.0230 & 1.1 & 5202600101--10 & NICER/XTI & +3 to +52 & < 2.4$\times 10^{-1}$  & \citet{2024Malyali}  \\ 
      & & & 00015054002--3 & Swift/XRT & --1 to +932 & < 1.42$\times 10^{-3}$ &  2022-08-29 \\
      & & & 00015054005--19 & & & &  \\
      & & & 00097629001--2 & & & &  \\
      & & & 0892200701 & XMM/EPIC-pn & +332 & < 2.83$\times 10^{-2}$ &  \\
      & & & 0892200801 & XMM/EPIC-pn & +516 & < 8.12$\times 10^{-2}$ &  \\ \hline
     AT~2022exr & 0.0960 & 5.8 & * & NICER/XTI & +144 to +728 & 1.84$\pm$0.05 & \citet{2022ATelGuolo} \\ 
      & & & 00015126010--22, & Swift/XRT & +138 to +954 & 6.00$\pm$0.93$\times 10^{-2}$ & 2022-08-22 \\ 
      & & & 00015348001--111, & & & &  \\
      & & & 00097598001--3 & & & &  \\
      & & & 0882591701 & XMM/EPIC-pn & +159 & $5.41_{-0.09}^{+0.05}\times 10^{-1}$ &  \\ 
      & & &  & XMM/EPIC-MOS1 &  & $9.15_{-0.06}^{+0.01}\times 10^{-2}$ &  \\ 
      & & &  & XMM/EPIC-MOS2 &  & $2.11_{-0.54}^{+0.16}\times 10^{-2}$ &  \\ 
      & & & 0882591801 & XMM/EPIC-pn & +185 & $6.21_{-0.16}^{+0.06}\times 10^{-1}$ &  \\ 
      & & &  & XMM/EPIC-MOS1 &  & $1.09_{-0.07}^{+0.02}\times 10^{-1}$ &  \\ 
      & & &  & XMM/EPIC-MOS2 &  & $1.13_{-0.09}^{+0.02}\times 10^{-1}$ &  \\ 
      & & & 0882592201 & XMM/EPIC-pn & +494 & < 5.49$\times 10^{-2}$ &  \\ \hline
     AT~2022hvp & 0.1200 & 1.0 & 5202770101--2 & NICER/XTI & --3 to --4 & < 3.5$\times 10^{-1}$ & ... \\
      & & & 00097600001--2 & Swift/XRT & +705 to + 840 & < 1.99$\times 10^{-3}$ &  \\
      & & & 0892200201 & XMM/EPIC & +147 & < 5.05$\times 10^{-2}$ &  \\
      & & & 0892201301 & XMM/EPIC & +147 & < 1.09$\times 10^{-2}$ &  \\ 
      & & & 0892200901 & XMM/EPIC & +197 & < 1.54$\times 10^{-2}$ &  \\ 
      & & & 0892201401 & XMM/EPIC & +197 & < 1.17$\times 10^{-2}$ & \\
      & & & 0892201001 & XMM/EPIC & +329 & < 1.07$\times 10^{-2}$ &  \\ \hline
     AT~2023clx & 0.0111 & 2.5 & 00015897001--6, & Swift/XRT & +2 to +367 & < 6.75$\times 10^{-4}$ &  \citet{2024Hoogendam} \\ 
      & & & 00015897008--11, &  &  &  & 2023-05-31\\
      & & & 00015897013--39 &  &  &  & \\ 
      & & & 0892201701 & XMM-EPIC & +289 & < 1.01$\times 10^{-2}$ &   \\ \hline
     AT~2023mhs & 0.0482 & 1.6 & 6204120101--22, & NICER/XTI & 0 to +360 & 3.57$\pm$0.04$\times 10^{-1}$ & ... \\
     & & & 7204120101--44 & & & & \\
     & & & 16109001--16, & Swift/XRT & --2 to +408 & < 2.08$\times 10^{-3}$ & \\  
     & & & 16109018--21, & & & &\\
     & & & 49471001--19, & & & & \\
     & & & 97606001--5 & & & & \\
     & & & 0935191201 & XMM/EPIC-pn & +358 & $7.94_{-0.08}^{+0.24}\times 10^{-2}$  &  \\ 
     & & & & XMM/EPIC-MOS1 &  & $1.52_{-0.04}^{+0.15}\times 10^{-2}$ &  \\ 
     & & & & XMM/EPIC-MOS2 &  & $1.35_{-0.04}^{+0.14}\times 10^{-2}$ &  \\ \hline  
     Sw 1644+57 & 0.354 & 1.6 & 13509001--19, & Swift/XRT & +2690 to +5134 & < 4.26$\times 10^{-3}$ & \cite{2021bCendes} \\ 
     & & & 13509021--33, & & & & 2018-08-06 \\ \hline
     Sw 2058+05 & 1.18 & 3.8 & ** & Swift/XRT & +4707 to +4776 & < 3.25$\times 10^{-3}$ & \citet{2015Pasham}\\
     & & & & & & & 2013-11-01 \\

\end{longtable}
\tablefoot{Columns: (1) TDE name, (2) redshift, (3) Galactic column density in units of $10^{20}$ cm$^{-2}$, (4) observation ID(s), (5) mission and instrument, (6) phase with respect to the observed optical maximum light, (7) count rate, (8) date of most recent published epoch and corresponding reference. For NICER, count rates are given in the 0.4--1.0 keV band, for Swift and XMM-Newton in the 0.3--1.0 keV band, and for Chandra in the 0.5--7.0 keV band. Swift J164449.3+573451 and Swift J2058+0516 have been abbreviated to Sw 1644+57 and Sw 2058+05.
\\
* Due to the large number of NICER observations of AT~2022exr, we only provide the date range in this table; we have considered all observations taken between 2022-08-28 and 2025-05-11.
\\
** Similarly for Swift J2058+0516, we have included Swift observations taken between 2018-08-06 and 2025-05-11.}
}

\end{appendix}

\end{document}